\documentclass[english]{JHEP3}
\usepackage[T1]{fontenc}
\usepackage[latin9]{inputenc}
\usepackage{array}
\usepackage{rotating}
\usepackage{amssymb}
\usepackage[numbers]{natbib}

\makeatletter

\providecommand{\tabularnewline}{\\}

\pdfoutput=1

\usepackage{amsmath,amssymb,amsthm}%,amscd}
\newcommand{\CB}{{\cal B}}

\newcommand{\CH}{{\cal H}}
\newcommand{\CI}{{\cal I}}

\newcommand{\CS}{{\cal S}}

\def\BN{{\mathbb N}}
\def\BZ{{\mathbb Z}}
\def\BR{{\mathbb R}}
\def\BC{{\mathbb C}}
\def\BP{{\mathbb P}}

\def\BS{{\mathbb S}}

\newcommand{\be}{\begin{equation}}
\newcommand{\ee}{\end{equation}}
\newcommand{\ba}{\begin{aligned}}
\newcommand{\ea}{\end{aligned}}
\newcommand{\bea}{\begin{eqnarray}}
\newcommand{\eea}{\end{eqnarray}}
\newcommand{\bean}{\begin{eqnarray*}}
\newcommand{\eean}{\end{eqnarray*}}

\def\r{\right\rangle}

\def\1{\mathbf{1}}
\def\0{|\1\r}
\def\im{{\mathbb{I}}{\mathrm{m}}}
\def\re{{\mathbb{R}}{\mathrm{e}}}

\newcommand{\rme}{{\rm e}}
\newcommand{\rmi}{{\rm i}}
\newcommand{\rmd}{{\rm d}}

\def\XXint#1#2#3{{\setbox0=\hbox{$#1{#2#3}{\int}$}
     \vcenter{\hbox{$#2#3$}}\kern-.5\wd0}}

\newdimen\tableauside\tableauside=1.0ex
\newdimen\tableaurule\tableaurule=0.4pt
\newdimen\tableaustep
\def\phantomhrule#1{\hbox{\vbox to0pt{\hrule height\tableaurule width#1\vss}}}
\def\phantomvrule#1{\vbox{\hbox to0pt{\vrule width\tableaurule height#1\hss}}}
\def\sqr{\vbox{%
  \phantomhrule\tableaustep
  \hbox{\phantomvrule\tableaustep\kern\tableaustep\phantomvrule\tableaustep}%
  \hbox{\vbox{\phantomhrule\tableauside}\kern-\tableaurule}}}
\def\squares#1{\hbox{\count0=#1\noindent\loop\sqr
  \advance\count0 by-1 \ifnum\count0>0\repeat}}
\def\tableau#1{\vcenter{\offinterlineskip
  \tableaustep=\tableauside\advance\tableaustep by-\tableaurule
  \kern\normallineskip\hbox
    {\kern\normallineskip\vbox
      {\gettableau#1 0 }%
     \kern\normallineskip\kern\tableaurule}%
  \kern\normallineskip\kern\tableaurule}}
\def\gettableau#1{\ifnum#1=0\let\next=\null\else
\squares{#1}\let\next=\gettableau\fi\next}
\preprint{
{\small{\texttt{arXiv:1308.1115[hep-th]}}}}

\title{Nonperturbative Ambiguities and the Reality of Resurgent Transseries}

\author{In\^es Aniceto and Ricardo Schiappa
\\
CAMGSD, Departamento de Matem\'atica, Instituto Superior T\'ecnico,\\
Av. Rovisco Pais 1, 1049--001 Lisboa, Portugal\\
\\
\email{ianiceto@math.ist.utl.pt}, \quad
\email{schiappa@math.ist.utl.pt}

}
\abstract{
In a wide range of quantum theoretical settings---from quantum mechanics to quantum field theory, from gauge theory to string theory---singularities in the complex Borel plane, usually associated to instantons or renormalons, render perturbation theory ill--defined as they give rise to nonperturbative ambiguities. These ambiguities are associated to choices of an integration contour in the resummation of perturbation theory, along (singular) Stokes directions in the complex Borel plane (rendering perturbative expansions non--Borel summable along any Stokes line). More recently, it has been shown that the proper framework to address these issues is that of resurgent analysis and transseries. In this context, the cancelation of all nonperturbative ambiguities is shown to be a consequence of choosing the transseries median resummation as the appropriate family of unambiguous real solutions along the coupling--constant real axis. While the median resummation is easily implemented for one--parameter transseries, once one considers more general multi--parameter transseries the procedure becomes highly dependent upon properly understanding Stokes transitions in the complex Borel plane. In particular, all Stokes coefficients must now be known in order to explicitly implement multi--parameter median resummations. In the cases where quantum--theoretical physical observables are described by resurgent functions and transseries, the methods described herein show how one may cancel nonperturbative ambiguities, and define these observables nonperturbatively starting out from perturbation theory. Along the way, structural results concerning resurgent transseries are also obtained.
}


\makeatother

\usepackage{babel}

\begin{document}

%%%%%%%%%%%%%%%%%%%%%%%%%%%%%%%%%%%%%%%%%%%%%%%%%%%%%%%%%%%%%%%%%
%%%%%%%%%%%%%%%%%%%%%%%%%%%%%%%%%%%%%%%%%%%%%%%%%%%%%%%%%%%%%%%%%

%%%%%%%%%%%%%%%%%%%%%%%%%%%%%%%%%%%%%%%%%%%%%%%%%%%%%%%%%%%%%%%%%
%%%%%%%%%%%%%%%%%%%%%%%%%%%%%%%%%%%%%%%%%%%%%%%%%%%%%%%%%%%%%%%%%

\vfill

\eject

\allowdisplaybreaks

%%%%%%%%%%%%%%%%%%%%%%%%%%%%%%%%%%%%%%%%%%%%%%%%%%%%%%%%%%%%%%%%%
%%%%%%%%%%%%%%%%%%%%%%%%%%%%%%%%%%%%%%%%%%%%%%%%%%%%%%%%%%%%%%%%%
\section{Introduction}
%%%%%%%%%%%%%%%%%%%%%%%%%%%%%%%%%%%%%%%%%%%%%%%%%%%%%%%%%%%%%%%%%
%%%%%%%%%%%%%%%%%%%%%%%%%%%%%%%%%%%%%%%%%%%%%%%%%%%%%%%%%%%%%%%%%

Perturbation theory is a fundamental tool of analysis when addressing non--trivial problems in quantum theories. One may find its very successful applications almost everywhere, \textit{e.g.}, from the computation of ground--state energies of anharmonic oscillators in quantum mechanics, to the computation of beta functions in quantum field theory; from the genus expansion of the bosonic string in flat spacetime, to the large $N$ expansion of nonabelian gauge theories. Unfortunately, except for  particular cases, these perturbative series expansions are often asymptotic: they have zero radius of convergence. As is rather well known, this occurs due to the existence of singularities in the complex Borel plane, usually associated to instantons \citep{ZinnJustin:1980uk} and renormalons \citep{Beneke:1998ui}.

In this context, how does one make sense out of perturbation theory? Let us denote by $F$ the quantity we wish to compute and by $z$ the perturbative parameter. Without loss of generality we consider that the perturbative expansion in $z$ takes place around $z \sim \infty$ and will denote the perturbative coefficients of $F$ by $F_g$. It does not matter if this is a ground--state energy, a beta function, the string free energy, or some large $N$ correlation function: our discussion is completely general within perturbation theory and we shall not have in mind any specific example. That an asymptotic series has zero radius of convergence simply means that its coefficients grow as $F_g \sim g!$. One may then use the Borel transform $\CB [F]$ to ``remove'' this factorial growth and, upon analytic continuation of the Borel transform to the full complex plane, define the resummation $\CS F$ of a perturbative expansion by its inverse Borel transform. Here is where a more serious problem arises: the inverse Borel transform is essentially a Laplace transform, which requires an integration contour in order to be properly defined. Now, if the required contour of integration meets a singularity in the  complex Borel plane, this whole construction seems to break down. Indeed, such singularity will create an ambiguity, the nonperturbative ambiguity, as one needs to decide how the integration contour will avoid it. Singularities in the complex Borel plane occur along Stokes lines, and perturbative expansions are thus said to be non--Borel summable along these lines. Let us assume that the singularity occurs on the positive real axis and that the physical set--up one is addressing concerns small positive real coupling--constant. Avoiding the singularity either from above or from below will thus necessarily induce an imaginary contribution, which is \textit{different} depending on how we chose to avoid it. This is the nature of the ambiguity. The reason why it is nonperturbative is simply due to the functional form of the inverse Borel transform; the ambiguity goes as $\sim \rme^{-z}$ (as the expansion is around $z \sim \infty$ this contribution is non--analytic). Of course the same situation will take place along any other Stokes line. As such, the lack of Borel summability seems to be a fatal problem as it renders perturbative expansions meaningless. In this way, if some nonperturbative definition is to be obtained starting out with perturbation theory, one must find a way to go beyond the usual perturbative expansion.

In this larger context one asks again, how does one make sense out of perturbation theory? As it turns out, instantons (or renormalons) are not only an apparent disease but they also carry along their own cure. One of the first examples which helped clarify the solution to the problem raised in the paragraph above was that of the quartic anharmonic oscillator in quantum mechanics, in particular the study of the large--order growth of the perturbative expansion associated to its ground--state energy \citep{Bender:1969si, Bender:1990pd}. One finds that the coefficients of this ground--state energy grow as $F_g \sim g!\, A^{-g}$, where $A$ is the (real) instanton action locating the instanton singularity in the complex Borel plane, \textit{i.e.}, there is a Borel singularity on the positive real axis. As mentioned above, resumming perturbation theory along the real axis, and avoiding the singularity either through the left or through the right, leads to a nonperturbative ambiguity. The solution arises once one realizes that it is not only the perturbative sector which has an ambiguity. In fact, if one considers some fixed multi--instanton sector, say the $n$--instantons sector, then it is also the case that the perturbative expansion around this sector, with coefficients $F^{(n)}_g$, will also be asymptotic with non--trivial large--order growth $F^{(n)}_g \sim g!\, n\, A^{-g}$. In other words, also any multi--instanton series will suffer from nonperturbative ambiguities. While this could seem to make the problem with perturbation theory even worse, specially recalling that in most cases there is an infinite number of instanton sectors, it was shown in \citep{Bogomolny:1980ur, ZinnJustin:1981dx, ZinnJustin:1982td} that, instead, these ambiguities in the instanton sectors are in fact the \textit{solution} to our problem\footnote{In order to be fully rigorous, a small clarification is needed. In the context of quantum mechanics or quantum field theories with degenerate potentials, one needs to include both instantons and anti--instantons. In these cases, a topological charge will specify different topological sectors: with instanton number $+1$ ($-1$) for each (anti--)instanton, these sectors are  then characterized by their total instanton number. Assuming independent expansions for each of these topological sectors, the perturbative series (a vanishing number of instantons or anti--instantons) will appear as the ``level zero'' of the topological sector with topological charge $0$. Other contributions to this sector are the $n$--instanton/anti--instanton levels, denoted by $\left[\mathcal{I}^{n}\bar{\mathcal{I}}^{n} \right]$ in \cite{Dunne:2012ae}. Of course other topological sectors will also have a corresponding ``level zero'' in their expansions, but one which will already have the appropriate number of instantons (and anti--instantons) corresponding to the required topological charge. The results presented in this paper are directly applicable to this class of problems, one just needs to be aware that by ``perturbative series'' we mean the level--zero of each topological sector, while the $n$--instanton sectors are the higher levels with the \textit{same} topological charge (but see also the ``resurgence triangle'' in \cite{Dunne:2012ae}).}. These references showed that, in the calculation of the ground--state energy of the double--well potential, the ambiguity in the two--instantons sector precisely \textit{cancels} the ambiguity in the perturbative expansion; the ambiguity in the three--instantons sector cancels the ambiguity in the one--instanton sector; and so on. In light of this result, if one considers that the expansion of the ground--state energy is not only given by the usual perturbative expansion, but rather it is to be considered as a sum over \textit{all} multi--instanton sectors---including all asymptotic expansions around these nonperturbative sectors---, then it is possible that the final answer is in fact \textit{real}\footnote{Of course in some problems, depending on the physics, one is actually looking for (unambiguous!) imaginary results, in order to describe instabilities, decay and so on. But this is not what we are discussing here.} and \textit{free} of any nonperturbative ambiguity, as long as nonperturbative ambiguities arising in different sectors all conspire to cancel each other off. This cancelation of ambiguities in anharmonic oscillators has been checked to a very high numerical precision in a large number of references; see, \textit{e.g.}, \citep{ZinnJustin:1983nr, Jentschura:2004ib, Jentschura:2004cg, Ambrozinski:2012zw} and references therein.

The cancelation of nonperturbative ambiguities we just reviewed is actually just scratching the surface of a larger structure behind perturbation theory: that of resurgent analysis and transseries (we refer the reader to, \textit{e.g.}, the reviews \citep{Candelpergher:1993np, Delabaere:1999ef, Seara:2003ss} on resurgent analysis and \citep{Edgar:2008ga} on transseries, and to \citep{Marino:2008ya, Aniceto:2011nu, Marino:2012zq} for introductions to resurgence and transseries within physical contexts). In fact, within the aforementioned set--up of anharmonic potentials in quantum mechanics, it was further noted that computing (real, unambiguous, well--defined) ground--state energies does not simply amount to a specific summation over all multi--instanton sectors, but there are also contributions involving logarithms of the (anharmonic) coupling--constant. Essentially, this means that the ground--state energy, as a function over the complex plane of the anharmonic coupling--constant, will generically have a complicated multi--sheeted analytical structure, with singularities, poles and branch--cuts, and cannot possibly be described by a simple power series. Instead, this power series needs to be augmented with different non--analytical terms in order to fully describe the complete solution to the considered problem, and this is what the transseries accomplishes. Furthermore, the transseries will precisely encode Stokes phenomena in order to properly construct our final solution. But there is still more to this structure: the many components giving rise to the transseries are not arbitrary; the transseries are also resurgent. This means that once we fix a particular perturbative or multi--instanton sector, and study the large--order behavior of perturbation theory around this sector, it will be the case that this large--order behavior will be precisely dictated by information from all other sectors. Reversely, encoded deep in the large--order data of some fixed semiclassical sector, lie all others (hence the name resurgence). Within the quantum mechanical context, WKB and Bohr--Sommerfeld methods were used in order to derive exact quantization conditions, directly for the energy eigenvalues, which may then be solved with the use of resurgent transseries \textit{ansatze}, see, \textit{e.g.}, \citep{Voros:1983, Delabaere:1990eh, Voros:1993an, Voros:1994, ZinnJustin:2003jt, Jentschura:2004jg, Jentschura:2004ib, Jentschura:2004cg, Dunne:2013ada} and references therein. These works have laid solid ground to the use of resurgent transseries within quantum mechanical settings.

While the solution to the nonperturbative ambiguity problem, described above, works nicely within quantum mechanics by making use of the transseries multi--instantons expansion, the question remains if it may be generalized to quantum field theory. In this context, instantons are not the worse singularities in the complex Borel plane. Renormalons pose much more problematic singularities as they are not only dominant as compared to instantons, but they also seem to lack a general description in terms of semiclassical data \citep{'tHooft:1977am}. In fact, the quantum mechanical solution will work generically, from quantum field theory through string theory, as long as all singularities in the complex Borel plane have a semiclassical description and, as such, may be incorporated into a resurgent transseries where all nonperturbative ambiguities may be canceled. Recently, in \citep{Argyres:2012vv, Argyres:2012ka, Dunne:2012ae, Dunne:2012zk}, it was shown that akin to instantons also renormalons may be described in terms of semiclassical data and that they may be used in order to cancel ambiguities of the perturbative expansion within gauge theories. This opens the door to defining quantum field theory and asymptotically free gauge theories nonperturbatively, starting out with their perturbative data and augmenting them into transseries involving both multi--instanton and multi--renormalon nonperturbative sectors. Note that the use of resurgent methods within quantum field theory had already been pointed out in \citep{Stingl:2002cb}, but it was not until the work in the aforementioned references that it became clear that also in quantum field theory one may generically cancel nonperturbative ambiguities in a fashion completely identical to the quantum mechanical one.

These ideas have also been extended into string theoretic settings, and towards the nonperturbative study of the large $N$ expansion. This was first pointed out within the study of large--order behavior in string theory and large $N$ random matrix models, see, \textit{e.g.}, \citep{Marino:2006hs, Marino:2007te, Marino:2008vx, Pasquetti:2009jg, Klemm:2010tm}, where it also became clear that the framework of resurgence and transseries is in fact the appropriate framework to address nonperturbative issues within these models \citep{Marino:2008ya, Garoufalidis:2010ya, Aniceto:2011nu, Schiappa:2013opa}. In this set--up, \citep{Marino:2008ya} considered a specific example addressing superstrings in two dimensions, as described by the Painlev\'e II equation. In this case, the computation of the string free energy displays a nonperturbative ambiguity which again may be canceled by higher--order multi--instanton effects, in a fashion completely identical to the one which already worked in both quantum mechanics and quantum field theory. It was in fact already suggested in \citep{Marino:2008ya} that this procedure is nothing but the implementation of the transseries median resummation and that this is the correct procedure which cancels nonperturbative ambiguities and allows for a construction of real solutions to the string theoretic free energy, along the string--coupling real axis.

All things considered, quite a few results seem to be transversal and applicable over a wide range of quantum theoretical settings. Most perturbative expansions have nonperturbative ambiguities which may be canceled by higher multi--instanton effects (or multi--renormalon effects, generalized multi--instantons effects \citep{Garoufalidis:2010ya, Aniceto:2011nu, Schiappa:2013opa, Basar:2013eka}, or more exotic saddles \citep{Cherman:2013yfa}). This further indicates that physical observables are not only given by a resummation of their perturbative expansions, but by adequate resummations of transseries, encoding the full (nonperturbative) semiclassical data concerning the problem at hand. That the same procedure works in so many different contexts is simply saying that what one is considering is a rather general solution within the resurgent transseries framework. In fact, as we shall make clear in this paper, cancelation of all nonperturbative ambiguities is achieved by considering the transseries median resummation as the correct resummation prescription. Intuitively, one may think of the median resummation as follows. Let us suppose there was a single pole along the real axis. Integrating either above or below the pole yields either a $+\rmi$ or $-\rmi$ contribution, and the median of these integrations precisely cancels the ambiguous imaginary part. Of course in more complicated settings the singularity structure is much more involved, with an infinite tower of multi--instanton contributions, but the main idea behind the median resummation is precisely to ensure that the ambiguous imaginary contributions cancel among all multi--instanton sectors. As we shall see, while this procedure is simple when considering a one--parameter transseries, it becomes much more intricate for multi--parameter transseries\footnote{In \citep{Dunne:2012ae} it was also noticed that the cancelation mechanism becomes highly non--trivial when different type of ambiguities need to cancel, and this was denoted as a set of ``confluence equations'' in that reference. We believe that all these cancelations are particular cases of general (multi--parameter) median resummations.}. In spite of this, we shall show how solutions can always be constructed, and how they are highly dependent upon the Stokes data of the problem (in fact, median resummations may be defined along any Stokes line---and we shall address them all).

Given that Borel resummation alone cannot properly define perturbation theory because of the nonperturbative ambiguities it faces along Stokes lines, the overall picture which we try to convey is that it is the median resummation of the resurgent transseries which is always the general, unambiguous nonperturbative answer one should consider. Of course, while this prescription is mathematically rather universal, extra work still has to be done to implement it in different physical settings. In fact, the transseries can only be made fully explicit once we have managed to identify the complete nonperturbative content of the theory in terms of semiclassical configurations\footnote{Note that by the use of resurgence and large--order analysis it might still be possible to identify the full nonperturbative content of some given theory, explicitly written in terms of semiclassical configurations, even though these semiclassical configurations may still lack a proper physical interpretation \citep{Garoufalidis:2010ya, Aniceto:2011nu, Schiappa:2013opa}.}. This is the physical problem which remains to be worked out in each concrete case. Nonetheless, it seems natural to assume that whatever quantity one aims at computing within quantum theoretical settings, it will always be described by resurgent functions and transseries. In this case, as described, perturbation theory may always be made meaningful and used to yield nonperturbative solutions to whatever initial question we had in mind.

%%%%%%%%%%%%%%%%%%%%%%%%%%%%%%%%%%%%%%%%%%%%%%%%%%%%%%%%%%%%%%%%%
%%%%%%%%%%%%%%%%%%%%%%%%%%%%%%%%%%%%%%%%%%%%%%%%%%%%%%%%%%%%%%%%%
\section{Nonperturbative Ambiguities and Real Transseries\label{sec:Nonpert-Ambiguity-real-transs}}
%%%%%%%%%%%%%%%%%%%%%%%%%%%%%%%%%%%%%%%%%%%%%%%%%%%%%%%%%%%%%%%%%
%%%%%%%%%%%%%%%%%%%%%%%%%%%%%%%%%%%%%%%%%%%%%%%%%%%%%%%%%%%%%%%%%

This paper is somewhat self--contained in the sense that we only require the reader to be familiar with sections 2 and 4 of \citep{Aniceto:2011nu}; most of our results will follow from there. Still, we shall begin by recalling precisely a few of the contents in those sections of \citep{Aniceto:2011nu} in order to set the stage, as the definition of median resummation may be immediately explained with just a few formulae.

\begin{figure}
\centering{}
\includegraphics[scale=0.6]{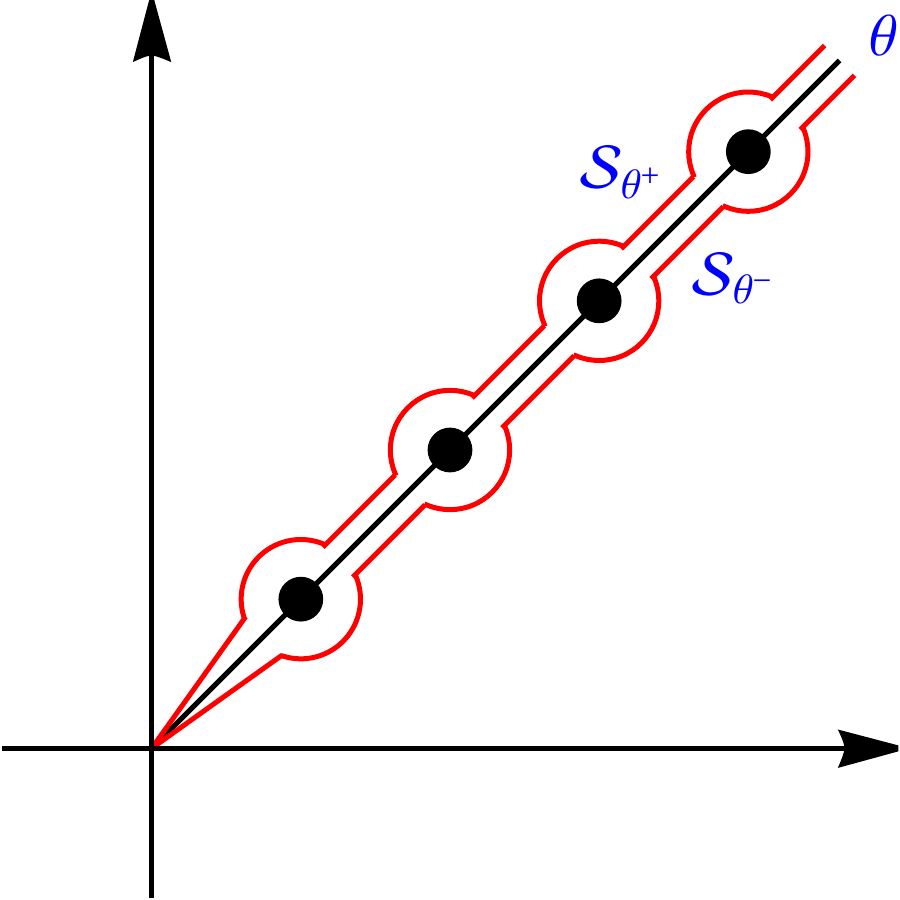}
\caption{Schematic representation of the median resummation as an ``average'' of lateral resummations. This ``average'' acts on full transseries and thus necessarily constraints the choice of transseries parameters.
\label{fig:median-resummation}}
\end{figure}

Consider a perturbative series around $z\sim\infty$,
\begin{equation}
F(z) \simeq \sum_{g=0}^{+\infty} \frac{F_{g}}{z^{g+1}}.
\label{eq:d}
\end{equation}
\noindent
This series is asymptotic with zero radius of convergence when its coefficients grow as $F_{g}\sim g!$. In order to extract information out of such asymptotic series, it is common to use Borel analysis. The Borel transform
\begin{equation}
\CB \left[ \frac{1}{z^{\alpha+1}} \right] (s) = \frac{s^{\alpha}}{\Gamma(\alpha+1)}\label{eq:boreltrans}
\end{equation}
\noindent
constructs the Borel transformed series, $\CB[F](s)$, with non--vanishing convergence radius and which may be analytically continued throughout $s\in\BC$. In order to associate a value to the divergent sum \eqref{eq:d}, and given a direction $\theta$ in the complex $s$--plane where $\CB[F](s)$ has no singularities, one may invert the Borel transform into the Borel resummation $\CS_{\theta}F(z)$ as
\begin{equation}
\CS_{\theta}F(z) = \int_{0}^{\rme^{\rmi\theta}\infty} \rmd s\,\CB[F](s)\,\rme^{-zs}.\end{equation}
\noindent
In principle, this would be the nonperturbative answer arising from the perturbative expansion. But if $\CB[F](s)$ has singularities\footnote{As discussed in \citep{Aniceto:2011nu}, we are only considering poles or logarithmic branch--cuts as singularities.} along the direction $\theta$, this singular direction becomes known as a Stokes line and the resummation is no longer possible as its integration contour just became ambiguous. We then need to define lateral Borel resummations, $\CS_{\theta^{\pm}}F(z)$, avoiding the singularities via the left or via the right, and leading to distinct (sectorial) resummations of our original asymptotic series (see figure \ref{fig:median-resummation}). In this language, the nonperturbative ambiguity is associated to having $\CS_{\theta^{+}} - \CS_{\theta^{-}} \neq 0$. However, the key point to stress is that these lateral Borel resumations are still \textit{related} via the Stokes automorphism $\underline{\mathfrak{S}}_{\theta}$ as follows:
\begin{equation}
\CS_{\theta^{+}} = \CS_{\theta^{-}} \circ \underline{\mathfrak{S}}_{\theta}.\label{eq:stokesauto}
\end{equation}
\noindent
In order to determine the Stokes automorphism one uses alien calculus, and we refer the reader to \citep{Aniceto:2011nu} for more details. In short, $\underline{\mathfrak{S}}_{\theta}$ may be computed in terms of the alien derivative, $\Delta_{\omega}$, a differential operator which essentially encodes the singular behavior of the Borel transform (\textit{i.e.}, it vanishes if evaluated at a regular point of $\CB[F](s)$). If the singular points along the $\theta$--direction are denoted by $\left\{ \omega_{\theta} \right\}$, one finds
\begin{equation}
\underline{\mathfrak{S}}_{\theta}= \exp \left\{ \sum_{\omega\in\left\{ \omega_{\theta}\right\} }\rme^{-\omega z}\Delta_{\omega}\right\}.
\label{eq:Stokes-aut-from-alien-derivatives}
\end{equation}
 
The proper use of alien calculus is made within the setting of transseries and resurgent functions. As we have explained in the introduction, transseries augment power series by the incorporation of non--analytic terms. For example, a one--parameter transseries is of the form
\be
F (z,\sigma) = \sum_{n=0}^{+\infty} \sigma^n F^{(n)} (z),
\label{eq:Foneparatransseries}
\ee
\noindent
where $F^{(0)} (z)$ is the (asymptotic) perturbative expansion one started out with, (\ref{eq:d}), $F^{(n)} (z)$ are (again asymptotic) multi--instanton sectors, and where $\sigma$ is a formal parameter counting the instanton number and selecting distinct nonperturbative completions to whatever problem one is addressing. Transseries may depend on multiple parameters; \textit{e.g.}, in the solution to the Painlev\'e I equation and the quartic matrix model in \citep{Garoufalidis:2010ya, Aniceto:2011nu}, or the solution to the Painlev\'e II equation in \citep{Schiappa:2013opa}, a two--parameters transseries was needed. With parameters $\sigma_{1}$ and $\sigma_{2}$, this was given by
\begin{equation}
F(z,\sigma_{1},\sigma_{2}) = \sum_{n=0}^{+\infty} \sum_{m=0}^{+\infty} \sigma_{1}^{n}\sigma_{2}^{m} F^{(n|m)}(z),
\label{eq:Ftransseries}
\end{equation}
\noindent
where again $F^{(0|0)}(z)$ is the formal asymptotic power series (\ref{eq:d}), and where the $F^{(n|m)}(z)$ are now generalized instanton contributions with generalized instanton actions $\pm A\in\mathbb{R}$, of the form
\begin{equation}
F^{(n|m)}(z) = \rme^{-(n-m)Az}\, \Phi_{(n|m)}(z),
\label{eq:IntroFn-def}
\end{equation}
\noindent
with $\Phi_{(n|m)}(z)$ perturbative expansions around each instanton sector,
\begin{equation}
\Phi_{(n|m)}(z) \simeq z^{-\beta_{nm}}\, \sum_{g=0}^{+\infty} \frac{F_{g}^{(n|m)}}{z^{g}}.
\label{eq:pert-instanton-expansions-Phi_n}
\end{equation}
\noindent
To be fully precise, the two--parameter transseries used in the aforementioned references also include logarithmic sectors due to resonance, in the sense that the asymptotic expansion \eqref{eq:pert-instanton-expansions-Phi_n} also includes a (finite) sum over powers of logarithms. In order not to clutter the discussion with unnecessary technicalities, and because these sectors are only relevant when the explicit asymptotic expansion \eqref{eq:pert-instanton-expansions-Phi_n} needs to be taken into account, we will not consider these sectors in the main text. Nevertheless, we do discuss them in appendix \ref{sec:App-Properties-of-med-resumm}. Note that a one--parameter transseries is recovered from \eqref{eq:Ftransseries} by setting $\sigma_{2}=0$. In this case we also define $F^{(n)}(z) \equiv F^{(n|0)}(z)$ and the same for $\Phi_{n}(z)$. In our cases of study we shall consider $\beta_{nm} = \left( n+m \right) \beta$ where $\beta$ is a rational number. For a general $\beta_{nm}$ one would have to address the problem case by case. 

More general multi--parameter transseries may be considered, \textit{e.g.}, with $k$ parameters and $k$ distinct instanton actions (real or not). In this case, (generalized) multi--instanton sectors are labeled by a set of integers $\boldsymbol{n} = \left( n_1, \ldots, n_k \right) \in \BN^k$ as $F^{(\boldsymbol{n})}$ and the whole structure is more involved. For our purposes, however, the aforementioned one and two parameter cases will suffice as they already illustrate the median resummation construction in great generality, both including multi--parameter transseries and the inclusion of generalized instanton sectors.

One reason why transseries are introduced is that their alien derivatives may be related to their common derivatives by a set of equations known as the bridge equations (implementing a ``bridge'' between alien and usual calculus; see, \textit{e.g.},  \citep{Aniceto:2011nu} for further details). This allows us to explicitly evaluate the Stokes automorphism (\ref{eq:Stokes-aut-from-alien-derivatives}) in terms of a set of constants which encode the nonperturbative information of the system one is trying to solve---these are the Stokes constants. For our example of a two--parameter transseries (\ref{eq:Ftransseries}), and as we are dealing with instanton actions $\pm A\in\mathbb{R}$, $\theta=0,\pi$ are singular directions of the Borel transform, \textit{i.e.}, they are Stokes lines. In fact, the Borel transform has poles at $s=\ell A$, for $\ell\in\mathbb{Z}\setminus\left\{ 0\right\}$ (in the one--parameter case, with $\sigma_{2}=0$, we find $\ell\le1$). In this case, the bridge equations then take the form\footnote{For the one--parameter case, one would obtain the same equation with
$\widetilde{S}_{\ell}=0$.}
\begin{equation}
\dot{\Delta}_{\ell A} F (z,\sigma_{1},\sigma_{2}) = S_{\ell} \left(\sigma_{1},\sigma_{2}\right) \frac{\partial F}{\partial\sigma_{1}} + \widetilde{S}_{\ell} \left(\sigma_{1},\sigma_{2}\right) \frac{\partial F}{\partial\sigma_{2}},
\label{eq:Bridge-eqs-2-param}
\end{equation}
\noindent
where $\dot{\Delta}_{\ell A} := \rme^{-\ell A}\, \Delta_{\ell A}$ is denoted as the pointed alien derivative. As to $S_{\ell} \left(\sigma_{1},\sigma_{2}\right)$ and $\widetilde{S}_{\ell} \left(\sigma_{1},\sigma_{2}\right)$, they have natural power series expansions and one mostly works with the respective coefficients instead; in this case with $S_{\ell}^{(k)}$ and $\widetilde{S}_{\ell}^{(k)}$ (but see \citep{Aniceto:2011nu} for these explicit expansions). It should be clear that, when inserting the transseries expansion (\ref{eq:Ftransseries}) into the above bridge equations (\ref{eq:Bridge-eqs-2-param}), one will find that the alien derivative of any given sector $\Delta_{\ell A} \Phi_{(n|m)}$ \textit{only} depends on \textit{other} sectors $\Phi_{(n'|m')}$, and this is in essence why the transseries are ``resurgent''.

A simple and probably familiar example occurs when restricting to the one--parameter case. Applying the Stokes automorphism (\ref{eq:Stokes-aut-from-alien-derivatives}) for the $\theta=0$ direction, and making use of the bridge equations, one finds
\begin{equation}
\CS_{0^+} F\left(z,\sigma\right) = \CS_{0^-} F\left(z,\sigma+S_{1}\right),
\label{eq:stokespheno}
\end{equation}
\noindent
where $S_{1}$ is a Stokes constant. This expression precisely describes Stokes phenomena of classical asymptotics within the resurgence framework---crossing the Stokes line at $\theta=0$ corresponds to a ``jump'' in the parameter $\sigma$, as governed by the Stokes constant $S_{1}$. In later sections we will discuss the role that each Stokes constant plays when crossing different Stokes lines. 

Nonperturbative ambiguities in the resummation of asymptotic series (and transseries) precisely arise along singular directions, \textit{i.e.}, directions along which the Stokes automorphism is non--trivial and where Stokes phenomena takes place. Their cancelation via the transseries median resummation must thus relate to the Stokes automorphism. For the moment, let us point out that there is a very simple argument to understand how this occurs. In most cases, when one writes the perturbative expansion \eqref{eq:d} it has real coefficients and addresses positive real coupling. But positive real coupling also corresponds to the $\theta=0$ Stokes line, where Stokes phenomena takes place as in \eqref{eq:stokespheno}. At the same time, in this set--up the ambiguities are purely imaginary. So, the cancelation of all ambiguities naturally translates to the construction of a real transseries along the $\theta=0$ Stokes line. In the one--parameter transseries this is just setting $\im\, F (z,\sigma)=0$. As first discussed in \citep{Marino:2008ya}, this condition is satisfied if and only if $\im\, \sigma = \frac{{\rmi}}{2}\, S_{1}$ (where $S_1 \in \rmi\BR$), with real instanton action, and where the real solution is then
\begin{equation}
F_{\mathbb{R}} (z,\sigma) = \mathcal{S}_{0^+} F \left(z,\sigma-\frac{1}{2}S_{1}\right) = \mathcal{S}_{0^-} F \left(z,\sigma+\frac{1}{2}S_{1}\right).
\end{equation}
\noindent
In this expression $\sigma\in\mathbb{R}$ and one uses \eqref{eq:stokespheno} in the second equality. But this real solution is precisely the median resummation. More generally, the median resummation of some given transseries along $\theta=0$ is defined as (see, \textit{e.g}, \citep{Delabaere:1999ef})
\begin{equation}
\CS_{0}^{\text{med}} := \CS_{0^{+}} \circ \underline{\mathfrak{S}}_{0}^{-1/2} = \CS_{0^{-}} \circ \underline{\mathfrak{S}}_{0}^{+1/2},
\label{eq:Median-sum-def}
\end{equation}
\noindent
where the last equality comes directly from (\ref{eq:stokesauto}). The reverse statement is also true: the median resummation of a transseries along the $\theta=0$ Stokes line yields a \textit{real} transseries at positive real coupling. In this way, $F_{\mathbb{R}} (z,\sigma) \equiv \CS_{0}^{\text{med}} F (z,\sigma)$.

This result may be understood as follows. Assume for the moment that all we had was the original perturbative expansion $F^{(0)}$. Because $\theta=0$ is a singular direction, $\mathcal{S}_{0^+} - \mathcal{S}_{0^-} \neq 0$ gives rise to a nonperturbative ambiguity and we may use these (distinct) lateral Borel resumations to naturally define the imaginary ambiguity as
\begin{equation}
\im\, F^{(0)} \equiv \frac{1}{2{\rmi}} \left( \mathcal{S}_{0^+} - \mathcal{S}_{0^-} \right) F^{(0)}.
\end{equation}
\noindent
Further defining the associated real contribution as
\begin{equation}
\re\, F^{(0)} \equiv \frac{1}{2} \left( \mathcal{S}_{0^+} + \mathcal{S}_{0^-} \right) F^{(0)},
\end{equation}
\noindent
we may rewrite the resummation of our original series as\footnote{Here we made a choice of lateral resummation $\mathcal{S}_{0^-}$. Analogous results would be achieved with $\mathcal{S}_{0^+}$ instead.}
\begin{equation}
\mathcal{S}_{0^-} F^{(0)} = \re\, F^{(0)} - {\rmi}\, \im\, F^{(0)}.
\label{eq:Imaginary-part-in-lateral-S-F0}
\end{equation}
\noindent
In order to construct an (unambiguous) real solution, we need to cancel the (ambiguous) imaginary part. In order to do that, let us better understand which exact contributions give rise to this second term. As shown in appendix \ref{sec:App-Formulae-One-Param-transs}, the imaginary contribution to any $F^{(n)}$ may be determined by simply using the fact that
\begin{equation}
\left( \mathcal{S}_{0^+} - \mathcal{S}_{0^-} \right) F^{(n)}(z) = - \mathcal{S}_{0^-} \circ \left( \mathbf{1} - \underline{\mathfrak{S}}_{0} \right) F^{(n)}(z).
\end{equation}
\noindent
This expression also makes it clear that the full content of the ambiguity is encoded in the Stokes automorphism. Then, using formulae from appendix \ref{sec:App-Formulae-One-Param-transs}, $\im\, F^{(0)}$ is given by
\begin{eqnarray}
2{\rmi}\, \im\, F^{(0)} &=& \sum_{\ell=1}^{+\infty} S_{1}^{\ell}\, \mathcal{S}_{0^-}F^{(\ell)} = \nonumber \\
&=& \sum_{\ell=1}^{+\infty} S_{1}^{\ell} \left( \re\, F^{(\ell)} - {\rmi}\, \im\, F^{(\ell)} \right) = \nonumber \\
&=& S_{1}\, \re\, F^{(1)} - \frac{1}{2}S_{1}^{3}\, \re\, F^{(3)} + S_{1}^{5}\, \re\, F^{(5)} + \cdots.
\end{eqnarray}
\noindent
Here we have rewritten the $\im\, F^{(\ell)}$ terms as real contributions, via \eqref{eq:App-1-param-im-part-F(n)}, to recursively write the ambiguity in $F^{(0)}$ as a multi--instanton expansion of real contributions. Now plugging this expansion back into \eqref{eq:Imaginary-part-in-lateral-S-F0}, one immediately finds that in order to cancel the above ambiguous imaginary term we need to add to the perturbative expansion at least contributions arising from the one--instanton sector,
\be
\frac{1}{2} S_{1}\, \mathcal{S}_{0^-} F^{(1)} = \frac{1}{2} S_{1}\, \re\, F^{(1)} - \frac{{\rmi}}{2}S_{1}\, \im\, F^{(1)}.
\ee
\noindent
This, in turn, will still contribute to an imaginary ambiguity but only at the \textit{next} order in instanton number. In this case, an ``improved'' version of \eqref{eq:Imaginary-part-in-lateral-S-F0} becomes
\begin{eqnarray}
\mathcal{S}_{0^-} \left( F^{(0)} + \frac{1}{2} S_{1}\, F^{(1)} \right) &=& \re\, F^{(0)} + \frac{1}{2} \left( \frac{1}{2} S_{1}^{3}\, \re\, F^{(3)} - S_{1}^{5}\, \re\, F^{(5)} + \cdots \right) - \frac{{\rmi}}{2} S_{1}\, \im\, F^{(1)} = \nonumber \\
&=& \re\, F^{(0)} - \frac{1}{2} S_{1}^{2}\, \re\, F^{(2)} + \frac{1}{4} S_{1}^{3}\, \re\, F^{(3)} + \frac{1}{2} S_{1}^{4}\, \re\, F^{(4)} + \cdots. 
\end{eqnarray}
\noindent
To obtain the second line above, we have again used (\ref{eq:App-1-param-im-part-F(n)}) to expand $\im\, F^{(1)}$. Continuing the iteration of this process, we quickly find that what one is constructing is a real transseries solution starting out from the perturbative expansion, and this is precisely the aforementioned median resummation of the transseries when $\sigma=0$, \textit{i.e.}, this process constructs
\begin{equation}
F_{\BR} (z,0) = \mathcal{S}_{0^-} \left( \sum_{n=0}^{+\infty} \frac{S_{1}^{n}}{2^{n}}\, F^{(n)} \right) = \mathcal{S}_{0^-} F \left(z,\frac{1}{2}S_{1}\right).
\end{equation}
\noindent
What this simple exercise has done is to mathematically formalize the procedure to cancel the nonperturbative ambiguities to \textit{all} orders, in both quantum mechanics and quantum field theory, which we have outlined in the introduction.

If one expands the median resummation $F_{\mathbb{R}}(z,\sigma)$ in powers of $\sigma$, and further rewrites all terms as explicitly real contributions (\textit{e.g.}, following the guidelines we describe at the end of appendix \ref{sec:App-Formulae-One-Param-transs}), it is simple to find
\begin{equation}
F_{\mathbb{R}} (z,\sigma) = \re\, F^{(0)} (z) + \sigma\, \re\, F^{(1)} (z) +\left( \sigma^{2}-\frac{1}{4}S_{1}^{2} \right) \re\, F^{(2)} (z) + \cdots.
\end{equation}
\noindent
This expansion shows that real solutions, where all ambiguities have been canceled, will always display multi--instanton corrections even if one sets $\sigma=0$. A similar construction can also be carried through for the two--parameter transseries, and we shall discuss it in detail in section \ref{sec:Nonp-Ambiguities-2-param}. For the moment, we just want to make clear that the explicit determination of the transseries median resummation gives us a very direct way to determine real solutions, without having to follow the more intricate recursive construction we outlined above. In particular, this is the simplest approach to canceling all nonperturbative ambiguities, in all multi--instanton sectors.

This procedure will work along \textit{any} Stokes line. Given some arbitrary singular direction $\theta$ it is natural to ask if it is always possible to find a median resummation prescription such that nonperturbative ambiguities cancel. Intuitively it is quite simple to realize this is true. Trivially writing
\be
\CS_{\theta^\pm} = \frac{1}{2} \left( \CS_{\theta^+}+\CS_{\theta^-} \right) \pm \frac{1}{2} \left( \CS_{\theta^+}-\CS_{\theta^-} \right),
\ee
\noindent
then if $\theta$ is a singular direction where $\underline{\mathfrak{S}}_{\theta} \neq \boldsymbol{1}$, one finds a nonperturbative ambiguity as $\CS_{\theta^+} \neq \CS_{\theta^-}$. Canceling the ambiguity entails setting $\CS_{\theta^+}-\CS_{\theta^-} \sim 0$ \textit{at the level of the transseries} (to stress this point we have used $\sim$ instead of an equal sign in the previous formula). This means that one also needs to define projections parallel and orthogonal to the direction $\theta$ for the transseries parameters $\sigma_i$, implemented as the operations $\re_\theta$ and $\im_\theta$. But this is simple to do, in which case the median resummation follows as (see figure \ref{fig:median-resummation})
\be
\CS_{\theta}^{\text{med}} \sim \frac{1}{2} \left( \CS_{\theta^+}+\CS_{\theta^-} \right).
\ee
\noindent
More precisely, note that along $\theta$ the Stokes automorphism is non--trivial and, making use of its definition in (\ref{eq:stokesauto}), one may always write
\begin{equation}
\CS_{\theta}^{\text{med}} := \CS_{\theta^{+}} \circ \underline{\mathfrak{S}}_{\theta}^{-\nu} = \CS_{\theta^{-}} \circ \underline{\mathfrak{S}}_{\theta}^{1-\nu}
\label{smednutheta}
\end{equation}
\noindent
for some yet undefined value of $\nu$; but where an appropriate value for $\nu$ will be equivalent to requiring that the transseries lateral Borel resummations coincide, \textit{i.e.}, $ \CS_{\theta^{+}}-\CS_{\theta^{-}} \sim 0$. Now, as discussed in appendix \ref{sec:App-Properties-of-med-resumm}, one may always rotate the singular direction $\theta$ in the complex Borel plane to the positive real axis, where, in the new variable, the median resummation and cancelation of nonperturbative ambiguities translate to a reality requirement: $\mathcal{H} F_{\BR} (z, \sigma) = F_{\BR} \left( z, \sigma \right)$, with $\mathcal{H} F \equiv \overline{F}$ the complex conjugation operator and $F_{\BR} (z,\sigma) = \CS_{0}^{\text{med}} F (z,\sigma)$. As we shall see, this will constrain $\sigma$ and naturally set $\nu = \frac{1}{2}$; in which case translating back to the original Stokes direction one has
\begin{equation}
\CS_{\theta}^{\text{med}} = \CS_{\theta^{+}} \circ \underline{\mathfrak{S}}_{\theta}^{-{1}/{2}} = \CS_{\theta^{-}} \circ \underline{\mathfrak{S}}_{\theta}^{{1}/{2}}.
\end{equation}

%%%%%%%%%%%%%%%%%%%%%%%%%%%%%%%%%%%%%%%%%%%%%%%%%%%%%%%%%%%%%%%%%
%%%%%%%%%%%%%%%%%%%%%%%%%%%%%%%%%%%%%%%%%%%%%%%%%%%%%%%%%%%%%%%%%

In the next two sections we shall analyze in detail both Stokes phenomena---at the origin of the ambiguity---and the implementation of the median resummation in the one--parameter transseries framework, on what concerns its two possible singular directions $\theta=0,\pi$. In particular, we shall discuss what are the consequences of requiring reality conditions and cancelation of nonperturbative ambiguities along these Stokes lines. Once this is understood, we then move to the two--parameters case (with generalized instanton sectors) where things get much more involved. Nonetheless, it is still possible to present all details and a final solution to the ambiguity problem: we again explain in detail the intricacies of Stokes phenomena, and how it is used to cancel ambiguities. Before concluding, we also have a section discussing how Stokes and anti--Stokes lines interplay with reality conditions and the monodromy of the transseries solution. Because there are many technical details, in order to keep a consistent and light line of thought throughout the paper we have packaged most technicalities into three (somewhat long) appendices. In the first of these appendices we have obtained many structural results concerning real resurgent transseries and their Stokes data, including cases with resonance and logarithms.

%%%%%%%%%%%%%%%%%%%%%%%%%%%%%%%%%%%%%%%%%%%%%%%%%%%%%%%%%%%%%%%%%
%%%%%%%%%%%%%%%%%%%%%%%%%%%%%%%%%%%%%%%%%%%%%%%%%%%%%%%%%%%%%%%%%
\section{Stokes Phenomena in One--Parameter Transseries}
%%%%%%%%%%%%%%%%%%%%%%%%%%%%%%%%%%%%%%%%%%%%%%%%%%%%%%%%%%%%%%%%%
%%%%%%%%%%%%%%%%%%%%%%%%%%%%%%%%%%%%%%%%%%%%%%%%%%%%%%%%%%%%%%%%%

Stokes lines create nonperturbative ambiguities for the resummation of perturbation theory, which means the first step in order to understand how to cancel these ambiguities is to understand exactly what occurs at those Stokes lines. We shall first focus upon one--parameter transseries, as in (\ref{eq:Foneparatransseries}) or by setting $\sigma_{2}=0$ in (\ref{eq:Ftransseries}) and just keeping $F^{(n|0)}(z)\equiv F^{(n)}(z)$. As shown in \cite{Aniceto:2011nu}, there are two singular directions in the Borel plane, $\theta=0$ and $\theta=\pi$, with distinct features. In the singular direction $\theta=0$ the Stokes automorphism, acting on the perturbative series $\Phi_{n}(z)$ defined in (\ref{eq:pert-instanton-expansions-Phi_n}), is given by
\begin{equation}
\underline{\mathfrak{S}}_{0}^{\nu} \Phi_{n} = \sum_{\ell=0}^{+\infty} \binom{n+\ell}{n} \left(\nu S_{1} \right)^{\ell} {\rme}^{- \ell A z}\, \Phi_{n+\ell}.
\end{equation}
\noindent
The derivation of this result (along with a few others in the following) may be found in appendix \ref{sec:App-Formulae-One-Param-transs}. This expression defines an arbitrary power, $\nu$, of the Stokes automorphism $\underline{\mathfrak{S}}_{0}$, with the usual Stokes automorphism obtained when setting $\nu=1$. On what concerns the transseries itself, this automorphism translates to a Stokes transition as
\begin{equation}
\underline{\mathfrak{S}}_{0}^{\nu}\, F \left( z, \sigma \right) = F \left( z, \sigma+\nu S_{1} \right).
\label{eq:Stokes-trans-anypower-zero}
\end{equation}
\noindent
For $\nu=1$ it is easy to see that this leads to the Stokes phenomenon (\ref{eq:stokespheno}), by using the original definition of the Stokes automorphism (\ref{eq:stokesauto}). This Stokes phenomenon associated to the $\theta=0$ Stokes line, $\CS_{0^{+}} F \left( z, \sigma \right) = \CS_{0^{-}} F \left( z, \sigma+S_{1} \right)$, is essentially realized by a ``jump'' of the transseries parameter, $\sigma \rightarrow \sigma + S_{1}$, which is governed by the Stokes constant $S_{1}$.

As we choose the other Stokes line, $\theta=\pi$, things will get a bit more intricate. The Stokes automorphism (or a general power $\nu$ thereof), acting on the perturbative series $\Phi_{n} (z)$, is now given by
\begin{equation}
\underline{\mathfrak{S}}_{\pi}^{\nu} \Phi_{n} = \sum_{\ell=0}^{n} \underline{\Sigma}_{\nu} \left( n, \ell \right) {\rme}^{\ell A z}\, \Phi_{n-\ell},
\end{equation}
\noindent
where $\underline{\Sigma}_{\nu}(n,\ell)$ is defined in appendix \ref{sec:App-Formulae-One-Param-transs}, equation (\ref{eq:AppA-coefficients-automorphism-at-pi-nu}). These coefficients now have a dependence on all the Stokes coefficients $S_{-\ell}$, for $\ell \ge 1$, thus becoming much more involved than the case of $\theta=0$, where only the one Stokes parameter $S_{1}$ came into play. On what concerns the transseries, the automorphism translates to a Stokes transition as
\begin{equation}
\underline{\mathfrak{S}}_{\pi}^{\nu} F \left( z, \sigma \right) = F \left( z, \BS^{(\nu)}_\pi (\sigma) \right).
\label{eq:Stokes-trans-anypower-pi}
\end{equation}
\noindent
As before, we may say that Stokes phenomenon is realized by having the transseries parameter ``jump'', only now this is no longer a simple shift; instead one finds
\begin{eqnarray}
\BS^{(\nu)}_\pi (\sigma) &=& \sum_{n=0}^{+\infty} \sigma^{n+1} \underline{\Sigma}_{\nu} (n+1,n) = 
\label{eq:Def-gsigma-at-pi} \\
&=&
\sigma + \sigma^{2} \nu S_{-1} + \sigma^{3} \left(\nu S_{-2} + \nu^{2} S_{-1}^{2} \right) + \sigma^{4}\left( \nu S_{-3} + \frac{5}{2} \nu^{2} S_{-1} S_{-2} + \nu^{3} S_{-1}^{3} \right) + \cdots. \nonumber 
\end{eqnarray}
\noindent
As just mentioned, supporting evidence and derivation of these results may be found in appendix \ref{sec:App-Formulae-One-Param-transs}. Consequently, Stokes phenomena associated to the $\theta=\pi$ singular Stokes line is naturally given by\footnote{We set $\nu=1$ and defined $\BS^{(1)}_\pi (\sigma) \equiv \BS_\pi (\sigma)$. To recover the full dependence in $\nu$ one simply multiplies each and every Stokes constant by $\nu$, in the series expansion.}
\begin{equation}
\CS_{\pi^{+}} F \left( z,\sigma \right) = \CS_{\pi^{-}} F \left( z,\BS_\pi (\sigma) \right).
\end{equation}

As we compare \eqref{eq:Stokes-trans-anypower-zero} and \eqref{eq:Stokes-trans-anypower-pi}, it would seem that Stokes phenomenon across the Stokes lines at $\theta=0$ and $\theta=\pi$ is completely different; one leads to a simple shift of $\sigma$, the other to a very intricate change in $\sigma$. This difference is essentially associated to the number of Stokes constants along each singular line---were we to compare the effect of \textit{single} Stokes constants, then the jump would always become apparent. To understand the effect of each Stokes constant in the jump, set to zero all Stokes constants except $S_{1}$ and one $S_{-\ell}$, for some fixed $\ell \ge 1$. As we shall show, in this case the Stokes constant $S_{-\ell}$ is responsible for a jump very similar to the jump $\sigma \rightarrow \sigma + S_{1}$ in the $\theta=0$ transition, but in this case it will be $\sigma^{-\ell}$ which jumps.

Going back to the bridge equations; for one--parameter transseries they may be simply obtained from (\ref{eq:Bridge-eqs-2-param}) by setting $\widetilde{S}_{\ell} = 0$, as
\begin{equation}
\dot{\Delta}_{kA} F (z,\sigma) = S_{k}(\sigma)\, \frac{\partial F}{\partial\sigma}.
\end{equation}
\noindent
The coefficients in the power series expansion of $S_{k}(\sigma)$, \textit{i.e.}, the Stokes constants, may be found for instance in section 2 of \citep{Aniceto:2011nu}; they are:
\begin{equation}
S_{k} (\sigma) = S_{k}\, \sigma^{1-k}, \qquad \forall\, k \le 1,\, k \ne 0.
\end{equation}
\noindent
In this case, the Stokes automorphism (\ref{eq:Stokes-aut-from-alien-derivatives}) along the $\theta=\pi$ direction follows as
\begin{equation}
\underline{\mathfrak{S}}_{\pi}^{\nu} F (z,\sigma) = \exp \left\{ \nu\, \sum_{k \ge 1} \dot{\Delta}_{-kA} \right\} F (z,\sigma) = \exp \left\{ \nu\, \sum_{k \ge 1} S_{-k}\, \sigma^{1+k}\, \frac{\partial}{\partial\sigma} \right\} F (z,\sigma).
\end{equation}
\noindent
If we restrict ourselves to a scenario where all Stokes constants vanish, $S_{-k}=0$ for $k \ne \ell$, except for $S_{1}$ and $S_{-\ell}$, and further defining a new variable $\tau_{(\ell)} := - \sigma^{-\ell}/\ell$, it immediately follows
\begin{equation}
\underline{\mathfrak{S}}_{\pi}^{\nu} F \left( z, \tau_{(\ell)} \right) = \exp \left\{ \nu\, S_{-\ell}\, \frac{\partial}{\partial\tau_{(\ell)}} \right\} F \left( z, \tau_{(\ell)} \right) = F \left( z, \tau_{(\ell)} + \nu\, S_{-\ell} \right).
\label{eq:Stokes-at-pi-change-of-var}
\end{equation}
\noindent
One finds that, in this particular case, the Stokes transition along $\theta=\pi$ may also be described by a simple ``jump'' of an adequate transseries parameter\footnote{In terms of the original variable, $\sigma$, this transition is given by $\underline{\mathfrak{S}}_{\pi} F (z,\sigma) = F \left( z, \left( \sigma^{-\ell} - \ell\,  S_{-\ell} \right)^{-1/\ell} \right)$.} and governed by the associated Stokes constant, as in
\begin{equation}
\sigma^{-\ell} \rightarrow \sigma^{-\ell} -\ell\, S_{-\ell}.
\end{equation}
\noindent
This has exactly the same form as the case for $\theta=0$ which we described above, and for which one would have $\ell=-1$. The role of each Stokes constant, $S_{\ell}$, by itself, is very similar; it is just associated to a ``jump'' of the corresponding power of the transseries parameter, $\sigma^{\ell}$, into $\sigma^{\ell}+\ell\, S_{\ell}$. The full set of Stokes constants will naturally lead to more intricate transitions, as described by the function $\BS_\pi (\sigma)$, but the building blocks of these transitions are simple to understand.

%%%%%%%%%%%%%%%%%%%%%%%%%%%%%%%%%%%%%%%%%%%%%%%%%%%%%%%%%%%%%%%%%
%%%%%%%%%%%%%%%%%%%%%%%%%%%%%%%%%%%%%%%%%%%%%%%%%%%%%%%%%%%%%%%%%
\section{Nonperturbative Ambiguities and One--Parameter Transseries}
%%%%%%%%%%%%%%%%%%%%%%%%%%%%%%%%%%%%%%%%%%%%%%%%%%%%%%%%%%%%%%%%%
%%%%%%%%%%%%%%%%%%%%%%%%%%%%%%%%%%%%%%%%%%%%%%%%%%%%%%%%%%%%%%%%%

Having understood Stokes phenomena/transitions, one may now proceed to address the nonperturbative ambiguity along either the $\theta=0$ or $\theta=\pi$ Stokes lines. Along the $\theta=0$ singular direction the nonperturbative ambiguity may be canceled by selecting transseries solutions obeying particular reality conditions. This ambiguity is
\begin{equation}
\left(\CS_{0^{+}}-\CS_{0^{-}}\right) F(z,\sigma) \ne 0.
\end{equation}
\noindent
A \textit{real} transseries solution will automatically have the ambiguity canceled. But there are physical examples, and even examples within the realm of ordinary differential equations, where one is interested in finding real solutions across the full real line in $z$, both positive and negative. In this case there is also an ambiguity at $\theta=\pi$ which needs to be canceled, \textit{i.e.}, one further needs
\begin{equation}
\left(\CS_{\pi^{+}}-\CS_{\pi^{-}}\right)F(z,\sigma)=0.
\end{equation}
\noindent
In this section we shall study the restrictions which arise from each of these conditions separately, $\theta=0$ and $\theta=\pi$, as well as from their eventual combination.

There is a crucial observation to be made at this point. The perturbative expansions of the type \eqref{eq:d} or \eqref{eq:pert-instanton-expansions-Phi_n}, with instanton factor as in \eqref{eq:IntroFn-def}, are in some sense ``special'': they have the most adequate form to simplify the calculations we address in this paper. However, experience from examples tells us that the variable $z$ appearing in the aforementioned expressions is usually not the variable one starts off with. Rather, given either a quantum theoretical problem with perturbative coupling $\kappa$, or some differential equation in the variable $\kappa$, one commonly has to do some (mild) rescaling $z=\kappa^{\alpha}$ in order to write a transseries with the precise structure as in \eqref{eq:Ftransseries}. In this case, one has to be careful with what it means to require reality of the transseries solution in the full real line---physically this would be the full real line in the original coupling $\kappa$, but it may differ from the reality requirements with respect to our ``working'' variable $z$. We will discuss this issue in detail later on, but let us point out for the moment that for real positive coupling, cancelation of ambiguities along the $\theta=0$ singular direction is somewhat insensitive to this issue. For negative real $\kappa$ things are slightly trickier as this may be a Stokes line, an anti--Stokes line, or none at all. In the following we shall assume that the relation $z=\kappa$ holds, where reality across the whole real line forces cancelation of ambiguities in both $\theta=0,\pi$ directions. How to disentangle these results in the case $z=\kappa^{\alpha}$ will then be addressed in section \ref{sec:Monodromy-and-couplings}.

%%%%%%%%%%%%%%%%%%%%%%%%%%%%%%%%%%%%%%%%%%%%%%%%%%%%%%%%%%%%%%%%%
\subsection*{Cancelation of the Nonperturbative Ambiguity Along $\theta=0$}
%%%%%%%%%%%%%%%%%%%%%%%%%%%%%%%%%%%%%%%%%%%%%%%%%%%%%%%%%%%%%%%%%

In appendix \ref{sec:App-Properties-of-med-resumm} we showed that the one Stokes constant associated to the $\theta=0$ Stokes line, in the one--parameter transseries setting, is purely imaginary, $S_{1}\in\rmi\mathbb{R}$. Using this information, let us next try to explicitly determine the median resummation along this Stokes line, as defined in \eqref{smednutheta}. Further using properties we gave in appendix \ref{sec:App-Properties-of-med-resumm} concerning complex conjugation and Stokes transitions, one first finds
\begin{equation}
\mathcal{H} F_{\BR} (z,\sigma) = \mathcal{H} \circ \CS_{0^{+}} \circ \underline{\mathfrak{S}}_{0}^{-\nu} F (z,\sigma) = \CS_{0^{-}} \circ \mathcal{H} \circ \underline{\mathfrak{S}}_{0}^{-\nu} F (z,\sigma).
\end{equation}
\noindent
But if the imaginary ambiguity canceled and we are left with a real solution, one must demand the transseries satisfies $\mathcal{H} F_{\BR}(z,\sigma)=F_{\BR}(z,\sigma)$. In this case via \eqref{smednutheta} one must have
\begin{equation}
\CS_{0^{-}} \circ \mathcal{H} \circ \underline{\mathfrak{S}}_{0}^{-\nu} F (z,\sigma) = \CS_{0^{-}} \circ \underline{\mathfrak{S}}_{0}^{1-\nu} F (z,\sigma),
\end{equation}
\noindent
implying that
\begin{equation}
\mathcal{H} \circ \underline{\mathfrak{S}}_{0}^{-\nu} F (z,\sigma) = \underline{\mathfrak{S}}_{0}^{1-\nu} F (z,\sigma).
\end{equation}
\noindent
Using the Stokes transition in (\ref{eq:Stokes-trans-anypower-zero}), we can rewrite this last equation as
\begin{equation}
F \left( z, \overline{\sigma}-\nu\, \overline{S}_{1} \right) = F \left( z, \sigma + \left( 1-\nu \right) S_{1} \right).
\end{equation}
\noindent
Finally recalling $\overline{S}_{1}=-S_{1}$, the reality condition requires
\begin{equation}
- 2 \rmi\, \im\, \sigma = \left(1-2\nu\right) S_{1}.
\end{equation}
\noindent
Note that $\nu$ is not fixed by reality. What this result shows is that one can, in principle, choose different prescriptions for the median resummation while still canceling the imaginary nonperturbative ambiguity and obtaining a real transseries solution along the $\theta=0$ direction. As shown above, different prescriptions simply translate to different imaginary parts of the transseries parameter $\sigma$. This will not change the final result and one is free to choose the ``natural'' prescription where $\sigma \in \BR$, corresponding to $\nu=1/2$. In fact, this particular prescription is the most common one in resurgent analysis, see, \textit{e.g.}, \citep{Delabaere:1999ef}, and is the one which was already mentioned in section \ref{sec:Nonpert-Ambiguity-real-transs}. Furthermore, this is the only prescription which verifies
\be
\CH \circ \CS^{\text{med}}_\theta = \CS^{\text{med}}_\theta \circ \CH.
\ee
\noindent
To summarize, using the median resummation as defined in \eqref{eq:Median-sum-def} for the direction $\theta=0$, the real transseries is given by
\begin{equation}
F_{\BR} \left(z,\sigma\right) = \CS_{0^{+}} F \left( z,\sigma-\frac{1}{2}S_{1} \right) = \CS_{0^{-}} F \left( z,\sigma+\frac{1}{2}S_{1} \right).
\label{eq:real-one-transs-zero}
\end{equation}
\noindent
This transseries obeys $\mathcal{H}F_{\BR}(z,\sigma)=F_{\BR}(z,\sigma)$ if and only if
\begin{equation}
\sigma \in \mathbb{R}, \qquad S_{1} \in \rmi\mathbb{R}.
\end{equation}

In section \ref{sec:Nonpert-Ambiguity-real-transs} we motivated the median resummation with an exercise of canceling ambiguities order by order in instanton number. Now that it should be clear this is the correct prescription, we may use its complete final expression to understand, iteratively to all orders, how the cancelation of the ambiguities occurs within the transseries (see \citep{Marino:2008ya} as well). In appendix \ref{sec:App-Formulae-One-Param-transs} we have discussed how to compute $\im\, F \left(z,\sigma\right)$, \eqref{eq-app-imFgeneral}, whose first few terms are given by (here we set $\sigma = \sigma_{\text{R}} + \rmi \sigma_{\text{I}}$)
\begin{eqnarray}
\im\, F \left( z,\sigma \right) &=& \im\, F^{(0)} + \sigma_{\text{I}}\, \re\, F^{(1)} + \sigma_{\text{R}}\, \im\, F^{(1)} + 2 \sigma_{\text{R}} \sigma_{\text{I}}\, \re\, F^{(2)} + \left( \sigma_{\text{R}}^{2} - \sigma_{\text{I}}^{2} \right) \im\, F^{(2)} + \nonumber \\
&&
+ \left( 3 \sigma_{\text{R}}^{2} \sigma_{\text{I}} - \sigma_{\text{I}}^{3} \right) \re\, F^{(3)} + \left( \sigma_{\text{R}}^{3} - 3 \sigma_{\text{R}} \sigma_{\text{I}}^{2} \right) \im\, F^{(3)} + \cdots.
\label{imfexpansion}
\end{eqnarray}
\noindent
This expression explicitly shows ambiguities arising in different perturbative and multi--instanton sectors, but because it includes both real and imaginary contributions it is still not very useful. Now recall that the ambiguities may be evaluated by the Stokes automorphism and, in particular, one may rewrite all these imaginary terms as expansions of real, higher--order nonperturbative sectors as in \eqref{eq:App-1-param-im-part-F(n)}. Once this is done, one may explicitly relate the real and imaginary terms in \eqref{imfexpansion} above. This is done in table \ref{tab:Cancelations-1-parameter}, where we have separately displayed the terms which contribute to each of the real and imaginary contributions appearing in $\im\, F \left(z,\sigma\right)$ above, and explain how they all cancel each other. The coefficients that appear in this table are the contributions associated to the expansion of the corresponding term as $\im\,  F^{(\ell)} \sim \sum_{a} C_{\ell}^{a}\, \re\, F^{(a)}$. Using \eqref{eq:App-1-param-im-part-F(n)}, the first few are given by
\begin{alignat}{4}
C_{0}^{1} &= \frac{1}{2\rmi}\, S_{1}, & C_{0}^{3} &= -\frac{1}{4\rmi} \left( S_{1} \right)^{3}, & C_{0}^{5} &= \frac{1}{2\rmi} \left( S_{1} \right)^{5}, \\
C_{1}^{2} &= \frac{1}{\rmi}\, S_{1}\, \sigma_{\text{R}}, & C_{1}^{4} &= - \frac{1}{\rmi} \left( S_{1} \right)^{3} \sigma_{\text{R}}, & C_{1}^{6} &= \frac{3}{\rmi} \left( S_{1} \right)^{5} \sigma_{\text{R}}, \\
C_{2}^{3} &= \frac{3}{2\rmi}\, S_{1} \left( \sigma_{\text{R}}^{2}-\sigma_{\text{I}}^{2} \right), & C_{2}^{5} &= - \frac{5}{2\rmi} \left( S_{1} \right)^{3} \left( \sigma_{\text{R}}^{2}-\sigma_{\text{I}}^{2} \right), & &\cdots, \\
C_{3}^{4} &= \frac{2}{\rmi}\, S_{1} \left( \sigma_{\text{R}}^{3}-3\sigma_{\text{R}}\sigma_{\text{I}}^{2} \right), \qquad & C_{3}^{6} &= - \frac{5}{\rmi} \left( S_{1} \right)^{3} \left( \sigma_{\text{R}}^{3}-3\sigma_{\text{R}}\sigma_{\text{I}}^{2} \right), \qquad & &\cdots.
\label{eq:Coefficients-of-expansion-imF-1-parameter}
\end{alignat}
\noindent
As we are rewriting $\im\, F \left(z,\sigma\right)$ solely in terms of real multi--instanton contributions, then in order to cancel the ambiguity each of the rows in table  \ref{tab:Cancelations-1-parameter} needs to add up to zero (a complete closed--form expression may be found in \eqref{eq:app1imFfullformula}---cancelation of the rows in table \ref{tab:Cancelations-1-parameter} translates to cancelation of all coefficients in \eqref{eq:app1imFfullformula}). Noting that for each fixed $\ell$ the first non--zero coefficient in the $\im\, F^{(\ell)}$ column is $C_{\ell}^{\ell+1}$, it follows that a fixed $a$ row $\re\, F^{(a)}$ only receives contributions from terms $\im\, F^{(\ell)}$ with $\ell<a$ (alongside with the natural term proportional to $\re\, F^{(a)}$). This truncation allows us to obtain a constraint for $\sigma_{\text{I}}$. For example, look at the first data--row of the table: adding the terms in $\im\, F \left(z,\sigma\right)$ proportional to $\re\, F^{(1)}$ yields
\begin{equation}
C_{0}^{1} + \sigma_{\text{I}}=0 \qquad \Leftrightarrow \qquad \sigma_{\text{I}} = \frac{\rmi}{2}S_{1}.
\end{equation}
\noindent
Without surprise this is the expected constraint in $\sigma$ for finding a real transseries solution \eqref{eq:real-one-transs-zero}, along the direction $\theta=0$. Also as discussed in appendix \ref{sec:App-Formulae-One-Param-transs}, but interesting to observe explicitly, is that given this sole constraint all other rows in our table automatically add up to zero; \textit{e.g.},  the rows corresponding to $\re\, F^{(2)}$ and $\re\, F^{(3)}$ can be easily seen to vanish, while the following rows would cancel with other terms which were not included in the table, and so on.

\begin{table}
\noindent
\begin{centering}
\begin{tabular}{|c|c|c|c|c|c|c|c|c}
\cline{2-8} 
\multicolumn{1}{c|}{{\scriptsize \begin{picture}(33,15)(0,0) 
\put(17,5){$\im\, F$} 
\put(4.6,10){\line(2,-1){33.3}} 
\put(-4,-5){} 
\end{picture}}}
& {\scriptsize $\im\, F^{(0)}$} & {\scriptsize $\sigma_{\text{I}}\, \re\, F^{(1)}$} & {\scriptsize $\sigma_{\text{R}}\, \im\, F^{(1)}$} & {\scriptsize $2\sigma_{\text{R}}\sigma_{\text{I}}\, \re\, F^{(2)}$} & {\scriptsize $\begin{array}{c} \left( \sigma_{\text{R}}^{2}-\sigma_{\text{I}}^{2} \right) \\ \times \im\, F^{(2)} \end{array}$} & {\scriptsize $\begin{array}{c} \left( 3\sigma_{\text{R}}^{2}\sigma_{\text{I}}-\sigma_{\text{I}}^{3} \right) \\ \times \re\, F^{(3)} \end{array}$} & {\scriptsize $\begin{array}{c} \left( \sigma_{\text{R}}^{3}-3\sigma_{\text{R}}\sigma_{\text{I}}^{2} \right) \\ \times \im\, F^{(3)} \end{array}$} & \tabularnewline
\cline{1-8} 
{\scriptsize $\re\, F^{(1)}$} & {\scriptsize $C_{0}^{1}$} & {\scriptsize $\sigma_{\text{I}}$} & {\scriptsize 0} & {\scriptsize -} & {\scriptsize 0} & {\scriptsize -} & {\scriptsize 0} & \tabularnewline
\cline{1-8} 
{\scriptsize $\re\, F^{(2)}$} & {\scriptsize 0} & {\scriptsize -} & {\scriptsize $C_{1}^{2}$} & {\scriptsize $2\sigma_{\text{R}}\sigma_{\text{I}}$} & {\scriptsize 0} & {\scriptsize -} & {\scriptsize 0} & \tabularnewline
\cline{1-8} 
{\scriptsize $\re\, F^{(3)}$} & {\scriptsize $C_{0}^{3}$} & {\scriptsize -} & {\scriptsize 0} & {\scriptsize -} & {\scriptsize $C_{2}^{3}$} & {\scriptsize $\left(3\sigma_{\text{R}}^{2}\sigma_{\text{I}}-\sigma_{\text{I}}^{3}\right)$} & {\scriptsize 0} & \tabularnewline
\cline{1-8} 
{\scriptsize $\re\, F^{(4)}$} & {\scriptsize 0} & {\scriptsize -} & {\scriptsize $C_{1}^{4}$} & {\scriptsize -} & {\scriptsize 0} & {\scriptsize -} & {\scriptsize $C_{3}^{4}$} & \tabularnewline
\cline{1-8} 
{\scriptsize $\re\, F^{(5)}$} & {\scriptsize $C_{0}^{5}$} & {\scriptsize -} & {\scriptsize 0} & {\scriptsize -} & {\scriptsize $C_{2}^{5}$} & {\scriptsize -} & {\scriptsize 0} & \tabularnewline
\cline{1-8} 
{\scriptsize $\re\, F^{(6)}$} & 0 & {\scriptsize -} & {\scriptsize $C_{1}^{6}$} & {\scriptsize -} & {\scriptsize -} & {\scriptsize -} & {\scriptsize $C_{3}^{6}$} & \tabularnewline
\cline{1-8}
\end{tabular}
\par
\end{centering}
\noindent
\begin{centering}
\par
\end{centering}
\centering{}
\caption{Cancelations of the first few terms contributing to the ambiguity of a one--parameter transseries. Each column corresponds to each term contributing to the full $\im\, F \left(z,\sigma\right)$ in \eqref{imfexpansion}. The ambiguities associated to each perturbative or multi--instanton sector are evaluated via \eqref{eq:App-1-param-im-part-F(n)} where we show these first contributions to the expansion $\im\, F^{(\ell)} \sim \sum_{a} C_{\ell}^{a}\, \re\, F^{(a)}$ up to $a=6$. In this rewriting, $\im\, F \left(z,\sigma\right)$ is expanded in real contributions from multi--instanton sectors. To cancel the ambiguity at the transseries level, the coefficients proportional to each $\re\, F^{(a)}$ need to cancel separately, \textit{i.e.}, each row in the table needs to add up to zero independently. The coefficients $C_{\ell}^{a}$ can be found in the text.
\label{tab:Cancelations-1-parameter}}
\end{table}

%%%%%%%%%%%%%%%%%%%%%%%%%%%%%%%%%%%%%%%%%%%%%%%%%%%%%%%%%%%%%%%%%
\subsection*{Real Transseries in the Real Line?}
%%%%%%%%%%%%%%%%%%%%%%%%%%%%%%%%%%%%%%%%%%%%%%%%%%%%%%%%%%%%%%%%%

Moving on, one might be interested in constructing real solutions not only along $\theta=0$ but along the full real line. In order to achieve this, the first step is to verify the reality constraints specifically associated to $\theta=\pi$. As always, we are considering a transseries with real asymptotic coefficients and real instanton action, and we have $\mathcal{H} \circ \CS_{\pi^{+}} = \CS_{\pi^{-}} \circ \mathcal{H}$. Using the median resummation with the Stokes transition across $\theta=\pi$ \eqref{eq:Stokes-trans-anypower-pi}, one determines a real transseries as\footnote{Note that, unlike the $\theta=0$ case, setting $\sigma=0$ now yields $\BS^{(\pm1/2)}_\pi (0) = 0$ and there would thus be no instanton corrections in \eqref{eq:real-one-transs-pi}. This is of course perfectly consistent as the perturbative expansion is \textit{not} ambiguous along $\theta=\pi$: in fact, it is only the asymptotic series around fixed multi--instanton sectors, starting at $n=2$ instantons, which will have singularities along $\theta=\pi$ (see also \cite{Aniceto:2011nu}).}
\begin{equation}
F_{\BR,\pi} \left(z,\sigma\right) = \CS_{\pi^{+}} F \left( z, \BS^{(-1/2)}_\pi (\sigma) \right) = \CS_{\pi^{-}} F \left( z, \BS^{(1/2)}_\pi (\sigma) \right),
\label{eq:real-one-transs-pi}
\end{equation}
\noindent
as long as
\begin{equation}
\sigma\in\mathbb{R}.
\end{equation}
\noindent
Indeed, it is simple to check that the reality condition $\mathcal{H}F_{\BR,\pi} (z,\sigma) = F_{\BR,\pi} (z,\sigma)$ is satisfied in this case, given the definition of $\BS^{(\nu)}_\pi (\sigma)$ in \eqref{eq:Def-gsigma-at-pi}, the coefficients $\underline{\Sigma}_{\nu}(n,m)$ defined in \eqref{eq:AppA-coefficients-automorphism-at-pi-nu}, and that we showed in appendix \ref{sec:App-Properties-of-med-resumm} that considering a transseries with real asymptotic coefficients, and taking $\beta=0$, the Stokes coefficients along $\theta=\pi$ are all purely imaginary,
\begin{equation}
S_{-\ell} \in \rmi \mathbb{R}, \quad \forall\,\ell\ge1.
\end{equation}

We can now ask what are the conditions to be met---if any---in order to have a real transseries solution for $z \in \BR$, \textit{i.e.}, along \textit{both} $\theta=0,\pi$ singular directions (this question often arises in contexts dealing with differential equations). In principle the answer is simple, one just needs to satisfy simultaneously the reality constraints we have already discussed, and connect them both together. Starting at $\theta=0^+$, the transseries parameter is fixed to have the structure in \eqref{eq:real-one-transs-zero} with real $\sigma$. Rotating counterclockwise, between $\theta=0^{+}$ and $\theta=\pi^{-}$ there are no Stokes lines where $\sigma$ could have jumped. On the other hand, at $\theta=\pi^{-}$, the transseries parameter is fixed to have the structure \eqref{eq:real-one-transs-pi} which must thus match \eqref{eq:real-one-transs-zero}. In this way, both reality constraints hold if one finds
\begin{equation}
\sigma_{\text{R}}- \frac{1}{2}S_{1} = \sum_{n=0}^{+\infty} \sigma_{\text{R}}^{n+1}\, \underline{\Sigma}_{1/2}(n+1,n).
\end{equation}
\noindent
However, this expression gives us a highly non--trivial relation between $\sigma_{\text{R}}$, $S_{1}$, and all the $S_{-k}$. Verifying if it allows for solutions is probably only possible within specific examples; as it might be verified with a finite number of Stokes constants, an infinite number of Stokes constants, or not verified at all. We tried to solve this constraint generically with a small finite number of (arbitrary) non--vanishing Stokes constants, and thus obtain general conditions for real solutions. Unfortunately we were not able to find any positive result along this line.

%%%%%%%%%%%%%%%%%%%%%%%%%%%%%%%%%%%%%%%%%%%%%%%%%%%%%%%%%%%%%%%%%
\subsection*{Inclusion of the $\beta$ Exponent}
%%%%%%%%%%%%%%%%%%%%%%%%%%%%%%%%%%%%%%%%%%%%%%%%%%%%%%%%%%%%%%%%%

Finally, one may be interested in dealing with more general asymptotic expansions within the transseries, namely with a non--trivial characteristic exponent $\beta$. In this case, let us consider a one--parameter transseries of the form
\begin{equation}
F(z,\sigma) = \sum_{n=0}^{+\infty} \sigma^{n}\, \rme^{-nAz}\, \Phi_{n}(z), \qquad \Phi_{n}(z) \simeq z^{-n\beta} \sum_{g=0}^{+\infty} \frac{F_{g}^{(n)}}{z^{g}},
\end{equation}
\noindent
where $\beta$ is a rational number. As in our previous analysis, we should be able to find a non--ambiguous result for the transseries out of median resummation. First notice that even with the extra factor of $\beta$, the same Stokes automorphisms and Stokes transitions hold. As expected, the expression for the median resummation is thus unchanged, and should be valid at both $\theta=0,\pi$. Now, for $\theta=0$, $z$ is real positive and the factor $z^{-\beta}$ induces no changes on our previous arguments. For $\theta=\pi$, on the other hand, we now have $z=\left|z\right|\rme^{-\rmi\pi}$ and the requirement of reality requires some more thought. If the asymptotic expansions in the transseries have real coefficients, as usual, it is not difficult to obtain
\begin{equation}
\mathcal{H} \circ \mathcal{S}_{\pi^{+}} \circ \underline{\mathfrak{S}}_{\pi}^{-1/2} F \left(z,\sigma\right) = \mathcal{H} \circ \mathcal{S}_{\pi^{+}} F \left(z,\mathbb{S}_{\pi}^{(-1/2)}\left(\sigma\right)\right) = \mathcal{S}_{\pi^{-}} F \left(z, \rme^{-2\pi\rmi\beta}\, \overline{\mathbb{S}_{\pi}^{(-1/2)}\left(\sigma\right)}\right).
\end{equation}
\noindent
Thus, enforcing reality $\mathcal{H} \circ \mathcal{S}^{\text{med}}_\pi F = \mathcal{S}^{\text{med}}_\pi F$ requires
\begin{equation}
\mathbb{S}_{\pi}^{(1/2)} \left(\sigma\right) = \rme^{-2\pi\rmi\beta}\, \overline{\mathbb{S}_{\pi}^{(-1/2)}\left(\sigma\right)}.
\end{equation}
\noindent
Making use of the constraints for the Stokes constants found in appendix \ref{sec:App-Properties-of-med-resumm},
\begin{equation}
\overline{S_{-\ell}^{(1+\ell)}} = - S_{-\ell}^{(1+\ell)}\, \rme^{-2\pi\rmi\ell\beta},
\end{equation}
\noindent
and using the explicit form of $\mathbb{S}_{\pi}^{(\nu)}\left(\sigma\right)$ computed in appendix \ref{sec:App-Formulae-One-Param-transs}, we find the following constraint for the transseries parameter $\sigma$,
\begin{equation}
\sigma = \overline{\sigma}\,\rme^{-2\pi\rmi\beta}.
\end{equation}
\noindent
This is just a generalization of the reality condition found for $\sigma$ when $\beta=0$. This argument is generalizable for resonant transseries including logarithmic sectors, as in appendix \ref{sec:App-Properties-of-med-resumm}.

%%%%%%%%%%%%%%%%%%%%%%%%%%%%%%%%%%%%%%%%%%%%%%%%%%%%%%%%%%%%%%%%%
%%%%%%%%%%%%%%%%%%%%%%%%%%%%%%%%%%%%%%%%%%%%%%%%%%%%%%%%%%%%%%%%%
\section{Stokes Phenomena in Two--Parameter Transseries}
%%%%%%%%%%%%%%%%%%%%%%%%%%%%%%%%%%%%%%%%%%%%%%%%%%%%%%%%%%%%%%%%%
%%%%%%%%%%%%%%%%%%%%%%%%%%%%%%%%%%%%%%%%%%%%%%%%%%%%%%%%%%%%%%%%%

As we said before, Stokes lines create nonperturbative ambiguities for the resummation of perturbation theory, which means the first step to understand how to cancel these ambiguities is to understand exactly what occurs along those directions. We shall now address the two--parameters case, in the form \eqref{eq:Ftransseries} and \eqref{eq:IntroFn-def}. Recall that the two--parameters transseries is a two--fold generalization of the one--parameter case discussed in the previous sections, in the sense that it involves more transseries parameters and, at the same time, it involves a generalized instanton sector. Due to the structure of instanton actions as $\pm A$, in this case the Stokes automorphism along the singular directions $\theta=0$ and $\theta=\pi$ is very symmetric; this can be seen in the formulae found in appendix \ref{sec:App-Formulae-for-Two--Parameters}. From equations \eqref{eq:Stokes-transition-at-zero-2-param-strange-variables} and \eqref{eq:Stokes-transition-at-pi-2-param-strange-variables}, we have
\begin{eqnarray}
\underline{\mathfrak{S}}_{0}^{\nu} F \left(z,\sigma_{1},\sigma_{2}\right) &=& F \left( z, \BS_{0,1}^{(\nu)} (\sigma_{1},\sigma_{2}), \BS_{0,2}^{(\nu)} (\sigma_{1},\sigma_{2}) \right),
\label{eq:Stokes-transition-zero-2-param} \\
\underline{\mathfrak{S}}_{\pi}^{\nu} F \left(z,\sigma_{1},\sigma_{2}\right) &=& F \left( z, \BS_{\pi,1}^{(\nu)} (\sigma_{1},\sigma_{2}), \BS_{\pi,2}^{(\nu)} (\sigma_{1},\sigma_{2}) \right),
\label{eq:Stokes-transition-pi-2-param}
\end{eqnarray}
\noindent
with $\BS_{\theta,i}^{(\nu)}$ defined in appendix \ref{sec:App-Formulae-for-Two--Parameters}. These expressions are much more intricate than in the one--parameter case. In order to better understand the nature of the Stokes transitions and what is the role of the two types of Stokes constants at play in this situation, $S_{\ell}$ and $\widetilde{S}_{\ell}$, we shall now specify particular cases where we set all Stokes constants to zero, except for a small set.

We shall use the definition of the Stokes automorphism \eqref{eq:Stokes-aut-from-alien-derivatives} via alien derivatives,
\begin{eqnarray}
\underline{\mathfrak{S}}_{0}^{\nu} F \left(z,\sigma_{1},\sigma_{2}\right) &=& \exp \left\{ \nu\, \sum_{\ell\ge1}\dot{\Delta}_{\ell A} \right\} F\left(z,\sigma_{1},\sigma_{2}\right), \\
\underline{\mathfrak{S}}_{\pi}^{\nu} F \left(z,\sigma_{1},\sigma_{2}\right) &=& \exp \left\{ \nu\, \sum_{\ell\ge1}\dot{\Delta}_{-\ell A} \right\} F\left(z,\sigma_{1},\sigma_{2}\right),
\end{eqnarray}
\noindent
and further use the expressions for the pointed alien derivatives (acting on the full transseries) given by the bridge equations \eqref{eq:Bridge-eqs-2-param}, to obtain differential operators directly acting on the transseries parameters, $\sigma_1, \sigma_2$, which may be easier to handle---at least in special cases. All we need are the expansions of the Stokes factors appearing in the bridge equations \eqref{eq:Bridge-eqs-2-param}, $S_{\ell}$ and $\widetilde{S}_{\ell}$, written in terms of Stokes constants. They are \citep{Aniceto:2011nu}
\begin{eqnarray}
S_{\ell} \left(\sigma_{1},\sigma_{2}\right) &=& \sum_{k=\max(0,\ell-1)}^{+\infty} S_{\ell}^{(k+1-\ell)} \sigma_{1}^{k+1-\ell} \sigma_{2}^{k},
\label{eq:Expansion-of-Sell-2-param} \\
\widetilde{S}_{\ell} \left(\sigma_{1},\sigma_{2}\right) &=& \sum_{k=\max(0,-\ell-1)}^{+\infty} \widetilde{S}_{\ell}^{(k+1+\ell)} \sigma_{1}^{k} \sigma_{2}^{k+1+\ell}.
\label{eq:Expansion-of-Stildeell-2-param}
\end{eqnarray}
\noindent
We now have all the required information to analyze Stokes phenomena for the following cases:

%%%%%%%%%%%%%%%%%%%%%%%%%%%%%%%%%%%%%%%%%%%%%%%%%%%%%%%%%%%%%%%%%
\subsection*{Case $\widetilde{S}_{\ell}^{(k)}=0$, $\forall \ell,k$}
%%%%%%%%%%%%%%%%%%%%%%%%%%%%%%%%%%%%%%%%%%%%%%%%%%%%%%%%%%%%%%%%%

This case is expected to be very similar to the one--parameter case studied in a previous section. In this situation, the Stokes automorphisms become
\begin{eqnarray}
\underline{\mathfrak{S}}_{0}^{\nu} F \left(z,\sigma_{1},\sigma_{2}\right) &=& \exp \left\{ \nu\, \sum_{\ell\ge1} \sum_{k=0}^{+\infty} S_{\ell}^{(k)} \sigma_{1}^{k} \sigma_{2}^{k+\ell-1} \frac{\partial}{\partial\sigma_{1}} \right\} F \left(z,\sigma_{1},\sigma_{2}\right), \\
\underline{\mathfrak{S}}_{\pi}^{\nu} F \left(z,\sigma_{1},\sigma_{2}\right) &=& \exp \left\{ \nu\, \sum_{\ell\ge1} \sum_{k=\ell+1}^{+\infty} S_{-\ell}^{(k)} \sigma_{1}^{k} \sigma_{2}^{k-\ell-1} \frac{\partial}{\partial\sigma_{1}} \right\} F \left(z,\sigma_{1},\sigma_{2}\right).
\end{eqnarray}
\noindent
We shall further restrict ourselves to a specific case where all Stokes constants are zero, $S_{\ell}^{(n)}=0$, except for two, $S_{-m}^{(a)}$ and $S_{\ell}^{(k)}$ for some fixed $\ell,m>0$ ($k\ge0$, $a\ge m+1$). To follow the strategy used in the one--parameter case, we want to find appropriate changes of variables, $\tau_{(k)}(\sigma_1)$, $\tau_{(a)}(\sigma_1)$, such that
\begin{equation}
\sigma_{1}^{r}\, \frac{\partial\tau_{(r)}}{\partial\sigma_{1}}=1.
\end{equation}
\noindent
In terms of these new variables we find
\begin{eqnarray}
\underline{\mathfrak{S}}_{0}^{\nu} F \left(z,\tau_{(k)},\sigma_{2}\right) &=& F \left( z, \tau_{(k)}+\nu\, S_{\ell}^{(k)}\sigma_{2}^{k+\ell-1}, \sigma_{2} \right), \\
\underline{\mathfrak{S}}_{\pi}^{\nu} F \left(z,\tau_{(a)},\sigma_{2}\right) &=& F \left( z, \tau_{(a)}+\nu\, S_{-m}^{(a)}\sigma_{2}^{a-m-1}, \sigma_{2} \right).
\end{eqnarray}
\noindent
It immediately follows that the action of the Stokes constants $S_{-m}^{(a)}$ and $S_{\ell}^{(k)}$ translates to shifts of the variables $\tau_{(r)}$, and that the Stokes transitions only affect the transseries parameter $\sigma_{1}$; the sector governed by the parameter $\sigma_{2}$ remains untouched. It is possible to be even more explicit in various different cases (setting $\nu=1$):
\begin{itemize}
\item $k=0$

One has $\tau_{(0)}=\sigma_{1}$, and the ``jumps'' in the original transseries variables $\sigma_{1}$, $\sigma_{2}$, are trivially given by
\begin{equation}
\sigma_{1} \rightarrow \sigma_{1} + S_{\ell}^{(0)} \sigma_{2}^{\ell-1}, \qquad \sigma_{2} \rightarrow \sigma_{2}.
\end{equation}
\noindent
Specifically for $\ell=1$ we find that the two sectors, $\sigma_{1}$ and $\sigma_{2}$, are completely decoupled.
\item $k=1$

One finds $\tau_{(1)} = \log \sigma_{1}$, and the transitions in the variables $\sigma_{1},\sigma_{2}$ become
\be
\sigma_{1} \rightarrow \sigma_{1}\, \rme^{S_{\ell}^{(1)}\sigma_{2}^{\ell}}, \qquad \sigma_{2} \rightarrow \sigma_{2}.
\ee
\item $k>1$, $a\ge m+1$

Choosing $r=k$ or $r=a$, we have $\tau_{(r)}=\sigma_{1}^{1-r}/(1-r)$. The corresponding transitions are, for $\theta=0$,
\begin{equation}
\sigma_{1}^{1-k} \rightarrow \sigma_{1}^{1-k} + \left(1-k\right) S_{\ell}^{(k)} \sigma_{2}^{k+\ell-1}, \qquad \sigma_{2} \rightarrow \sigma_{2},
\end{equation}
\noindent
and, for $\theta=\pi$,
\begin{equation}
\sigma_{1}^{1-a} \rightarrow \sigma_{1}^{1-a} + \left(1-a\right) S_{-m}^{(a)} \sigma_{2}^{a-m-1}, \qquad \sigma_{2} \rightarrow \sigma_{2}.
\end{equation}
\noindent
Note that for $a=m+1$, we again find a decoupling of the two sectors, $\sigma_{1}$, $\sigma_{2}$.
\end{itemize}

%%%%%%%%%%%%%%%%%%%%%%%%%%%%%%%%%%%%%%%%%%%%%%%%%%%%%%%%%%%%%%%%%
\subsection*{Case $S_{\ell}^{(k)}=0$, $\forall \ell,k$}
%%%%%%%%%%%%%%%%%%%%%%%%%%%%%%%%%%%%%%%%%%%%%%%%%%%%%%%%%%%%%%%%%

We are now interested in analyzing the role of the ``symmetric'' Stokes constants $\widetilde{S}_{\ell}^{(k)}$ within the Stokes transitions. In this case, the Stokes automorphisms become
\begin{eqnarray}
\underline{\mathfrak{S}}_{0}^{\nu} F \left(z,\sigma_{1},\sigma_{2}\right) &=& \exp \left\{ \nu\, \sum_{\ell\ge1} \sum_{k=\ell+1}^{+\infty} \widetilde{S}_{\ell}^{(k)} \sigma_{1}^{k-\ell-1} \sigma_{2}^{k} \frac{\partial}{\partial\sigma_{2}} \right\} F \left(z,\sigma_{1},\sigma_{2}\right), \\
\underline{\mathfrak{S}}_{\pi}^{\nu} F \left(z,\sigma_{1},\sigma_{2}\right) &=& \exp \left\{ \nu\, \sum_{\ell\ge1} \sum_{k=0}^{+\infty} \widetilde{S}_{-\ell}^{(k)} \sigma_{1}^{k+\ell-1} \sigma_{2}^{k} \frac{\partial}{\partial\sigma_{2}} \right\} F \left(z,\sigma_{1},\sigma_{2}\right).
\end{eqnarray}
\noindent
Similarly to the previous case, we will address the situation where all Stokes constants are zero, $\widetilde{S}_{\ell}^{(n)}=0$, except for two, $\widetilde{S}_{-m}^{(a)}$ and $\widetilde{S}_{\ell}^{(k)}$ for some fixed $\ell,m>0$ ($k\ge\ell+1$, $a\ge0$). As before, we want to find appropriate changes of variables $\gamma_{(k)}(\sigma_{2})$, $\gamma_{(a)}(\sigma_{2})$, such that
\begin{equation}
\sigma_{2}^{r}\, \frac{\partial\gamma_{(r)}}{\partial\sigma_{2}} = 1.
\end{equation}
\noindent
In terms of these new variables we find
\begin{eqnarray}
\underline{\mathfrak{S}}_{0}^{\nu} F \left(z,\sigma_{1},\gamma_{(k)}\right) &=& F \left( z, \sigma_{1}, \gamma_{(k)}+\nu\,\widetilde{S}_{\ell}^{(k)}\sigma_{1}^{k-\ell-1} \right), \\
\underline{\mathfrak{S}}_{\pi}^{\nu} F \left(z,\sigma_{1},\gamma_{(a)}\right) &=& F \left( z, \sigma_{1}, \gamma_{(a)}+\nu\,\widetilde{S}_{-m}^{(a)}\sigma_{1}^{a+m-1} \right).
\end{eqnarray}
\noindent
Again, the action of the Stokes constants $\widetilde{S}_{-m}^{(a)}$, $\widetilde{S}_{\ell}^{(k)}$ translates to shifts of the variables $\gamma_{(r)}$, and the Stokes transitions only affect the transseries parameter $\sigma_{2}$; the sector governed by the parameter $\sigma_{1}$ remains untouched.  The explicit changes of variables $\gamma_{(r)}$ have the exact same form as the previous case.

%%%%%%%%%%%%%%%%%%%%%%%%%%%%%%%%%%%%%%%%%%%%%%%%%%%%%%%%%%%%%%%%%
\subsection*{Case $S_{\ell}^{(k)}$ and $\widetilde{S}_{\ell'}^{(k')}$ Non--Zero}
%%%%%%%%%%%%%%%%%%%%%%%%%%%%%%%%%%%%%%%%%%%%%%%%%%%%%%%%%%%%%%%%%

Let us finally consider the case of having two non--vanishing Stokes constants of different ``type'', $S_{\ell}^{(k)}$ and $\widetilde{S}_{\ell'}^{(k')}$. The only non--trivial case is when either $\ell$ and $\ell'$ are both positive, or both negative. If one is positive and the other negative, then the results are given by the expressions already found in the previous cases: for the transition at $\theta=0$ and $\theta=\pi$, respectively,
\begin{enumerate}
\item If $S_{\ell}^{(k)}, \widetilde{S}_{-m}^{(a)} \ne 0$ ($\ell,m>0$) we have
\begin{eqnarray}
\underline{\mathfrak{S}}_{0}^{\nu} F \left(z,\tau_{(k)},\sigma_{2}\right) &=& F \left( z, \tau_{(k)}+\nu\,S_{\ell}^{(k)}\sigma_{2}^{k+\ell-1}, \sigma_{2} \right), \\
\underline{\mathfrak{S}}_{\pi}^{\nu} F \left(z,\sigma_{1},\gamma_{(a)}\right) &=& F \left( z, \sigma_{1}, \gamma_{(a)}+\nu\,\widetilde{S}_{-m}^{(a)}\sigma_{1}^{a+m-1} \right).
\end{eqnarray}
\item If $S_{-m}^{(a)}, \widetilde{S}_{\ell}^{(k)} \ne 0$ ($\ell,m>0$) we have
\begin{eqnarray}
\underline{\mathfrak{S}}_{0}^{\nu} F \left(z,\sigma_{1},\gamma_{(k)}\right) &=& F \left( z, \sigma_{1}, \gamma_{(k)}+\nu\,\widetilde{S}_{\ell}^{(k)}\sigma_{1}^{k-\ell-1} \right), \\
\underline{\mathfrak{S}}_{\pi}^{\nu} F \left(z,\tau_{(a)},\sigma_{2}\right) &=& F \left( z, \tau_{(a)}+\nu\,S_{-m}^{(a)}\sigma_{2}^{a-m-1}, \sigma_{2} \right).
\end{eqnarray}
\end{enumerate}

The first non--trivial case is then if $S_{\ell}^{(k)}, \widetilde{S}_{\ell'}^{(k')} \ne 0$ ($\ell,\ell'>0$, $k\ge0$, $k'\ge\ell'+1$). In this case, the transition at $\theta=\pi$ will be trivial, $\underline{\mathfrak{S}}_{\pi}^{\nu} F \left(z,\sigma_{1},\sigma_{2}\right) = F \left(z,\sigma_{1},\sigma_{2}\right)$. At $\theta=0$, on the other hand, the Stokes transition is non--trivial. Here, the Stokes automorphism will be given by
\begin{equation}
\underline{\mathfrak{S}}_{0}^{\nu} F \left(z,\sigma_{1},\sigma_{2}\right) = \exp \left\{ \nu\, S_{\ell}^{(k)} \sigma_{2}^{k+\ell-1} \sigma_{1}^{k} \frac{\partial}{\partial\sigma_{1}} + \nu\, \widetilde{S}_{\ell'}^{(k')} \sigma_{1}^{k'-\ell'-1} \sigma_{2}^{k'} \frac{\partial}{\partial\sigma_{2}} \right\} F \left(z,\sigma_{1},\sigma_{2}\right).
\end{equation}
\noindent
The general strategy to solve this problem has the same flavor as in the previous cases. We want to find two new independent variables, $x(\sigma_{1},\sigma_{2})$ and $y(\sigma_{1},\sigma_{2})$, that are two independent solutions of the following differential equation for $f(\sigma_{1},\sigma_{2})$,
\begin{equation}
\nu\, S_{\ell}^{(k)} \sigma_{2}^{k+\ell-1} \sigma_{1}^{k} \frac{\partial f}{\partial\sigma_{1}} + \nu\, \widetilde{S}_{\ell'}^{(k')} \sigma_{1}^{k'-\ell'-1} \sigma_{2}^{k'} \frac{\partial f}{\partial\sigma_{2}} = \mathrm{constant}.
\label{eq:Stokes-aut-for zero-solving-for-new-variables}
\end{equation}
\noindent
If this can be done, then the Stokes automorphism becomes again a simple shift of the appropriate variables,
\begin{equation}
\underline{\mathfrak{S}}_{0}^{\nu} F \left(z,x,y\right) = \exp \left\{ \frac{\partial}{\partial x} + \frac{\partial}{\partial y} \right\} F \left(z,x,y\right) = F \left(z,x+1,y+1\right).
\label{eq:Stokes-aut-at-zero-in-simple-variables}
\end{equation}
\noindent
Let us be more explicit in various different cases:
\begin{itemize}
\item $k=0$, $\ell=1$, $k'=\ell'+1\ge2$, $S_{1}^{(0)}$, $\widetilde{S}_{\ell'}^{(\ell'+1)} \ne 0$

In this case the sectors corresponding to the two transseries parameters $\sigma_{1}$ and $\sigma_{2}$ decouple, and the Stokes automorphism simply becomes
\begin{eqnarray}
\underline{\mathfrak{S}}_{0}^{\nu} F \left(z,\sigma_{1},\sigma_{2}\right) &=& \exp \left\{ \nu\, S_{1}^{(0)} \frac{\partial}{\partial\sigma_{1}} \right\} \exp \left\{ \nu\, \widetilde{S}_{\ell'}^{(\ell'+1)} \sigma_{2}^{\ell'+1} \frac{\partial}{\partial\sigma_{2}} \right\} F\left(z,\sigma_{1},\sigma_{2}\right).
\end{eqnarray}
\noindent
If one implements the change of variables $\widetilde{\sigma}_{2}=\sigma_{2}^{-\ell'}$, it follows
\begin{equation}
\underline{\mathfrak{S}}_{0}^{\nu} F \left(z,\sigma_{1},\widetilde{\sigma}_{2}\right) = F \left( z, \sigma_{1}+\nu\, S_{1}^{(0)}, \widetilde{\sigma}_{2}-\nu\,\ell'\,\widetilde{S}_{\ell'}^{(\ell'+1)} \right).
\end{equation}
\noindent
In the original variables, crossing the $\theta=0$ Stokes line thus corresponds to yet another shift of their adequate combination. One has
\begin{equation}
\sigma_{1} \rightarrow \sigma_{1}+\nu\,S_{1}^{(0)}, \qquad \sigma_{2}^{-\ell'} \rightarrow \sigma_{2}^{-\ell'}-\nu\,\ell'\,\widetilde{S}_{\ell'}^{(\ell'+1)}.
\end{equation}
\item $k=0$, $\ell=1$, $k'=\ell'+\alpha$, $\alpha\ge2$, $S_{1}^{(0)}$, $\widetilde{S}_{\ell'}^{(\ell'+\alpha)} \ne 0$

In this case the decoupling of the two sectors, $\sigma_{1}$ and $\sigma_{2}$, will no longer occur. Following the strategy outlined above, let us find the new variables, $x$ and $y$, which obey the differential equation \eqref{eq:Stokes-aut-for zero-solving-for-new-variables}. They are
\begin{eqnarray}
x & = & \frac{\sigma_{1}}{\nu S_{1}^{(0)}}, \\
y & = & \frac{\sigma_{1}}{\nu S_{1}^{(0)}}-\frac{\sigma_{2}^{-(\ell'+\alpha-1)}}{\ell'+\alpha-1}-\widetilde{S}_{\ell'}^{(\ell'+\alpha)}\frac{\sigma_{1}^{\alpha}}{\alpha S_{1}^{(0)}}.
\end{eqnarray}
\noindent
In these variables the Stokes automorphism has the simple form \eqref{eq:Stokes-aut-at-zero-in-simple-variables}. In the original variables we find
\begin{equation}
\sigma_{1} \rightarrow \sigma_{1}+\nu\,S_{1}^{(0)}, \quad \sigma_{2}^{-(\ell'+\alpha-1)} \rightarrow \sigma_{2}^{-(\ell'+\alpha-1)} - \left(\ell'+\alpha-1\right)\frac{\widetilde{S}^{(\ell'+\alpha)}}{\alpha\,S_{1}^{(0)}} \left( \left(\sigma_{1}+\nu\,S_{1}^{(0)}\right)^{\alpha}-\sigma_{1}^{\alpha} \right).
\end{equation}
\item General

In the general case we want to solve
\begin{equation}
\underline{\mathfrak{S}}_{0}^{\nu} F \left(z,\sigma_{1},\sigma_{2}\right) = \exp \left\{ C\, \sigma_{1}^{a_{1}}\sigma_{2}^{a_{2}} \frac{\partial}{\partial\sigma_{1}} + \widetilde{C}\, \sigma_{1}^{b_{1}}\sigma_{2}^{b_{2}} \frac{\partial}{\partial\sigma_{2}} \right\} F\left(z,\sigma_{1},\sigma_{2}\right),
\end{equation}
\noindent
where we defined $C=\nu\,S_{\ell}^{(k)}$, $a_{1}=k$, $a_{2}=k+\ell-1$, and $\widetilde{C}=\nu\,\widetilde{S}_{\ell'}^{(k')}$, $b_{1}=k'-\ell'-1$, $b_{2}=k'$. Let us first find a change of variables to $\rho_{1}$ and $\rho_{2}$ such that
\begin{equation}
\frac{\partial\rho_{1}}{\partial\sigma_{1}}=\sigma_{1}^{-a_{1}}, \qquad \frac{\partial\rho_{2}}{\partial\sigma_{2}}=\sigma_{2}^{-b_{2}}.
\end{equation}
\noindent
The solution to these equations is simply
\begin{eqnarray}
\sigma_{1}=\left(\left(1-a_{1}\right)\rho_{1}\right)^{\frac{1}{1-a_{1}}}, \quad a_{1}\ne1, & \qquad & \sigma_{1}=\rme^{\rho_{1}}, \quad a_{1}=1, \\
\sigma_{2}=\left(\left(1-b_{2}\right)\rho_{2}\right)^{\frac{1}{1-b_{2}}}, \quad b_{2}\ne1, & \qquad & \sigma_{2}=\rme^{\rho_{2}}, \quad b_{2}=1.
\end{eqnarray}
\noindent
We shall continue this example under the assumption that $a_{1},b_{2}\ne1$. The other possibilities may be solved in the same way, with similar results. Thus, if $a_{1},b_{2}\ne1$, we now try to find yet another set of independent variables, $x(\rho_{1},\rho_{2})$ and $y(\rho_{1},\rho_{2})$, such that they solve 
\be
C\left(\left(1-b_{2}\right)\rho_{2}\right)^{\frac{a_{2}}{1-b_{2}}}\, \frac{\partial f}{\partial\rho_{1}} + \widetilde{C}\left(\left(1-a_{1}\right)\rho_{1}\right)^{\frac{b_{1}}{1-a_{1}}} \frac{\partial f}{\partial\rho_{2}}=1,
\ee
\noindent
for $f=x$ and $f=y$. Using the definitions
\begin{eqnarray}
C_{1} & = & C\,(1-b_{2})^{k_{1}}, \qquad k_{1}=\frac{a_{2}}{1-b_{2}}, \\
C_{2} & = & \widetilde{C}\,(1-a_{1})^{k_{2}}, \qquad k_{2}=\frac{b_{1}}{1-a_{1}},
\end{eqnarray}
\noindent
the solution to this equation is finally given by
\begin{eqnarray}
x\left(\rho_{1},\rho_{2}\right) &=& y\left(\rho_{1},\rho_{2}\right) - C_{2} \left(1+k_{1}\right) \rho_{1}^{1+k_{2}} + C_{1} \left(1+k_{2}\right) \rho_{2}^{1+k_{1}}, \\
y\left(\rho_{1},\rho_{2}\right) &=& \frac{\left(1+k_{2}\right) \rho_{1}\rho_{2}}{C_{1} \left(1+k_{2}\right) \rho_{2}^{k_{1}+1} - C_{2} \left(1+k_{1}\right) \rho_{1}^{k_{2}+1}} \times \\
&&
\left. \times \,_{2}F_{1} \left[ 1, \frac{1}{1+k_{1}}+\frac{1}{1+k_{2}}, 1+\frac{1}{1+k_{2}} \right| \frac{C_{2} \left(1+k_{1}\right) \rho_{1}^{1+k_{2}}}{C_{2} \left(1+k_{1}\right) \rho_{1}^{1+k_{2}} - C_{1} \left(1+k_{2}\right) \rho_{2}^{1+k_{1}}}\right]. \nonumber
\end{eqnarray}
\noindent
Here, $\left. _{2}F_{1} \left[ a, b, c \right| z \right]$ is the hypergeometric function. These expressions define our original variables $\sigma_{1}$ and $\sigma_{2}$, implicitly as functions of $x$ and $y$. As mentioned above, in terms of these new variables the Stokes automorphism has again the very simple action \eqref{eq:Stokes-aut-at-zero-in-simple-variables}.
\end{itemize}
\noindent
If we had considered $\ell$ and $\ell'$ both negative instead, completely analogue results would follow.

As compared to the one--parameter case, the structure of Stokes transitions is now much more involved. Even finding ``good'' variables where the Stokes transition can be seen as a ``jump'' of an appropriate transseries parameter is not an easy task. Altogether, the full set of Stokes constants lead to highly non--trivial Stokes phenomena, as described by the functions $\BS_{\theta,i}^{(\nu)}(\sigma_{1},\sigma_{2})$.

%%%%%%%%%%%%%%%%%%%%%%%%%%%%%%%%%%%%%%%%%%%%%%%%%%%%%%%%%%%%%%%%%
%%%%%%%%%%%%%%%%%%%%%%%%%%%%%%%%%%%%%%%%%%%%%%%%%%%%%%%%%%%%%%%%%
\section{Nonperturbative Ambiguities and Two--Parameter Transseries\label{sec:Nonp-Ambiguities-2-param}}
%%%%%%%%%%%%%%%%%%%%%%%%%%%%%%%%%%%%%%%%%%%%%%%%%%%%%%%%%%%%%%%%%
%%%%%%%%%%%%%%%%%%%%%%%%%%%%%%%%%%%%%%%%%%%%%%%%%%%%%%%%%%%%%%%%%

We are finally ready to understand how nonperturbative ambiguities cancel in a two--parameters transseries, also with generalized instantons. This should set the ground to understand, within specific examples, how nonperturbative ambiguities will always cancel when dealing with multi--parameter transseries, with or without generalized instantons, as long as one considers the appropriate resummation prescription along the associated Stokes lines. Having understood Stokes phenomena/transitions, we may turn to the cancelation of the ambiguities. The overall strategy parallels what we worked out in the one--parameter case, although now formulae (as well as the cancelation mechanism itself) are much more involved.

The median resummation at $\theta=0$ obeys the same properties as in the one--parameter case. Once again we have
\begin{equation}
F_{\BR}=\mathcal{S}^{\text{med}}_0 F = \mathcal{S}_{0^+} \circ \underline{\mathfrak{S}}_{0}^{-\nu} F = \mathcal{S}_{0^-} \circ \underline{\mathfrak{S}}_{0}^{1-\nu} F.
\end{equation}
\noindent
As usual, canceling the ambiguity along this Stokes line translates to setting up a real function, $F_{\BR}$, which has to obey $\mathcal{H} F_{\BR} = F_{\BR}$. This means the allowed values for $\nu$ must be such that
\begin{equation}
\mathcal{H} \circ \underline{\mathfrak{S}}_{0}^{-\nu} F = \underline{\mathfrak{S}}_{0}^{1-\nu} F.
\end{equation}
\noindent
Recalling the expression for this Stokes transition, given in \eqref{eq:Stokes-transition-zero-2-param}, one immediately sees that one needs to determine the complex conjugate of $\BS_{0,i}^{(\nu)}(\sigma_{1},\sigma_{2})$. But as a consequence of the constraints found for the Stokes constants in appendix \ref{sec:App-Properties-of-med-resumm}, \textit{i.e.}, $S_{\ell}^{(k)}, \widetilde{S}_{\ell}^{(\ell+k)} \in {\rmi} \mathbb{R}$ for any $\ell,k\ge0$, and the definition of $\BS_{0,i}^{(\nu)}$ in appendix \ref{sec:App-Formulae-for-Two--Parameters}, it is straightforward to write
\begin{equation}
\mathcal{H}\, \BS_{0,i}^{(\nu)}(\sigma_{1},\sigma_{2}) = \BS_{0,i}^{(-\nu)} (\overline{\sigma}_{1},\overline{\sigma}_{2}).
\end{equation}
\noindent
The reality condition given above thus becomes
\begin{equation}
F \left( z, \BS_{0,1}^{(\nu)} (\overline{\sigma}_{1},\overline{\sigma}_{2}), \BS_{0,2}^{(\nu)} (\overline{\sigma}_{1},\overline{\sigma}_{2}) \right) = F \left( z, \BS_{0,1}^{(1-\nu)} (\sigma_{1},\sigma_{2}), \BS_{0,2}^{(1-\nu)} (\sigma_{1},\sigma_{2}) \right).
\end{equation}
\noindent
This is obeyed for every sector $(n|m)$ if $\nu=1/2$ and
\begin{equation}
\sigma_{1},\sigma_{2}\in\mathbb{R},
\end{equation}
\noindent
as in the one--parameter case. In this way, we can finally write the unambiguous real solution given by the median resummation as
\begin{eqnarray}
\label{eq:Median-resumm-2-param-alpha}
F_{\BR} \left(z,\sigma_{1},\sigma_{2}\right) &=& \sum_{n,m=0}^{+\infty} \sigma_{1}^{n}\sigma_{2}^{m}\, \BP_{0,-1/2}^{(n|m)}(\sigma_{1},\sigma_{2})\, \mathcal{S}_{0^+} F^{(n|m)}(z) = \\
&=& \sum_{n,m=0}^{+\infty} \sigma_{1}^{n}\sigma_{2}^{m}\, \BP_{0,+1/2}^{(n|m)} (\sigma_{1},\sigma_{2})\, \mathcal{S}_{0^-} F^{(n|m)}(z),
\label{eq:Median-resumm-2-param}
\end{eqnarray}
\noindent
with $\BP_{0,\nu}^{(n|m)}$ defined in \eqref{eq:App-Stokes-at-zero-2-param-extra}. Given the constraints on the Stokes constants $S_{-\ell}^{(k+\ell)}$, $\widetilde{S}_{-\ell}^{(k)}$, with $\ell\ge1$, $k\ge0$, found in appendix \ref{sec:App-Properties-of-med-resumm}, one can write the very same equations for the direction $\theta=\pi$, and in this way obtain the median resummation along this other Stokes line.

The main question we want to explicitly address in this section is whether the median resummation prescription just presented is canceling all possible nonperturbative ambiguities which appear in this larger, two--parameter transseries setting; and how exactly is it doing so. An expansion for the ambiguity was implicitly given in appendix \ref{sec:App-Formulae-for-Two--Parameters}, equation \eqref{eq:App-imaginary-amb-for-2-param-transseries}, which we write here at first few orders,
\begin{eqnarray}
\im\, F \left( z,\sigma_1,\sigma_2 \right) &=& \im\, F^{(0|0)} + \im\, \sigma_{1}\, \re\, F^{(1|0)} + \re\, \sigma_{1}\, \im\, F^{(1|0)} + \im\, \sigma_{1}^{2}\, \re\, F^{(2|0)} + \nonumber \\
&&
+ \re\, \sigma_{1}^{2}\, \im\, F^{(2|0)} + \im\, \sigma_{2}\, \re\, F^{(0|1)} + \re\, \sigma_{2}\, \im\, F^{(0|1)} + \nonumber \\
&&
+ \im \left(\sigma_{1}\sigma_{2}\right) \re\, F^{(1|1)} + \re \left(\sigma_{1}\sigma_{2}\right) \im\, F^{(1|1)} + \cdots.
\label{eq:ambiguity-imF-2-parameter}
\end{eqnarray}
\noindent
As usual, each sector has a (perturbative) ambiguity, and we will now analyze how the cancelation of these terms occurs so that the final transseries is ambiguity--free. In equation \eqref{eq:App-Sp-Sm-of-F_(l|m)} we see how to write the imaginary contributions $\im\, F^{(n|m)}$, or ambiguities, of a given sector in terms of higher sectors. In particular for $F^{(0|0)}$ we have
\begin{eqnarray}
2 \rmi\, \im\, F^{(0|0)} &=& S_{1}^{(0)}\, \re\, F^{(1|0)} + \left( S_{1}^{(0)} \right)^{2} \re\, F^{(2|0)} - \rmi S_{1}^{(0)}\, \im\, F^{(1|0)} - \rmi \left( S_{1}^{(0)} \right)^{2} \im\, F^{(2|0)} + \cdots = \nonumber \\
&=& S_{1}^{(0)}\, \re\, F^{(1|0)} - \frac{1}{2} \left(S_{1}^{(0)}\right)^{3} \re\, F^{(3|0)} + \mathcal{O}({\text{5-inst}}),
\end{eqnarray}
\noindent
where we used that
\begin{eqnarray}
2\rmi\, \im\, F^{(1|0)} &=& 2 S_{1}^{(0)}\, \re\, F^{(2|0)} - 2\rmi S_{1}^{(0)}\, \im\, F^{(2|0)} + \cdots, \\
2\rmi\, \im\, F^{(2|0)} &=& 3\, \re\, F^{(3|0)} + \cdots.
\end{eqnarray}
\noindent
This expansion should be familiar to us. Indeed, the contributions to the above imaginary ambiguity, arising from the multi--instanton series $F^{(n|0)}$, are completely equivalent to what we have already seen in the one--parameter case. In particular, if we had no other terms dependent upon $\sigma_{2}$, the very same\footnote{The choice of signs $\pm$ depends on the choice of lateral Borel resummation one is looking at; either $\mathcal{S}_{\pm}$.} solution we had before, $\rmi\, \im\, \sigma_{1} = \pm\frac{1}{2} S_{1}^{(0)}$, would cancel the ambiguity. However, as we take the full two--parameters transseries into consideration this is no longer true: for example, one finds that in order to cancel the term $S_{1}^{(0)}\, \re\, F^{(1|0)}$ appearing in $\im\, F^{(0|0)}$ above, we need not only contributions from $\im\, F^{(n|0)}$ (equivalently to the one--parameter case) but also several other terms need to contribute from mixed sectors in the transseries.

A schematic view of the cancelations occurring in the two--parameters situation, needed in order to remove the nonperturbative ambiguity and give rise to a median resummed real transseries, can be found in table \ref{tab:Cancelations-2-parameters} (as one compares this to table \ref{tab:Cancelations-1-parameter} do note that for reasons of space we have organized table \ref{tab:Cancelations-2-parameters} as the ``transpose'' of table \ref{tab:Cancelations-1-parameter}; other than that it is in fact useful to compare them both). The results in this table were obtained from the properties and expressions found in appendix \ref{sec:App-Formulae-for-Two--Parameters}. The rows correspond to the terms appearing in the expansion of $\im\, F \left(z,\sigma_{1},\sigma_{2}\right)$ given in \eqref{eq:ambiguity-imF-2-parameter} above. For these, each separate ambiguity contributing to this expansion, of the form $\im\, F^{(n|m)}$, is then expanded in higher (real) multi--instanton contributions as $\im F^{(n|m)} \sim \sum_{a,b} C_{(n|m)}^{(a|b)}\, \re\, F^{(a|b)}$, and we present the first contributions along the row. In particular, the listed coefficients $C_{(\ell|m)}^{(a|b)}$, associated to each real term shown on the top row, are functions of the Stokes constants and can be found in expressions (\ref{eq:Coefficients-of-expansion-imF00-2-parameter}--\ref{eq:Coefficients-of-expansion-imF22-2-parameter}).
\begin{table}
\noindent \begin{centering}
\begin{tabular}{|c|>{\centering}p{0.03\columnwidth}|>{\centering}p{0.03\columnwidth}|>{\centering}p{0.03\columnwidth}|>{\centering}p{0.03\columnwidth}|>{\centering}p{0.03\columnwidth}|>{\centering}p{0.03\columnwidth}|>{\centering}p{0.03\columnwidth}|>{\centering}p{0.03\columnwidth}|>{\centering}p{0.03\columnwidth}|>{\centering}p{0.03\columnwidth}|>{\centering}p{0.03\columnwidth}|>{\centering}p{0.03\columnwidth}|>{\centering}p{0.03\columnwidth}|>{\centering}p{0.03\columnwidth}|>{\centering}p{0.03\columnwidth}|}
\hline 
\begin{sideways}
{\scriptsize $\re\, F^{(4|2)}$}
\end{sideways} & \begin{sideways}
{\scriptsize -}
\end{sideways} & \begin{sideways}
{\scriptsize -}
\end{sideways} & \begin{sideways}
{\scriptsize -}
\end{sideways} & \begin{sideways}
{\scriptsize -}
\end{sideways} & \begin{sideways}
{\scriptsize -}
\end{sideways} & \begin{sideways}
{\scriptsize -}
\end{sideways} & \begin{sideways}
{\scriptsize -}
\end{sideways} & \begin{sideways}
{\scriptsize -}
\end{sideways} & \begin{sideways}
{\scriptsize -}
\end{sideways} & \begin{sideways}
{\scriptsize -}
\end{sideways} & \begin{sideways}
{\scriptsize 0}
\end{sideways} & \begin{sideways}
{\scriptsize -}
\end{sideways} & \begin{sideways}
{\scriptsize -}
\end{sideways} & \begin{sideways}
{\scriptsize -}
\end{sideways} & \begin{sideways}
{\scriptsize $C_{(1|2)}^{(4|2)}$}
\end{sideways}\tabularnewline
\hline 
\begin{sideways}
{\scriptsize $\re\, F^{(3|2)}$}
\end{sideways} & \begin{sideways}
{\scriptsize -}
\end{sideways} & \begin{sideways}
{\scriptsize -}
\end{sideways} & \begin{sideways}
{\scriptsize -}
\end{sideways} & \begin{sideways}
{\scriptsize -}
\end{sideways} & \begin{sideways}
{\scriptsize -}
\end{sideways} & \begin{sideways}
{\scriptsize -}
\end{sideways} & \begin{sideways}
{\scriptsize -}
\end{sideways} & \begin{sideways}
{\scriptsize -}
\end{sideways} & \begin{sideways}
{\scriptsize -}
\end{sideways} & \begin{sideways}
{\scriptsize -}
\end{sideways} & \begin{sideways}
{\scriptsize $C_{(0|2)}^{(3|2)}$}
\end{sideways} & \begin{sideways}
{\scriptsize -}
\end{sideways} & \begin{sideways}
{\scriptsize -}
\end{sideways} & \begin{sideways}
{\scriptsize -}
\end{sideways} & \begin{sideways}
{\scriptsize 0}
\end{sideways}\tabularnewline
\hline 
\begin{sideways}
{\scriptsize $\re\, F^{(2|2)}$}
\end{sideways} & \begin{sideways}
{\scriptsize -}
\end{sideways} & \begin{sideways}
{\scriptsize -}
\end{sideways} & \begin{sideways}
{\scriptsize -}
\end{sideways} & \begin{sideways}
{\scriptsize -}
\end{sideways} & \begin{sideways}
{\scriptsize -}
\end{sideways} & \begin{sideways}
{\scriptsize -}
\end{sideways} & \begin{sideways}
{\scriptsize -}
\end{sideways} & \begin{sideways}
{\scriptsize -}
\end{sideways} & \begin{sideways}
{\scriptsize -}
\end{sideways} & \begin{sideways}
{\scriptsize -}
\end{sideways} & \begin{sideways}
{\scriptsize 0}
\end{sideways} & \begin{sideways}
{\scriptsize -}
\end{sideways} & \begin{sideways}
{\scriptsize -}
\end{sideways} & \begin{sideways}
{\scriptsize -}
\end{sideways} & \begin{sideways}
{\scriptsize $C_{(1|2)}^{(2|2)}$}
\end{sideways}\tabularnewline
\hline 
\begin{sideways}
{\scriptsize $\re\, F^{(1|2)}$}
\end{sideways} & \begin{sideways}
{\scriptsize -}
\end{sideways} & \begin{sideways}
{\scriptsize -}
\end{sideways} & \begin{sideways}
{\scriptsize -}
\end{sideways} & \begin{sideways}
{\scriptsize -}
\end{sideways} & \begin{sideways}
{\scriptsize -}
\end{sideways} & \begin{sideways}
{\scriptsize -}
\end{sideways} & \begin{sideways}
{\scriptsize -}
\end{sideways} & \begin{sideways}
{\scriptsize -}
\end{sideways} & \begin{sideways}
{\scriptsize -}
\end{sideways} & \begin{sideways}
{\scriptsize -}
\end{sideways} & \begin{sideways}
{\scriptsize $C_{(0|2)}^{(1|2)}$}
\end{sideways} & \begin{sideways}
{\scriptsize -}
\end{sideways} & \begin{sideways}
{\scriptsize -}
\end{sideways} & \begin{sideways}
{\scriptsize $\im(\sigma_{1}\sigma_{2}^{2})$}
\end{sideways} & \begin{sideways}
{\scriptsize 0}
\end{sideways}\tabularnewline
\hline 
\begin{sideways}
{\scriptsize $\re\, F^{(0|2)}$}
\end{sideways} & \begin{sideways}
{\scriptsize -}
\end{sideways} & \begin{sideways}
{\scriptsize -}
\end{sideways} & \begin{sideways}
{\scriptsize -}
\end{sideways} & \begin{sideways}
{\scriptsize -}
\end{sideways} & \begin{sideways}
{\scriptsize -}
\end{sideways} & \begin{sideways}
{\scriptsize -}
\end{sideways} & \begin{sideways}
{\scriptsize -}
\end{sideways} & \begin{sideways}
{\scriptsize -}
\end{sideways} & \begin{sideways}
{\scriptsize -}
\end{sideways} & \begin{sideways}
{\scriptsize $\im\,\sigma_{2}^{2}$}
\end{sideways} & \begin{sideways}
{\scriptsize 0}
\end{sideways} & \begin{sideways}
{\scriptsize -}
\end{sideways} & \begin{sideways}
{\scriptsize -}
\end{sideways} & \begin{sideways}
{\scriptsize -}
\end{sideways} & \begin{sideways}
{\scriptsize 0}
\end{sideways}\tabularnewline
\hline 
\begin{sideways}
{\scriptsize $\re\, F^{(4|1)}$}
\end{sideways} & \begin{sideways}
{\scriptsize -}
\end{sideways} & \begin{sideways}
{\scriptsize -}
\end{sideways} & \begin{sideways}
{\scriptsize -}
\end{sideways} & \begin{sideways}
{\scriptsize -}
\end{sideways} & \begin{sideways}
{\scriptsize 0}
\end{sideways} & \begin{sideways}
{\scriptsize -}
\end{sideways} & \begin{sideways}
{\scriptsize -}
\end{sideways} & \begin{sideways}
{\scriptsize -}
\end{sideways} & \begin{sideways}
{\scriptsize $C_{(1|1)}^{(4|1)}$}
\end{sideways} & \begin{sideways}
{\scriptsize -}
\end{sideways} & \begin{sideways}
{\scriptsize $C_{(0|2)}^{(4|1)}$}
\end{sideways} & \begin{sideways}
{\scriptsize -}
\end{sideways} & \begin{sideways}
{\scriptsize 0}
\end{sideways} & \begin{sideways}
{\scriptsize -}
\end{sideways} & \begin{sideways}
{\scriptsize $C_{(1|2)}^{(4|1)}$}
\end{sideways}\tabularnewline
\hline 
\begin{sideways}
{\scriptsize $\re\, F^{(3|1)}$}
\end{sideways} & \begin{sideways}
{\scriptsize -}
\end{sideways} & \begin{sideways}
{\scriptsize -}
\end{sideways} & \begin{sideways}
{\scriptsize -}
\end{sideways} & \begin{sideways}
{\scriptsize -}
\end{sideways} & \begin{sideways}
{\scriptsize $C_{(0|1)}^{(3|1)}$}
\end{sideways} & \begin{sideways}
{\scriptsize -}
\end{sideways} & \begin{sideways}
{\scriptsize -}
\end{sideways} & \begin{sideways}
{\scriptsize -}
\end{sideways} & \begin{sideways}
{\scriptsize 0}
\end{sideways} & \begin{sideways}
{\scriptsize -}
\end{sideways} & \begin{sideways}
{\scriptsize $C_{(0|2)}^{(3|1)}$}
\end{sideways} & \begin{sideways}
{\scriptsize -}
\end{sideways} & \begin{sideways}
{\scriptsize $C_{(2|1)}^{(3|1)}$}
\end{sideways} & \begin{sideways}
{\scriptsize -}
\end{sideways} & \begin{sideways}
{\scriptsize $C_{(1|2)}^{(3|1)}$}
\end{sideways}\tabularnewline
\hline 
\begin{sideways}
{\scriptsize $\re\, F^{(2|1)}$}
\end{sideways} & \begin{sideways}
{\scriptsize -}
\end{sideways} & \begin{sideways}
{\scriptsize -}
\end{sideways} & \begin{sideways}
{\scriptsize -}
\end{sideways} & \begin{sideways}
{\scriptsize -}
\end{sideways} & \begin{sideways}
{\scriptsize 0}
\end{sideways} & \begin{sideways}
{\scriptsize -}
\end{sideways} & \begin{sideways}
{\scriptsize -}
\end{sideways} & \begin{sideways}
{\scriptsize -}
\end{sideways} & \begin{sideways}
{\scriptsize $C_{(1|1)}^{(2|1)}$}
\end{sideways} & \begin{sideways}
{\scriptsize -}
\end{sideways} & \begin{sideways}
{\scriptsize $C_{(0|2)}^{(2|1)}$}
\end{sideways} & \begin{sideways}
{\scriptsize $\im(\sigma_{1}^{2}\sigma_{2})$}
\end{sideways} & \begin{sideways}
{\scriptsize 0}
\end{sideways} & \begin{sideways}
{\scriptsize -}
\end{sideways} & \begin{sideways}
{\scriptsize $C_{(1|2)}^{(2|1)}$}
\end{sideways}\tabularnewline
\hline 
\begin{sideways}
{\scriptsize $\re\, F^{(1|1)}$}
\end{sideways} & \begin{sideways}
{\scriptsize -}
\end{sideways} & \begin{sideways}
{\scriptsize -}
\end{sideways} & \begin{sideways}
{\scriptsize -}
\end{sideways} & \begin{sideways}
{\scriptsize -}
\end{sideways} & \begin{sideways}
{\scriptsize $C_{(0|1)}^{(1|1)}$}
\end{sideways} & \begin{sideways}
{\scriptsize -}
\end{sideways} & \begin{sideways}
{\scriptsize -}
\end{sideways} & \begin{sideways}
{\scriptsize $\im(\sigma_{1}\sigma_{2})$}
\end{sideways} & \begin{sideways}
{\scriptsize 0}
\end{sideways} & \begin{sideways}
{\scriptsize -}
\end{sideways} & \begin{sideways}
{\scriptsize $C_{(0|2)}^{(1|1)}$}
\end{sideways} & \begin{sideways}
{\scriptsize -}
\end{sideways} & \begin{sideways}
{\scriptsize 0}
\end{sideways} & \begin{sideways}
{\scriptsize -}
\end{sideways} & \begin{sideways}
{\scriptsize $C_{(1|2)}^{(1|1)}$}
\end{sideways}\tabularnewline
\hline 
\begin{sideways}
{\scriptsize $\re\, F^{(0|1)}$}
\end{sideways} & \begin{sideways}
{\scriptsize -}
\end{sideways} & \begin{sideways}
{\scriptsize -}
\end{sideways} & \begin{sideways}
{\scriptsize -}
\end{sideways} & \begin{sideways}
{\scriptsize $\im\,\sigma_{2}$}
\end{sideways} & \begin{sideways}
{\scriptsize 0}
\end{sideways} & \begin{sideways}
{\scriptsize -}
\end{sideways} & \begin{sideways}
{\scriptsize -}
\end{sideways} & \begin{sideways}
{\scriptsize -}
\end{sideways} & \begin{sideways}
{\scriptsize 0}
\end{sideways} & \begin{sideways}
{\scriptsize -}
\end{sideways} & \begin{sideways}
{\scriptsize $C_{(0|2)}^{(0|1)}$}
\end{sideways} & \begin{sideways}
{\scriptsize -}
\end{sideways} & \begin{sideways}
{\scriptsize 0}
\end{sideways} & \begin{sideways}
{\scriptsize -}
\end{sideways} & \begin{sideways}
{\scriptsize 0}
\end{sideways}\tabularnewline
\hline 
\begin{sideways}
{\scriptsize $\re\, F^{(4|0)}$}
\end{sideways} & \begin{sideways}
{\scriptsize 0}
\end{sideways} & \begin{sideways}
{\scriptsize -}
\end{sideways} & \begin{sideways}
{\scriptsize $C_{(1|0)}^{(4|0)}$}
\end{sideways} & \begin{sideways}
{\scriptsize -}
\end{sideways} & \begin{sideways}
{\scriptsize $C_{(0|1)}^{(4|0)}$}
\end{sideways} & \begin{sideways}
{\scriptsize -}
\end{sideways} & \begin{sideways}
{\scriptsize 0}
\end{sideways} & \begin{sideways}
{\scriptsize -}
\end{sideways} & \begin{sideways}
{\scriptsize $C_{(1|1)}^{(4|0)}$}
\end{sideways} & \begin{sideways}
{\scriptsize -}
\end{sideways} & \begin{sideways}
{\scriptsize $C_{(0|2)}^{(4|0)}$}
\end{sideways} & \begin{sideways}
{\scriptsize -}
\end{sideways} & \begin{sideways}
{\scriptsize $C_{(2|1)}^{(4|0)}$}
\end{sideways} & \begin{sideways}
{\scriptsize -}
\end{sideways} & \begin{sideways}
{\scriptsize $C_{(1|2)}^{(4|0)}$}
\end{sideways}\tabularnewline
\hline 
\begin{sideways}
{\scriptsize $\re\, F^{(3|0)}$}
\end{sideways} & \begin{sideways}
{\scriptsize $C_{(0|0)}^{(3|0)}$}
\end{sideways} & \begin{sideways}
{\scriptsize -}
\end{sideways} & \begin{sideways}
{\scriptsize 0}
\end{sideways} & \begin{sideways}
{\scriptsize -}
\end{sideways} & \begin{sideways}
{\scriptsize $C_{(0|1)}^{(3|0)}$}
\end{sideways} & \begin{sideways}
{\scriptsize -}
\end{sideways} & \begin{sideways}
{\scriptsize $C_{(2|0)}^{(3|0)}$}
\end{sideways} & \begin{sideways}
{\scriptsize -}
\end{sideways} & \begin{sideways}
{\scriptsize $C_{(1|1)}^{(3|0)}$}
\end{sideways} & \begin{sideways}
{\scriptsize -}
\end{sideways} & \begin{sideways}
{\scriptsize $C_{(0|2)}^{(3|0)}$}
\end{sideways} & \begin{sideways}
{\scriptsize -}
\end{sideways} & \begin{sideways}
{\scriptsize $C_{(2|1)}^{(3|0)}$}
\end{sideways} & \begin{sideways}
{\scriptsize -}
\end{sideways} & \begin{sideways}
{\scriptsize $C_{(1|2)}^{(3|0)}$}
\end{sideways}\tabularnewline
\hline 
\begin{sideways}
{\scriptsize $\re\, F^{(2|0)}$}
\end{sideways} & \begin{sideways}
{\scriptsize 0}
\end{sideways} & \begin{sideways}
{\scriptsize -}
\end{sideways} & \begin{sideways}
{\scriptsize $C_{(1|0)}^{(2|0)}$}
\end{sideways} & \begin{sideways}
{\scriptsize -}
\end{sideways} & \begin{sideways}
{\scriptsize $C_{(0|1)}^{(2|0)}$}
\end{sideways} & \begin{sideways}
{\scriptsize $\im\,\sigma_{1}^{2}$}
\end{sideways} & \begin{sideways}
{\scriptsize 0}
\end{sideways} & \begin{sideways}
{\scriptsize -}
\end{sideways} & \begin{sideways}
{\scriptsize $C_{(1|1)}^{(2|0)}$}
\end{sideways} & \begin{sideways}
{\scriptsize -}
\end{sideways} & \begin{sideways}
{\scriptsize $C_{(0|2)}^{(2|0)}$}
\end{sideways} & \begin{sideways}
{\scriptsize -}
\end{sideways} & \begin{sideways}
{\scriptsize $C_{(2|1)}^{(2|0)}$}
\end{sideways} & \begin{sideways}
{\scriptsize -}
\end{sideways} & \begin{sideways}
{\scriptsize $C_{(1|2)}^{(2|0)}$}
\end{sideways}\tabularnewline
\hline 
\begin{sideways}
{\scriptsize $\re\, F^{(1|0)}$}
\end{sideways} & \begin{sideways}
{\scriptsize $C_{(0|0)}^{(1|0)}$}
\end{sideways} & \begin{sideways}
{\scriptsize $\im\,\sigma_{1}$}
\end{sideways} & \begin{sideways}
{\scriptsize 0}
\end{sideways} & \begin{sideways}
{\scriptsize -}
\end{sideways} & \begin{sideways}
{\scriptsize $C_{(0|1)}^{(1|0)}$}
\end{sideways} & \begin{sideways}
{\scriptsize -}
\end{sideways} & \begin{sideways}
{\scriptsize 0}
\end{sideways} & \begin{sideways}
{\scriptsize -}
\end{sideways} & \begin{sideways}
{\scriptsize $C_{(1|1)}^{(1|0)}$}
\end{sideways} & \begin{sideways}
{\scriptsize -}
\end{sideways} & \begin{sideways}
{\scriptsize $C_{(0|2)}^{(1|0)}$}
\end{sideways} & \begin{sideways}
{\scriptsize -}
\end{sideways} & \begin{sideways}
{\scriptsize 0}
\end{sideways} & \begin{sideways}
{\scriptsize -}
\end{sideways} & \begin{sideways}
{\scriptsize $C_{(1|2)}^{(1|0)}$}
\end{sideways}\tabularnewline
\hline 
\multicolumn{1}{c|}{\begin{sideways}
{\scriptsize \begin{picture}(45,20)(0,0) 
\put(15,12){} 
\put(3.6,15){\line(2,-1){40.3}} 
\put(2,-2){$\mathbb{I}\mathrm{m}$ $F$} 
\end{picture}}
\end{sideways}} & \begin{sideways}
{\scriptsize $\im\, F^{(0|0)}$}
\end{sideways} & \begin{sideways}
{\scriptsize $\begin{array}{c}
\im\,\sigma_{1}\\
\times\re\, F^{(1|0)}
\end{array}$}
\end{sideways} & \begin{sideways}
{\scriptsize $\begin{array}{c}
\re\,\sigma_{1}\\
\times\im\, F^{(1|0)}
\end{array}$}
\end{sideways} & \begin{sideways}
{\scriptsize $\begin{array}{c}
\im\,\sigma_{2}\\
\times\re\, F^{(0|1)}
\end{array}$}
\end{sideways} & \begin{sideways}
{\scriptsize $\begin{array}{c}
\re\,\sigma_{2}\\
\times\im\, F^{(0|1)}
\end{array}$}
\end{sideways} & \begin{sideways}
{\scriptsize $\begin{array}{c}
\im\,\sigma_{1}^{2}\\
\times\re\, F^{(2|0)}
\end{array}$}
\end{sideways} & \begin{sideways}
{\scriptsize $\begin{array}{c}
\re\,\sigma_{1}^{2}\\
\times\im\, F^{(2|0)}
\end{array}$}
\end{sideways} & \begin{sideways}
{\scriptsize $\begin{array}{c}
\im(\sigma_{1}\sigma_{2})\\
\times\re\, F^{(1|1)}
\end{array}$}
\end{sideways} & \begin{sideways}
{\scriptsize $\begin{array}{c}
\re(\sigma_{1}\sigma_{2})\\
\times\im\, F^{(1|1)}
\end{array}$}
\end{sideways} & \begin{sideways}
{\scriptsize $\begin{array}{c}
\im\,\sigma_{2}^{2}\\
\times\re\, F^{(0|2)}
\end{array}$}
\end{sideways} & \begin{sideways}
{\scriptsize $\begin{array}{c}
\re\,\sigma_{2}^{2}\\
\times\im\, F^{(0|2)}
\end{array}$}
\end{sideways} & \begin{sideways}
{\scriptsize $\begin{array}{c}
\im(\sigma_{1}^{2}\sigma_{2})\\
\times\re\, F^{(2|1)}
\end{array}$}
\end{sideways} & \begin{sideways}
{\scriptsize $\begin{array}{c}
\re(\sigma_{1}^{2}\sigma_{2})\\
\times\im\, F^{(2|1)}
\end{array}$}
\end{sideways} & \begin{sideways}
{\scriptsize $\begin{array}{c}
\im(\sigma_{1}\sigma_{2}^{2})\\
\times\re\, F^{(1|2)}
\end{array}$}
\end{sideways} & \begin{sideways}
{\scriptsize $\begin{array}{c}
\re(\sigma_{1}\sigma_{2}^{2})\\
\times\im\, F^{(1|2)}
\end{array}$}
\end{sideways}\tabularnewline
\cline{2-16} 
\end{tabular}
\par\end{centering}
\caption{Cancelation of the first terms in the nonperturbative ambiguity of the two--parameter transseries. Each row corresponds to a term contributing to $\im\, F$, as can
be seen in \eqref{eq:ambiguity-imF-2-parameter}, and the first contributions to the expansion $\im\, F^{(n|m)} \sim \sum_{a,b} C_{(n|m)}^{(a|b)}\, \re\, F^{(a|b)}$ are explicitly shown, up to $a=4$, $b=2$. To cancel the transseries nonperturbative ambiguity, the coefficients associated to each independent $\re\, F^{(a|b)}$ need to add up to zero, \textit{i.e.}, each column needs to cancel separately. Note that the - appearing in this table means that the associated $\re\, F^{(a|b)}$ term is not expected to appear in that expansion (as opposed to the $0$, corresponding to a coefficient actually being zero). The coefficients $C_{(\ell|m)}^{(a|b)}$ can be found in  (\ref{eq:Coefficients-of-expansion-imF00-2-parameter}--\ref{eq:Coefficients-of-expansion-imF12-2-parameter}).
\label{tab:Cancelations-2-parameters}}
\end{table}

To cancel the nonperturbative ambiguity of the two--parameters transseries, we need that all coefficients associated to each real term $\re\, F^{(a|b)}$ add up to zero. In other words, each column in the table needs to be canceled separately. Unfortunately, unlike in the one--parameter case, the contributions to each column no longer truncate. This implies that we cannot in general solve the constraints for $\sigma_{1}$ and $\sigma_{2}$ in terms of closed--form expressions (except in simple cases, \textit{e.g.}, setting $\sigma_{2}=0$ gives back the one--parameter case previously studied).

Nonetheless, these constraints should be compatible with finding a real transseries, as given by the median resumation \eqref{eq:Median-resumm-2-param-alpha} or \eqref{eq:Median-resumm-2-param}. How may we explicilty check this? Recall that in this case the median resumation is given by 
\begin{equation}
F_{\BR} \left(z,\sigma_{1},\sigma_{2}\right) = \mathcal{S}_{0^+} F \left( z, \mathbb{S}_{0,1}^{(-1/2)}(\sigma_{1},\sigma_{2}), \mathbb{S}_{0,2}^{(-1/2)}(\sigma_{1},\sigma_{2}) \right),
\end{equation}
\noindent
where, along $\theta=0^{+}$, we find the new transseries parameters $\widetilde{\sigma}_{1}$ and $\widetilde{\sigma}_{2}$, defined in terms of the two ``old'' real parameters $\sigma_{1},\sigma_{2}\in\mathbb{R}$ by the Stokes transitions
\begin{equation}
\widetilde{\sigma}_{i} = \mathbb{S}_{0,i}^{(-1/2)}\left(\sigma_{1},\sigma_{2}\right), \qquad i=1,2.
\end{equation}
\noindent
The main point now is that the constraints given by the cancelation of the ambiguity should be automatically satisfied by the parameters $\widetilde{\sigma}_{i}$ just introduced, \textit{i.e.}, if we were to take each column of table \ref{tab:Cancelations-2-parameters}, with its infinite set of contributions, and evaluate it at the values $\widetilde{\sigma}_{i}$, we would find that all the contribution would add up to zero as expected. Consider a very concrete example and look at the cancelations that must occur in the column corresponding to $\re\, F^{(1|0)}$. In this case, the constraint from ambiguity cancelation will read
\begin{equation}
\im\,\widetilde{\sigma}_{1} + C_{(0|0)}^{(1|0)} + C_{(0|1)}^{(1|0)} + C_{(1|1)}^{(1|0)} + C_{(0|2)}^{(1|0)} + C_{(1|2)}^{(1|0)} + C_{(2|2)}^{(1|0)} + \cdots = 0,
\label{eq:constraint-cancelation-F10}
\end{equation}
\noindent
with the coefficients\footnote{In the constraint we have also included the contribution $C_{(2|2)}^{(1|0)}$ coming from $\re\left(\sigma_{1}^{2}\sigma_{2}^{2}\right)\im\, F^{(2|2)} = \sum_{a,b}C_{(2|2)}^{(a|b)}\, \re\, F^{(a|b)}$, whose first non--zero coefficients $C_{(2|2)}^{(a|b)}$ may be found in (\ref{eq:Coefficients-of-expansion-imF22-2-parameter-alpha}--\ref{eq:Coefficients-of-expansion-imF22-2-parameter}), with $a\le4,b\le2$.} $C_{(a|b)}^{(0|1)}$ evaluated at the values $\widetilde{\sigma}_{1}, \widetilde{\sigma}_{2}$. Now using our results in appendix \ref{sec:App-Formulae-for-Two--Parameters}, we can explicitly expand these ``new'' parameters as power series in the ``old'' real parameters $\sigma_{1},\sigma_{2}\in\BR$. In terms of their real and imaginary parts, $\widetilde{\sigma}_{i}=\widetilde{\sigma}_{i,\text{R}} + \rmi\, \widetilde{\sigma}_{i,\text{I}}$, we find\footnote{Note that in these expansions we have presented all orders of $\sigma_{1}$ for each order of $\sigma_{2}$. Also, to determine the real and imaginary parts of the parameters, recall that $\sigma_{1},\sigma_{2}\in\mathbb{R}$ and $S_{a}^{(b)},\widetilde{S}_{a}^{(b)}\in{\rm i}\mathbb{R}$.}
\begin{eqnarray}
\widetilde{\sigma}_{1,\text{R}} &=& \frac{1}{8} \sigma_{2}\, S_{1}^{(0)} S_{1}^{(1)} + \frac{1}{48}\sigma_{2}^{2} \left( 6\, S_{2}^{(0)} \left( S_{1}^{(1)} + \widetilde{S}_{1}^{(2)} \right) + 6\, S_{1}^{(0)} S_{2}^{(1)} \right) + \nonumber \\
&&
+ \sigma_{1} \left( 1 + \frac{1}{8} \sigma_{2}^{2} \left( S_{1}^{(0)} S_{1}^{(2)} + S_{1}^{(1)} \left( S_{1}^{(1)} + \widetilde{S}_{1}^{(2)} \right) \right) \right) + \mathcal{O}(\sigma_{2}^{3}), \\
\rmi\, \widetilde{\sigma}_{1,\text{I}} &=& - \frac{1}{2} S_{1}^{(0)} - \frac{1}{2} \sigma_{2}\, S_{2}^{(0)} - \frac{1}{2} \sigma_{1} \sigma_{2} \left( S_{1}^{(1)} + \sigma_{2}\, S_{2}^{(1)} \right) - \frac{1}{2} \sigma_{1}^{2} \sigma_{2}^{2}\, S_{1}^{(2)} - \nonumber \\
&&
- \frac{1}{48} \sigma_{2} \left( 2 \left( S_{1}^{(0)} \right)^{2} S_{1}^{(2)} + 24\, S_{3}^{(0)} + S_{1}^{(0)} S_{1}^{(1)} \left( S_{1}^{(1)} + 2\, \widetilde{S}_{1}^{(2)} \right) \right) + \mathcal{O}(\sigma_{2}^{3}), \\
\widetilde{\sigma}_{2,\text{R}} &=& \sigma_{2} + \frac{1}{8} \sigma_{2}^{3} \left( 2 \left( \widetilde{S}_{1}^{(2)} \right)^{2} + S_{1}^{(0)} \widetilde{S}_{1}^{(3)} \right) + \mathcal{O}(\sigma_{2}^{4}), \\
\rmi\, \widetilde{\sigma}_{2,\text{I}} &=& -\frac{1}{2} \sigma_{2}^{2}\, \widetilde{S}_{1}^{(2)} - \frac{1}{2} \sigma_{1} \sigma_{2}^{3}\, \widetilde{S}_{1}^{(3)} - \frac{1}{2} \sigma_{2}^{3}\, \widetilde{S}_{2}^{(3)} + \mathcal{O}(\sigma_{2}^{4}). 
\end{eqnarray}
\noindent
Plugging these expansions back into the constraint in \eqref{eq:constraint-cancelation-F10}, we find that it is indeed satisfied up to order $\sigma_{1}^{2} \sigma_{2}^{2}$, precisely as expected. To explicitly see the cancelation working at higher orders, one would have to include the next contributions to $\re\, F^{(1|0)}$. Note that expanding the ambiguity of each distinct sector as $\im\, F^{(n|m)} \sim \sum_{a,b} C_{(n|m)}^{(a|b)}\, \re\, F^{(a|b)}$, it is not difficult to see that the coefficients $C_{(n|m)}^{(a|b)}$ are non--zero only if $b\le m$ and $a\ge\ell+1+b-m$ (and $a,b\ge0$). Consequently, we find that all terms $\im\, F^{(n|m)}$ with $n\le m$ will contribute to $\re\, F^{(1|0)}$.

As far as the other columns in table \ref{tab:Cancelations-2-parameters} are concerned, the coefficients written in the table are enough to see the precise same type of cancelations occur for $\re\, F^{(0|1)}$, $\re\, F^{(1|1)}$, $\re\, F^{(2|1)}$, $\re\, F^{(0|2)}$, and $\re\, F^{(1|2)}$. The other cases only cancel as one considers extra contributions which were not explicitly written down in this table. In conclusion, we see that the nonperturbative ambiguity is canceled also in the present two--parameters setting, albeit in a much more intricate way than what happened in the one--parameter case we addressed earlier.

Having explicitly shown the cancelation of nonperturbative ambiguities within the context of a two--parameters transseries, the one thing left to do is to give an idea of the resulting expansion of the answer (of the median resummation). All one has to do is to use the result in \eqref{eq:Median-resumm-2-param-alpha} or \eqref{eq:Median-resumm-2-param}, expand it in powers of the transseries parameters $\sigma_{1},\sigma_{2}\in\mathbb{R}$, and write the result in terms of real higher--instanton contributions. By using the expansions presented in table \ref{tab:Cancelations-2-parameters}, this expansion will now have contributions arising from all sectors, in the form
\begin{eqnarray}
F_{\BR} \left(z,\sigma_{1},\sigma_{2}\right) &=& \re\, F^{(0|0)} + \left( \sigma_{1} - \frac{\sigma_{2}}{8}\, S_{1}^{(0)} S_{1}^{1} + \cdots \right) \re\, F^{(1|0)} + \left( \sigma_{2} + \cdots \right) \re\, F^{(0|1)} + \nonumber \\
&+&
\label{ReF2pts}
\left( \sigma_{1} \sigma_{2} - \frac{\sigma_{2}^{2}}{8}\, S_{1}^{(0)} \left( S_{1}^{(1)} + 2\, \widetilde{S}_{1}^{(2)} \right) + \cdots \right) \re\, F^{(1|1)} + \\
&+&
\left( \sigma_{1}^{2} - \frac{1}{4} \left( S_{1}^{(0)} \right)^{2} - \frac{\sigma_{2}}{4}\, S_{1}^{(0)} \left( 2\, S_{2}^{(0)} + 3 \sigma_{1}\, S_{1}^{(1)} \right) + \cdots \right) \re\, F^{(2|0)} + \nonumber \\
&+&
\left( \sigma_{2}^{2} + \cdots \right) \re\, F^{(0|2)} + \left( \frac{\sigma_{2}}{4} \left( 4 \sigma_{1}^{2} - \left( S_{1}^{(0)} \right)^{2} \right) + \cdots \right) \re\, F^{(2|1)} + \nonumber \\
&+&
\left( \sigma_{1} \sigma_{2}^{2} + \cdots \right) \re\, F^{(1|2)}+ \sigma_{2}^{2} \left( \sigma_{1}^{2} - \frac{1}{4}\left(S_{1}^{(0)}\right)^{2} + \cdots \right) \re\, F^{(2|2)} + \cdots. \nonumber
\end{eqnarray}
\noindent
Let us make a few remarks. First, using a symbolic computation program it is automatic to include further terms in this expansion, with a whole lot more Stokes constants appearing. We have just included a few terms in order to give a general idea of the final expression; including more terms would make the expression too cumbersome. Second, this discussion shows how generalized instanton sectors are not only crucial in order to cancel the ambiguities of our two--parameters transseries, but they also play a definite role in the final (real) solution. However, it is already clear from the terms displayed above that if we take $\sigma_2=0$ we recover the one--parameter case. This is also to be expected when constructing a real solution along $\theta=0$ as we expect to have a natural mechanism to remove any exponential enhanced contributions along this direction. Where these terms should always be non--trivial is when addressing the median resummation along $\theta=\pi$. In this case, the analogue of \eqref{ReF2pts} is obtained from this equation by changing $m\leftrightarrow n$ in $F^{(n|m)}$, $\sigma_1 \leftrightarrow \sigma_2$, and $S_{\ell},\,\widetilde{S}_{\ell}$ with $\widetilde{S}_{-\ell}, S_{-\ell}$. Now, by setting $\sigma_1=0$ one constructs a real solution along $\theta=\pi$ without exponential large contributions along this direction. Note that this discussion followed without including logarithmic sectors due to resonance in the asymptotic expansions of the (mixed) nonperturbative sectors (but see appendix \ref{sec:App-Properties-of-med-resumm}). While along $\theta=0$ not much will change, it would be very interesting to analyze these expressions along $\theta=\pi$ when one further includes these sectors; but we leave this for future work.

%%%%%%%%%%%%%%%%%%%%%%%%%%%%%%%%%%%%%%%%%%%%%%%%%%%%%%%%%%%%%%%%%
%%%%%%%%%%%%%%%%%%%%%%%%%%%%%%%%%%%%%%%%%%%%%%%%%%%%%%%%%%%%%%%%%
\section{Monodromy of the Solution and Reality Conditions\label{sec:Monodromy-and-couplings}}
%%%%%%%%%%%%%%%%%%%%%%%%%%%%%%%%%%%%%%%%%%%%%%%%%%%%%%%%%%%%%%%%%
%%%%%%%%%%%%%%%%%%%%%%%%%%%%%%%%%%%%%%%%%%%%%%%%%%%%%%%%%%%%%%%%%

Earlier we mentioned that if ambiguities arise along different directions in the complex plane one might be interested in canceling all such ambiguities; for instance if looking for globally well--defined solutions in the complex plane. Canceling ambiguities along both $\theta=0$ and $\theta=\pi$ entails finding real transseries solutions along the real line. However, we also mentioned that in many specific cases there is a difference between the ``physical'' perturbative coupling, $\kappa$, and the ``working'' variable we use, $z$, in the form of a rescaling $z=\kappa^{\alpha}$. This means that what one means by the ``real line'' is different in $\kappa$ and $z$ coordinates. We have previously discussed what this means for $z$; in this section we want to understand what it means to find real solutions for real coupling $\kappa$. For real \textit{positive} coupling, canceling the nonperturbative imaginary ambiguity along the $\theta=0$ singular direction is enough, but reality along negative real $\kappa$ will in general differ from canceling the ambiguity along the singular $\theta=\pi$ direction in the $z$--plane.

To be fully precise, the singular directions $\theta=0,\pi$ arise not in the $z$--plane, but in the Borel $s$--plane; these are the directions where the Borel transform has singularities. As discussed in section \ref{sec:Nonpert-Ambiguity-real-transs}, the return from the Borel to the $z$--plane is implemented by a Laplace transform
\begin{equation}
\CS_{\theta} F(z) = \int_{0}^{\rme^{\rmi\theta}\infty}\rmd s\,\CB[F](s)\,\rme^{-zs}.\end{equation}
\noindent
In the $z$--plane we find Stokes and anti--Stokes lines, but in this coordinate the structure one finds is essentially equivalent to the one in the complex Borel plane: the Stokes lines at $\theta=0,\pi$ remain in the same place; the anti--Stokes lines will be along some ray in the $z$--plane such that (previously) exponentially suppressed  contributions to the transseries become of order one.

But what happens in the ``physical'' variable $\kappa=z^{1/\alpha}$? When $\alpha$ is a rational number the Stokes lines $\theta=0,\pi$ will spread in the complex plane: the positive real line will still be at $\theta=0$, but $\theta=\pi$ will be at an angle, dictated by $\alpha$. This will also dictate whether the negative real line is a Stokes line, an anti--Stokes line, or neither. If it is a Stokes line it is an open problem to construct a real solution in the full real axis: two conditions must be met simultaneously and one needs to check if this is possible or not. If the negative real axis is an anti--Stokes line then real solutions are possible; one such example is the Airy function which we will address below.

Let us understand better where the Stokes transitions occur in the ``physical'' variable $\kappa$. For that define
\begin{equation}
z=\kappa^{\alpha}, \qquad \alpha=\frac{n}{m}\in\mathbb{Q},
\end{equation}
\noindent
where we assumed that $\alpha$ is written in irreducible form. Moreover, define the two variables as
\begin{equation}
z = \left|z\right| \rme^{\rmi\theta_{z}}, \qquad \kappa = \left|\kappa\right| \rme^{\rmi\theta_{\kappa}}.
\end{equation}
\noindent
As one rotates the argument of $z$ in the complex plane $0\le\theta_{z}\le2\pi$ one crosses two Stokes lines, at $\theta_{z}=0,\pi$, which correspond to the exact same singular directions in the complex Borel plane. Then, in the ``physical'' variable $\kappa$ one finds the following unfolding,
\begin{eqnarray}
\mbox{Stokes line at }\theta_{z}=0 & \quad \Rightarrow \quad & \mbox{Stokes lines at }\theta_{\kappa}=\frac{m}{n}\left(0+2\pi p\right), \quad p=0,\ldots,n-1, \\
\mbox{Stokes line at }\theta_{z}=\pi & \quad \Rightarrow \quad & \mbox{Stokes lines at }\theta_{\kappa}=\frac{m}{n}\left(\pi+2\pi p\right), \quad p=0,\ldots,n-1.
\end{eqnarray}
\noindent
In general we find $2n$ Stokes lines in the $\kappa$--plane, with the ``type'' of transition alternating between $\underline{\mathfrak{S}}_{0}$ and $\underline{\mathfrak{S}}_{\pi}$. The denominator $m$ of $\alpha$ defines the number of full rotations the argument of $\kappa$ has to undergo when $\theta_{z}$ gives one full rotation in the complex plane. When $m$ is even, the Stokes lines corresponding to different transition ``types'' will fall on top of each other, and we will have $2n$ Stokes transitions but only $n$ different directions. 

We can also see where the anti--Stokes lines lie. These lines are defined as the lines in the complex $z$ or $\kappa$ plane where the contributions of both positive and negative exponentials in the transseries \eqref{eq:Ftransseries} contribute at the same
order,
\begin{equation}
\rme^{-Az} \sim \rme^{Az}.
\end{equation}
\noindent
As $A$ is real this will happen if and only if $\re\left(z\right)=0$, which implies $\theta_{z}=\frac{\pi}{2},\frac{3\pi}{2}$; exactly in--between each two Stokes lines. In the $\kappa$--plane, one has $\re\left(\kappa^\alpha\right)=0$, which corresponds to having $\cos\left(\alpha\,\theta_{\kappa}\right)=0$. These lines then fall on
\begin{equation}
\theta_{\kappa}=\frac{m}{n}\left(\frac{\pi}{2}+\pi p\right), \qquad p=0,\ldots,2n-1.
\end{equation}
\noindent
Again, there are $2n$ anti--Stokes lines, which fall exactly in--between each two Stokes lines.

%%%%%%%%%%%%%%%%%%%%%%%%%%%%%%%%%%%%%%%%%%%%%%%%%%%%%%%%%%%%%%%%%
\begin{figure}
\centering{}
\includegraphics[scale=0.4]{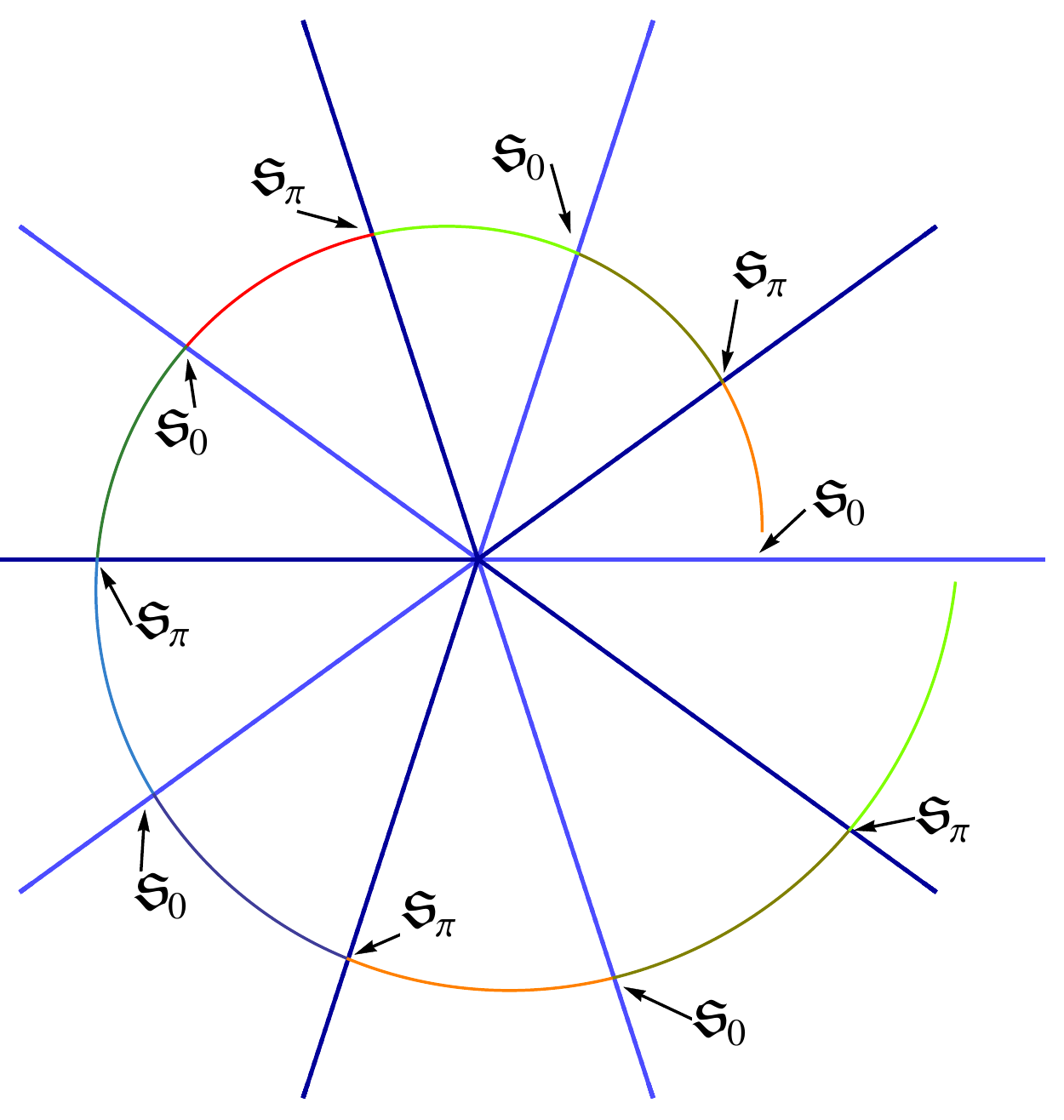}\hspace{10pt}\includegraphics[scale=0.4]{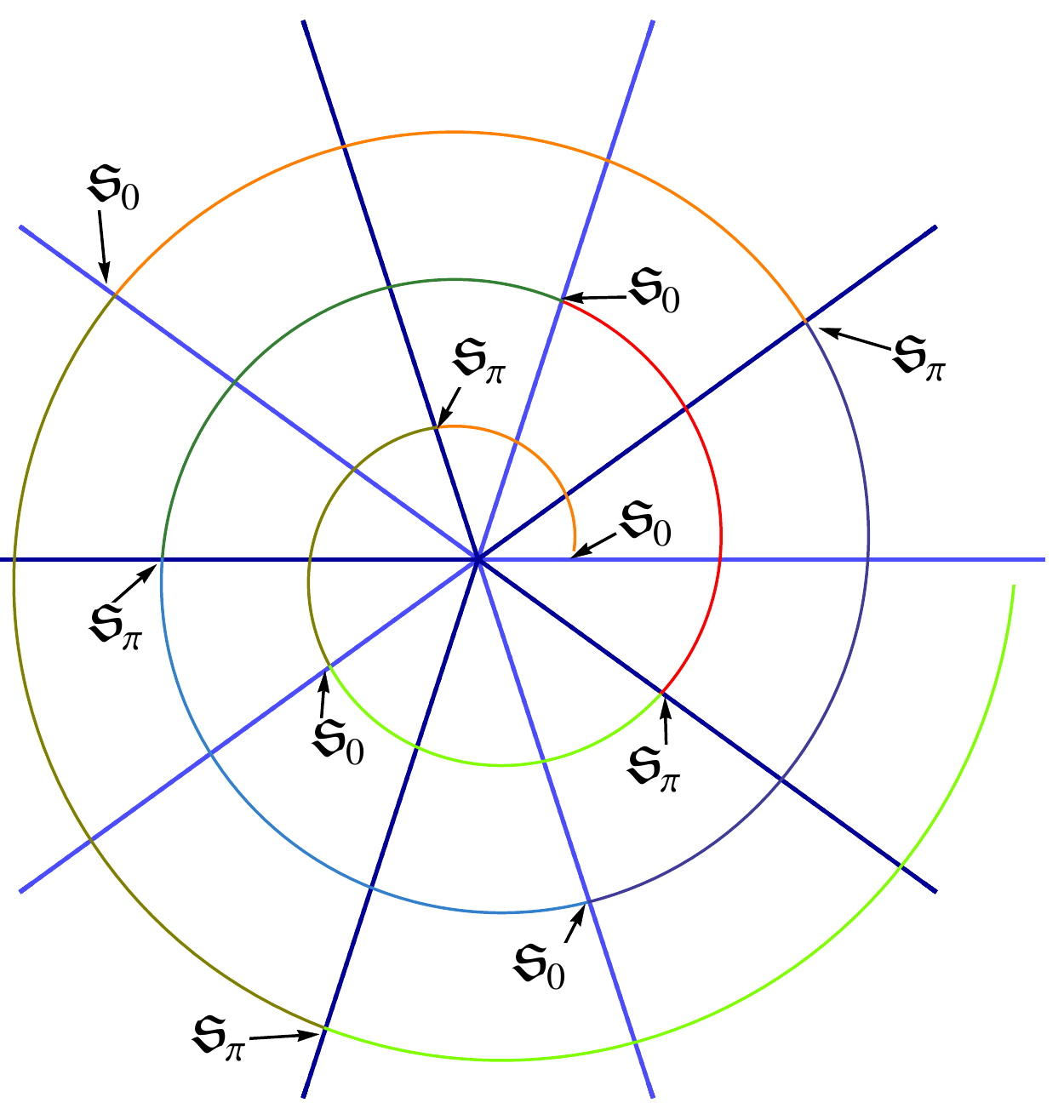}\hspace{10pt}\includegraphics[scale=0.4]{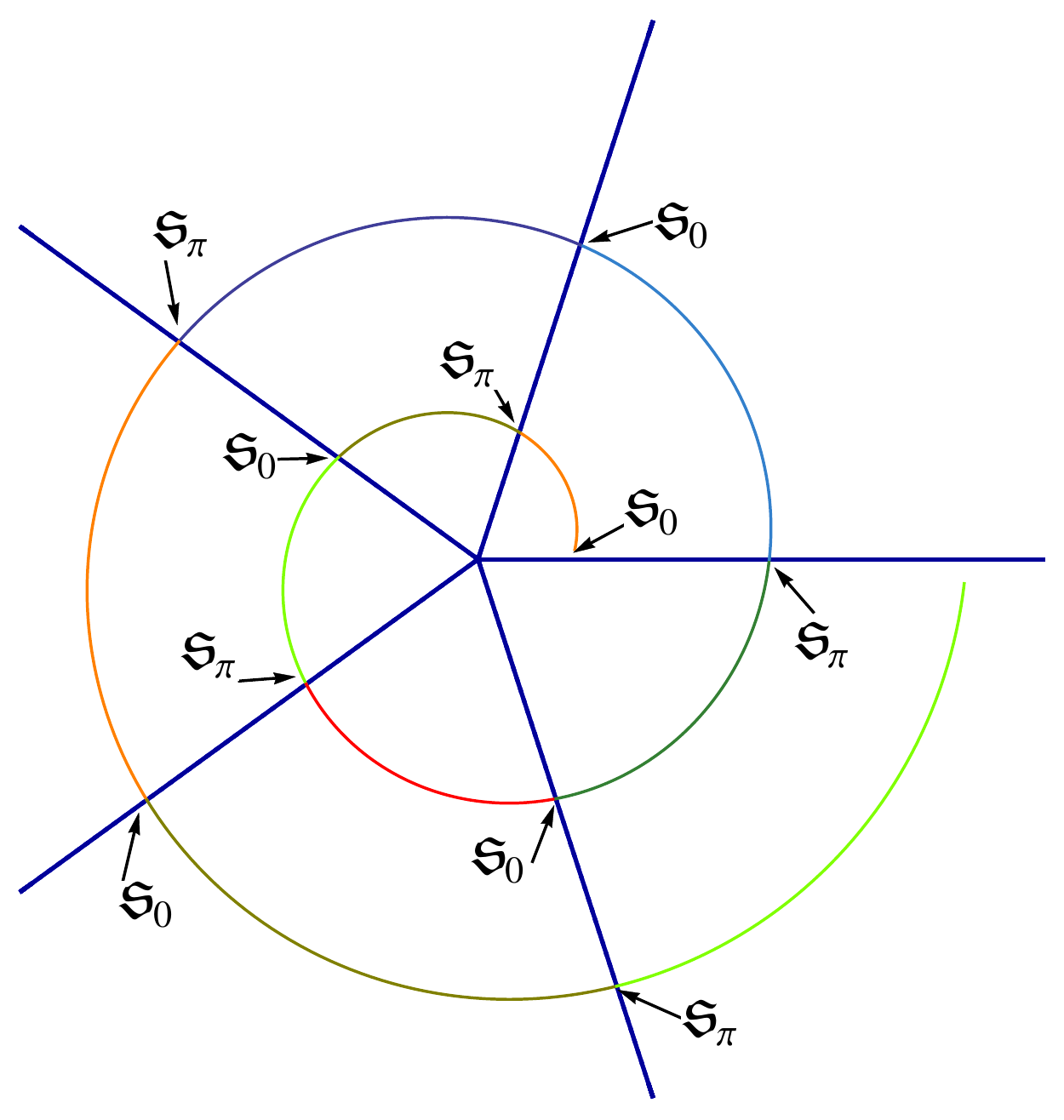}
\caption{Stokes transitions for different relations between the ``working'' variable $z$ and the ``physical'' coupling $\kappa$: from left to right we have $z=\kappa^{5}$, $z=\kappa^{5/3}$ and $z=\kappa^{5/2}$, respectively. For each case, we show the different successive transitions taking place as we rotate the argument $\theta_{\kappa}$ of the ``physical'' coupling $\kappa=\left|\kappa\right|\rme^{\rmi\theta_{\kappa}}$. The colored spiral represents all the transitions: the color changes whenever a transition occurs. The light blue lines correspond to transitions of the ``type'' $\underline{\mathfrak{S}}_{0}$ while the dark blue lines correspond to $\underline{\mathfrak{S}}_{\pi}$. In the last image they are on top of each other.
\label{fig:Stokes-transitions-in-coupling-variable}}
\end{figure}
%%%%%%%%%%%%%%%%%%%%%%%%%%%%%%%%%%%%%%%%%%%%%%%%%%%%%%%%%%%%%%%%%

In figure \ref{fig:Stokes-transitions-in-coupling-variable} we illustrate this unfolding in the $\kappa$--plane, for three different values of $\alpha$ representative of the properties described above (see also figure \ref{fig:transitions-for-Airy}). The anti--Stokes lines are not represented simply to make the figure easier to read, but they will always fall exactly in--between the Stokes lines. Concerning the cases illustrated in the figure, the two first cases have Stokes lines in the whole real axis, and only in the case $z=\kappa^{5/2}$ will we find the negative real axis being an anti--Stokes line. In general, the positive real axis will always be a Stokes line, while the negative real axis can be a Stokes or an anti--Stokes line, or neither. If $m$ is odd, the negative real axis will be a Stokes line, while if $m$ is even two cases may happen: if $m=4\ell+2$, $\ell\in\mathbb{N}$ the negative real axis is an anti--Stokes line, while if $m=4\ell$, $\ell\in\mathbb{N}$ the directions of anti--Stokes lines fall on top of the Stokes lines (but as one rotates around the complex $\kappa$--plane they alternate), and the negative real axis is neither a Stokes nor anti--Stokes line.

%%%%%%%%%%%%%%%%%%%%%%%%%%%%%%%%%%%%%%%%%%%%%%%%%%%%%%%%%%%%%%%%%
\subsection*{Computing the Monodromy}
%%%%%%%%%%%%%%%%%%%%%%%%%%%%%%%%%%%%%%%%%%%%%%%%%%%%%%%%%%%%%%%%%

In the usual ``working'' variable $z$ we can ask how many Stokes lines we cross, at $\theta=0$ and $\theta=\pi$, until we reach back our starting point. This is essentially a statement concerning the evaluation of the monodromy, defined by
\begin{equation}
\mathcal{\mathfrak{M}} := \underline{\mathfrak{S}}_{\pi} \circ \underline{\mathfrak{S}}_{0},
\end{equation}
\noindent
where we want to compute the value of $n$ such that
\begin{equation}
\mathfrak{M}^{n}F \left(z,\sigma_{1},\sigma_{2}\right) = F \left(z,\sigma_{1},\sigma_{2}\right),
\end{equation}
\noindent
\textit{i.e.}, find the order of the (finite) cyclic group describing the monodromy group in each case. Note that because Borel resummation occurs sectorially, along angular regions in the complex plane glued together by the Stokes automorphism, it is important to know the monodromy group if one is to fully construct the Riemann surface corresponding to the solution to the problem at hand. One might think that this problem is also intimately related to the fact that one has two variables at play, the ``physical'' variable $\kappa$ and the ``working'' variable $z$, and this certainly plays a role as we disentangle one coordinate into the other. Nonetheless, it is the Stokes constants which play a prominent role in the evaluation of the monodromy operator.

For the one--parameter transseries, and using expressions from appendix \ref{sec:App-Formulae-One-Param-transs}, the monodromy operator is given by
\begin{equation}
\mathfrak{M}\, F \left(z,\sigma\right) = \underline{\mathfrak{S}}_{\pi} F \left( z, \sigma+S_{1} \right) = F \left( z, \BS_{\pi} \left(\sigma+S_{1}\right) \right) \equiv F \left( z, \BS_{\pi} \circ \BS_{0} \left(\sigma\right) \right),
\end{equation}
\noindent
where we used
\begin{equation}
\BS_{0} (\sigma) = \sigma+S_{1}, \qquad \BS_{\pi} (\sigma) = \sum_{n=0}^{+\infty} \sigma^{n+1} \underline{\Sigma} (n+1,n),
\end{equation}
\noindent
and where $\underline{\Sigma}(n+1,n)$ is defined in appendix \ref{sec:App-Formulae-One-Param-transs}.

For the two--parameters transseries, the explicit monodromy operator because more cumbersome. Using the Stokes transitions listed in appendix \ref{sec:App-Formulae-for-Two--Parameters}, in particular expressions \eqref{eq:Stokes-transition-at-zero-2-param-strange-variables} and \eqref{eq:Stokes-transition-at-pi-2-param-strange-variables}, we can write
\begin{eqnarray}
\mathfrak{M}\, F \left(z,\sigma_{1},\sigma_{2}\right) &=& \underline{\mathfrak{S}}_{\pi}  F \left( z, \BS_{0,1} (\sigma_{1},\sigma_{2}), \BS_{0,2} (\sigma_{1},\sigma_{2}) \right) = \\
&=& F \left( z, \BS_{\pi,1} \left( \BS_{0,1} (\sigma_{1},\sigma_{2}), \BS_{0,2} (\sigma_{1},\sigma_{2}) \right), \BS_{\pi,2} \left( \BS_{0,1} (\sigma_{1},\sigma_{2}), \BS_{0,2} (\sigma_{1},\sigma_{2}) \right) \right),
\nonumber
\end{eqnarray}
\noindent
where $\BS_{\theta,i} \equiv \BS_{\theta,i}^{(\nu=1)}$ and
\begin{eqnarray}
\BS_{\theta,1} (\sigma_{1},\sigma_{2}) &=& \sigma_{1}\, \BP_{\theta,1}^{(1|0)} (\sigma_{1},\sigma_{2}), \\
\BS_{\theta,2} (\sigma_{1},\sigma_{2}) &=& \sigma_{2}\, \BP_{\theta,1}^{(0|1)} (\sigma_{1},\sigma_{2}),
\end{eqnarray}
\noindent
and the functions $\BP_{\theta,1}^{(n|m)}(\sigma_{1},\sigma_{2})$ are also given in appendix \ref{sec:App-Formulae-for-Two--Parameters}.

%%%%%%%%%%%%%%%%%%%%%%%%%%%%%%%%%%%%%%%%%%%%%%%%%%%%%%%%%%%%%%%%%
\subsection*{Example: The Airy Function}
%%%%%%%%%%%%%%%%%%%%%%%%%%%%%%%%%%%%%%%%%%%%%%%%%%%%%%%%%%%%%%%%%

In order to understand exactly how to build a real solution along the ``physical'' $\kappa$ real line, and what exactly is the role that the monodromy plays, let us illustrate the above set--up within a specific example; that of the well--known Airy function. Let us begin by quickly recalling the resurgent analysis of the Airy function, which has been thoroughly studied in the literature, see, \textit{e.g.}, \citep{Delabaere:2006ed, Marino:2012zq}. The solutions to the Airy differential equation 
\begin{equation}
\mathfrak{Z}''(\kappa) - \kappa\, \mathfrak{Z}(\kappa)=0
\end{equation}
\noindent
are given in integral form by
\begin{equation}
\mathfrak{Z}_{\gamma} (\kappa) = \frac{1}{2\pi\rmi}\, \int_{\gamma} \rmd u\, \rme^{V(u)}, \qquad V(u) = \kappa\, u - \frac{u^{3}}{3},
\end{equation}
\noindent
where the contour $\gamma$ is chosen such that the integral converges. There are two homologically independent choices of $\gamma$, giving the two independent solutions to the above differential equation usually denoted by $\mathcal{Z}_{\text{Ai}}$ and $\mathcal{Z}_{\text{Bi}}$. The transseries solution to the Airy equation can then be written with two parameters as
\begin{equation}
\mathcal{Z} \left(\kappa,\sigma_{1},\sigma_{2}\right) = \sigma_{1}\, \mathcal{Z}_{\text{Ai}}(\kappa) + \sigma_{2}\, \mathcal{Z}_{\text{Bi}}(\kappa),
\label{eq:Airy-function-transseries-ansatz}
\end{equation}
\noindent
with the two solutions defined asymptotically for $\kappa\gg1$ as
\begin{eqnarray}
\mathcal{Z}_{\text{Ai}} (\kappa) &=& \frac{1}{2\sqrt{\pi}\kappa^{1/4}}\, \rme^{-\frac{1}{2}A\, \kappa^{3/2}}\, \Phi_{-1/2}(\kappa), 
\label{eq:Airy-Asymptotics-Ai} \\
\mathcal{Z}_{\text{Bi}} (\kappa) &=& \frac{1}{2\sqrt{\pi}\kappa^{1/4}}\, \rme^{+\frac{1}{2}A\, \kappa^{3/2}}\, \Phi_{+1/2}(\kappa).
\label{eq:Airy-Asymptotics-Bi}
\end{eqnarray}
\noindent
In the above expressions the instanton action is $A=4/3$ and the asymptotic perturbative series are given by
\begin{equation}
\Phi_{\pm1/2}(\kappa) \simeq \sum_{n=0}^{+\infty} (\mp1)^{n}\, a_{n}\, \kappa^{-\frac{3}{2}n}.
\end{equation}
\noindent
The coefficients $a_{n}$ can be easily found via the original differential equation, and are such that the Borel transforms of the perturbative expansions are precisely given by the hypergeometric functions
\begin{equation}
\CB\left[\Phi_{\pm1/2}\right](s)=\pm\frac{5}{48}\,_{2}F_{1}\left(\frac{7}{6},\frac{11}{6},2\Big|\pm\frac{s}{A}\right).
\end{equation}
\noindent
This means one will find singularities at $s=\pm A$ in the Borel plane, for $\Phi_{\pm1/2}$, respectively. Consequently, there are two singular directions $\theta=0,\pi$. We can determine the Stokes automorphisms along these singular directions by first noting that the alien derivatives are
\begin{equation}
\Delta_{\pm A} \Phi_{\pm1/2}=+S_{\pm1}\,\Phi_{\mp1/2}, \qquad \Delta_{\pm A}\Phi_{\mp1/2}=0,
\end{equation}
\noindent
where the Stokes constants\footnote{In the notation used in appendix \ref{sec:App-Formulae-for-Two--Parameters}, these Stokes constants correspond to $S_{1} \equiv S_{1}^{(0)}$ and $S_{-1}\equiv\widetilde{S}_{-1}^{(0)}$.} are given by $S_{\pm1}=-\rmi$. The Stokes automorphisms in the relevant directions will follow by first performing a simple  change of variables, $z=\kappa^{3/2}$. In this new ``working'' variable, one has a one--to--one correspondence between singular directions in the Borel $s$--plane and Stokes lines in the $z$--plane. Then we can easily find the Stokes automorphisms straight from definition \eqref{eq:Stokes-aut-from-alien-derivatives}. In the $\theta=0$ direction one has
\begin{equation}
\underline{\mathfrak{S}}_{0} \mathcal{Z}_{\text{Ai}} \left(z\right) = \mathcal{Z}_{\text{Ai}} \left(z\right), \qquad \underline{\mathfrak{S}}_{0} \mathcal{Z}_{\text{Bi}} \left(z\right) = \mathcal{Z}_{\text{Bi}} \left(z\right) + S_{1}\, \mathcal{Z}_{\text{Ai}} \left(z\right),
\end{equation}
\noindent
while in the $\theta=\pi$ direction one finds
\begin{equation}
\underline{\mathfrak{S}}_{\pi} \mathcal{Z}_{\text{Bi}} \left(z\right) = \mathcal{Z}_{\text{Bi}} \left(z\right), \qquad \underline{\mathfrak{S}}_{\pi} \mathcal{Z}_{\text{Ai}} \left(z\right) = \mathcal{Z}_{\text{Ai}} \left(z\right) + S_{-1}\, \mathcal{Z}_{\text{Bi}} \left(z\right).
\end{equation}
\noindent
The Stokes transitions at the level of the transseries $\mathcal{Z} \left(z,\sigma_{1},\sigma_{2}\right)$, occurring in the directions $\theta=0,\pi$, are finally
\begin{eqnarray}
\underline{\mathfrak{S}}_{0}^{\nu} \mathcal{Z} (z,\sigma_{1},\sigma_{2}) &=& \mathcal{Z} \left( z, \sigma_{1} + \nu\, S_{1}\, \sigma_{2}, \sigma_{2} \right), \\
\underline{\mathfrak{S}}_{\pi}^{\nu} \mathcal{Z} (z,\sigma_{1},\sigma_{2}) &=& \mathcal{Z} \left( z, \sigma_{1}, \sigma_{2} + \nu\, S_{-1}\, \sigma_{1} \right).
\end{eqnarray}

\begin{figure}
\begin{centering}
\includegraphics[scale=0.5]{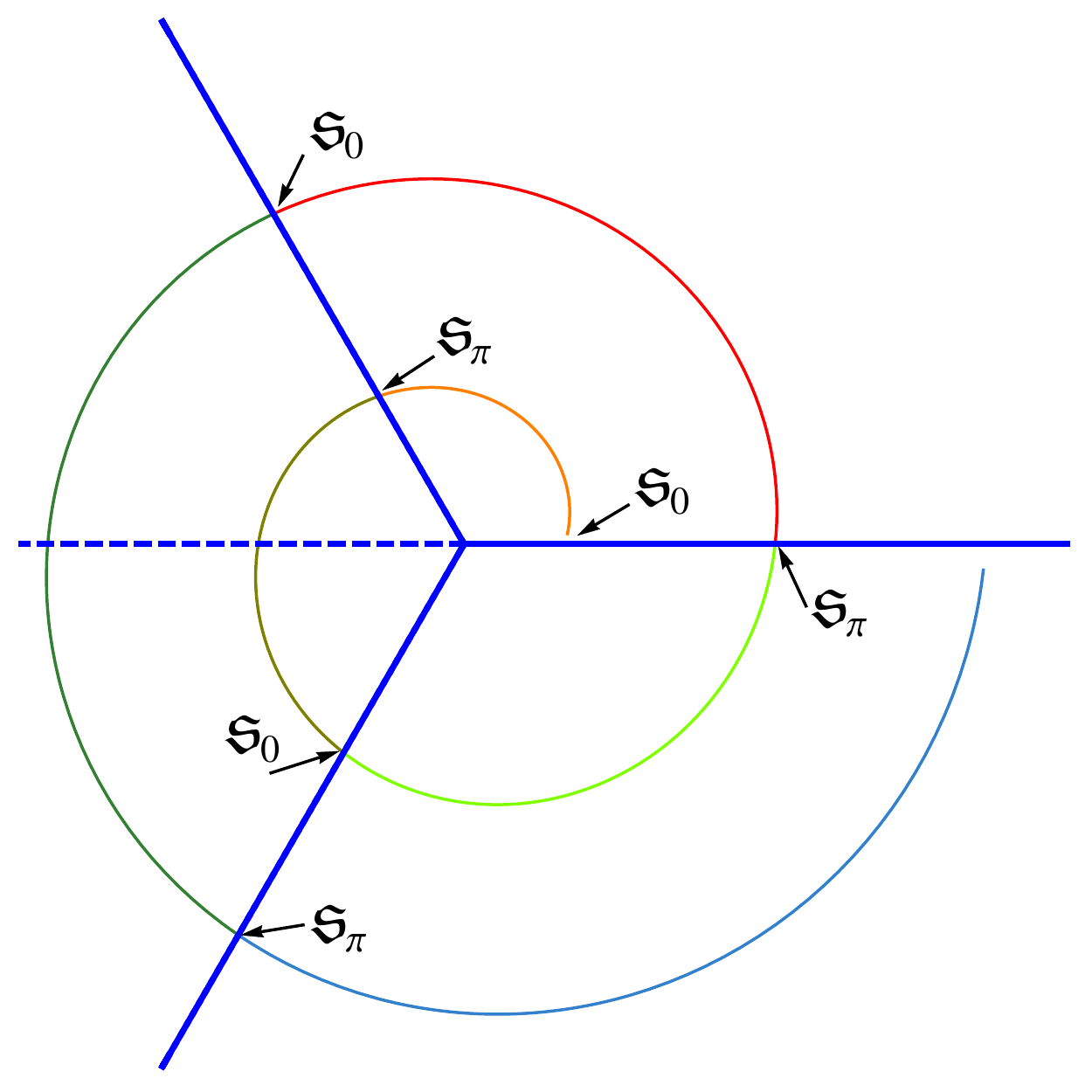}
\caption{Different successive transitions for the Airy function transseries, as one rotates the argument of the original ``physical'' variable $\kappa$. The Stokes lines are represented in thick blue, while in dashed blue is the negative real axis corresponding to an anti--Stokes line. Different colors are shown when there is a crossing of a Stokes line.
\label{fig:transitions-for-Airy}}
\par\end{centering}
\end{figure}

This very simple example now illustrates many features we discussed earlier. In the $z$--plane there are the usual two singular directions where Stokes phenomena takes place. But in the original variable these Stokes lines will unfold into extra Stokes lines. As discussed at the beginning of this section, in the original variable $\kappa=z^{2/3}$ one has the following unfolding,
\begin{eqnarray}
\mbox{Stokes line at }\arg z=0 & \quad \Rightarrow \quad & \mbox{Stokes lines at }\arg \kappa=0,\frac{4\pi}{3},\frac{2\pi}{3}, \\
\mbox{Stokes line at }\arg z=\pi & \quad \Rightarrow \quad & \mbox{Stokes lines at }\arg \kappa=\frac{2\pi}{3},0,\frac{4\pi}{3}.
\end{eqnarray}
\noindent
There are three Stokes directions in the $\kappa$--plane, occurring at $\arg \kappa \equiv \theta_{\kappa}=0,\frac{2\pi}{3},\frac{4\pi}{3}$; but in fact we will need to cross each of these lines at least twice in order to account for all possible Stokes phenomena in this problem, and the ``type'' of transition will alternate from $\underline{\mathfrak{S}}_{0}$ to $\underline{\mathfrak{S}}_{\pi}$. In figure \ref{fig:transitions-for-Airy} we have displayed the succession of different Stokes transitions taking place in the original variable $\kappa$, as we change the argument $\arg \kappa\in\left(0,4\pi\right)$.

%%%%%%%%%%%%%%%%%%%%%%%%%%%%%%%%%%%%%%%%%%%%%%%%%%%%%%%%%%%%%%%%%
%%%%%%%%%%%%%%%%%%%%%%%%%%%%%%%%%%%%%%%%%%%%%%%%%%%%%%%%%%%%%%%%%

Next, we would like to construct a real solution to the Airy equation, across the whole real line and in the original variable $\kappa$. At $\theta_{\kappa}=0$ there is a Stokes line, and an associated nonperturbative ambiguity which needs to be canceled to obtain a real transseries solution. As stated before, this cancelation is given by the median resummation, as
\begin{eqnarray}
\mathcal{Z}_{\mathbb{R}} \left(z,\sigma_{1},\sigma_{2}\right) &=& \mathcal{S}_{0}^{{\rm med}} \mathcal{Z}\left(z,\sigma_{1},\sigma_{2}\right) = \mathcal{S}_{0^{+}} \circ \underline{\mathfrak{S}}_{0}^{-1/2} \mathcal{Z}\left(z,\sigma_{1},\sigma_{2}\right) = \nonumber \\
&=& \mathcal{S}_{0^{+}} \mathcal{Z}\left(z,\sigma_{1}-\frac{1}{2}S_{1}\sigma_{2},\sigma_{2}\right),
\end{eqnarray}
\noindent
where $\sigma_{1},\sigma_{2}\in\mathbb{R}$. The particular case of $\sigma_{2}=0$ and $\sigma_{1}=1$ will then correspond to a real solution for $\theta_{\kappa}=0$, in particular to a very well--known real solution given by (where $z=\kappa^{3/2}$)
\begin{equation}
\mathcal{Z}_{\mathbb{R}} \left(z,1,0\right) = \mathcal{S}_{0^{+}} \mathcal{Z}\left(z,1,0\right) = \mathcal{S}_{0^{+}}\mathcal{Z}_{\text{Ai}}\left(z\right).
\end{equation}
\noindent
Let us start with this (real) solution along the positive real line and ask if one can also find a real solution along the negative real line $\arg \kappa = \pi$. Note that the negative real line in the original variable $\kappa$ does not correspond to a Stokes line and, thus, there will be no ambiguity along this direction. In figure \ref{fig:transitions-for-Airy} we show how rotating $\theta_{\kappa}$ from $0$ to $\pi$, the Stokes line at $\theta_{\kappa}=2\pi/3$ is crossed with the ``type'' of Stokes transition $\underline{\mathfrak{S}}_{\pi}$ (which takes place at $\arg z=\pi$). The transition in this singular direction is given by
\begin{eqnarray}
\CS_{\theta_{\kappa}=\left(2\pi/3\right)^{+}} \mathcal{Z}\left(z,1,0\right) \Big|_{z=\kappa^{3/2}} &=& \CS_{\theta_{\kappa}=\left(2\pi/3\right)^{-}} \underline{\mathfrak{S}}_{\pi} \mathcal{Z}\left(z,1,0\right) \Big|_{z=\kappa^{3/2}} = \nonumber \\
&=& \CS_{\theta_{\kappa}=\left(2\pi/3\right)^{-}} \mathcal{Z} \left( z=\kappa^{3/2}, 1, S_{-1} \right) = \nonumber \\
&=& \CS_{\theta_{\kappa}=\left(2\pi/3\right)^{-}} \mathcal{Z}_{\text{Ai}} \left(\kappa\right) - \rmi\, \CS_{\theta_{\kappa}=\left(2\pi/3\right)^{-}} \mathcal{Z}_{\text{Bi}} \left(\kappa\right).
\end{eqnarray}
\noindent
Having crossed the Stokes line, the solution for $\theta_{\kappa}=\pi$ is
\begin{equation}
\CS_{\theta_{\kappa}=\pi} \mathcal{Z} \left(\left|\kappa\right|\rme^{\rmi\pi}, 1, S_{-1} \right) = \CS_{\theta_{\kappa}=\pi} \mathcal{Z}_{\text{Ai}} \left( \left|\kappa\right|\rme^{\rmi\pi} \right) - \rmi\, \CS_{\theta_{\kappa}=\pi} \mathcal{Z}_{\text{Bi}} \left( \left|\kappa\right|\rme^{\rmi\pi} \right).
\end{equation}
\noindent
The reality of the solution along $\arg \kappa=\pi$ now follows. We analyze it in the asymptotic regime using the perturbative expansions \eqref{eq:Airy-Asymptotics-Ai} and \eqref{eq:Airy-Asymptotics-Bi}, with $\kappa=\left|\kappa\right|\rme^{\rmi\pi}$ and choosing a particular branch for the square roots\footnote{The possible choices are $\left(\rme^{\rmi\pi}\right)^{3/2} = \pm\rmi$ and $\left(\rme^{\rmi\pi}\right)^{1/4} = \frac{1}{\sqrt{2}}\left(1\mp\rmi\right)$. One choice will yield a real solution while the other will yield a purely imaginary one (and proportional to each other).} we have
\begin{equation}
\mathcal{Z} \left( \left|\kappa\right|\rme^{\rmi\pi}, 1, -\rmi \right) \simeq \sum_{n=0}^{+\infty} a_{n}\, C_{n} \left(-1\right)^{\left[3n/2\right]} \cos \left( \Theta + \left(-1\right)^{n} \frac{\pi}{4} \right),
\end{equation}
\noindent
where $\left[\bullet\right]$ is the integer part, the $a_{n}$ are the coefficients from the asymptotics, and
\begin{equation}
C_{n} = \frac{\left|\kappa\right|^{-3n/2}}{\sqrt{\pi}\left|\kappa\right|^{1/4}}, \qquad \Theta = - \frac{1}{2} A \left|\kappa\right|^{3/2}.
\end{equation}
\noindent
As expected in this familiar and very simple example, one indeed finds a real solution also along the negative real axis and, consequently, the transseries for the Airy function, \eqref{eq:Airy-function-transseries-ansatz}, with $\sigma_{1}=1$ and $\sigma_{2}=0$ defines a real transseries along the whole real line in the $\kappa$--plane.

Finally, we can compute the monodromy of the full Airy transseries. Given all the results above (and noting that $S_{1}S_{-1}=-1$) it is very straightforward to check that the monodromy action translates to
\begin{equation}
\mathfrak{M}\, \mathcal{Z} \left(z,\sigma_{1},\sigma_{2}\right) = \mathcal{Z} \left( z, \sigma_{1}+S_{1}\,\sigma_{2}, S_{-1}\,\sigma_{1} \right).
\end{equation}
\noindent
Applying the monodromy repeatedly we find that $\mathfrak{M}^{3}\, \mathcal{Z} \left(z,\sigma_{1},\sigma_{2}\right) = \mathcal{Z} \left(z,-\sigma_{1},-\sigma_{2}\right)$, and that 
\begin{equation}
\mathfrak{M}^{6} = \mathbf{1}.
\end{equation}
\noindent
It is important to note that this result is dictated by the structure of Stokes constants in the problem. Indeed, if one were just to think about the relation between $z$ and $\kappa$, figure \ref{fig:transitions-for-Airy} would show how rotating twice in $\kappa$ seems to bring us back to the starting point. But this would incorrectly imply that $\mathfrak{M}^{3} = \mathbf{1}$. The Stokes constants computing the monodromy tell us that one rather has to rotate four times in the ``physical'' variable $\kappa$ to return to the starting point.

%%%%%%%%%%%%%%%%%%%%%%%%%%%%%%%%%%%%%%%%%%%%%%%%%%%%%%%%%%%%%%%%%
%%%%%%%%%%%%%%%%%%%%%%%%%%%%%%%%%%%%%%%%%%%%%%%%%%%%%%%%%%%%%%%%%
\section{Outlook}
%%%%%%%%%%%%%%%%%%%%%%%%%%%%%%%%%%%%%%%%%%%%%%%%%%%%%%%%%%%%%%%%%
%%%%%%%%%%%%%%%%%%%%%%%%%%%%%%%%%%%%%%%%%%%%%%%%%%%%%%%%%%%%%%%%%

We have seen that while perturbation theory may lead to the appearance of nonperturbative ambiguities, it also contains, in itself, the proper prescription to cancel these ambiguities and become well--defined in a wide range of (quantum theoretical) problems. The greater the number of semiclassical saddles, instantons, renormalons or other more exotic saddles, the more complicated it is to write down the cancelation but in principle it always occurs in the precise same way: via the median resummation. It would be interesting to apply this general prescription across many different settings where perturbation theory plays prominent roles, from quantum mechanics to quantum field theory, from string theory to large $N$ gauge theories. Physical observables in these theories will generically be described by resurgent functions and transseries and, according to what we have explained in this work, these observables may be defined nonperturbatively starting out with perturbation theory and applying median resummation. The difficulty of explicitly implementing the actual calculation will naturally differ from problem to problem, in the form of properly identifying all relevant saddles and interpreting them physically, thus the obvious interest of seeing these methods applied over a wide range of concrete examples.

%%%%%%%%%%%%%%%%%%%%%%%%%%%%%%%%%%%%%%%%%%%%%%%%%%%%%%%%%%%%%%%%%
\begin{figure}
\centering{}
\includegraphics[scale=0.7]{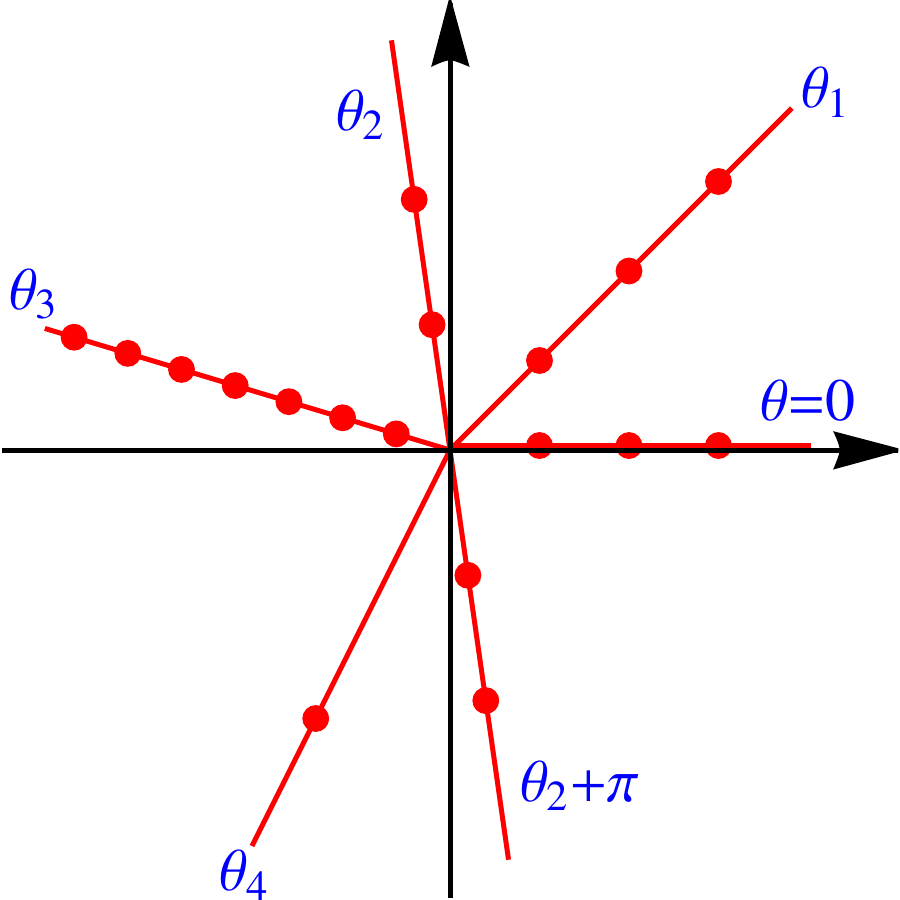}
\caption{The general global picture in the complex Borel plane, with many different Stokes lines characterized by distinct singularity structures. All these singularities will contribute to large--order resurgent asymptotics, and global definitions will require canceling all associated nonperturbative ambiguities.
\label{fig:global-picture}}
\end{figure}
%%%%%%%%%%%%%%%%%%%%%%%%%%%%%%%%%%%%%%%%%%%%%%%%%%%%%%%%%%%%%%%%%

In this work we have addressed both one and two--parameter transseries, where the two--parameters transseries also included generalized instanton sectors. In particular, the transseries with two parameters involved instanton actions $\pm A$, which is a familiar setting we have addressed in other, related, cases \citep{Aniceto:2011nu, Schiappa:2013opa}. However, this selection was solely due to practical purposes (it makes the discussion proceed along familiar ground, where many formulae are already available). We have not completely discussed the logarithmic sectors and resonance which also appear in this set--up, and it would be interesting to fully address these sectors in future work. In spite of this, we believe the general lesson should be clear and, in more complicated problems, one should follow along the same lines now applied to transseries with many parameters and many distinct instanton actions (or renormalons, generalized instantons, maybe even generalized renormalons and more). More specifically, when considering general actions, the Stokes automorphism $\underline{\mathfrak{S}}_\theta$ will be non--trivial across other directions than $0$ or $\pi$, and this will entail iterating our analysis along these new directions. In fact, ambiguities should be canceled along all possible Stokes lines if the function we are looking for is to be well--defined everywhere in the complex plane (see figure \ref{fig:global-picture}).

It is interesting to note that some ``part'' of the transseries parameters was used in the cancelation of nonperturbative ambiguities (as they took specified values under median resummation; \textit{e.g.}, in the one--parameter example the imaginary part of $\sigma$ was fixed). Whatever freedom is left can be seen to play the role of a theta--like QCD angle, or, in the differential equation context, of a parameterization of possible boundary conditions. In this case, globally defined solutions to the differential equation at stake will be obtained by canceling all possible ambiguities in the complex plane. It is not to exclude that transseries parameters may be further constrained in this way: requirements of global definition may fix the remaining freedom by imposing extra constraints (much like in our discussions of constructing real transseries along the full real line).

As we briefly commented in the introduction, in quantum mechanical problems with degenerate vacua \citep{Bogomolny:1980ur, ZinnJustin:1981dx, ZinnJustin:1982td}, and also in gauge field theories \citep{Dunne:2012ae}, one needs to consider topological charge and introduce a theta--angle $\Theta$. The effect of this theta--angle will be that exponential suppressed terms will acquire a phase: schematically, one will now find exponential corrections of the type $\rme^{-nAz}\, \rme^{\rmi \left(n-2k\right) \Theta}$, where $k=0,\ldots,n$. Each topological sector, defined by a specific phase, will have contributions from $m_{\mathcal{I}}$ instantons and $m_{\bar{\mathcal{I}}}$ anti--instantons, such that $m_{\mathcal{I}}-m_{\bar{\mathcal{I}}}=n-2k$ and the total number of instantons and anti--instantons will give the instanton level $n=m_{\mathcal{I}}+m_{\bar{\mathcal{I}}}$.

Perturbative expansions are independent of $\Theta$, so that sectors with different phases will not mix. As such, for each of these sectors we shall assume to have an independent transseries. Let us first analyze the topological sector with no theta--angle dependence. This topological sector occurs when $n-2k \equiv m_{\mathcal{I}}-m_{\bar{\mathcal{I}}}=0$. As $k$ needs to be an integer, only multi--instantons with $n=2\ell$ even will contribute. These contributions will thus arise from levels with the same number of instantons and anti--instantons $\left[\mathcal{I}^{\ell} \bar{\mathcal{I}}^{\ell} \right]$. The lowest level (the least suppressed contribution) of this topological sector occurs at $n=0$ where we find the usual perturbative series; the next level contributing to the transseries will be the instanton/anti--instanton sector $\left[\mathcal{I}\bar{\mathcal{I}}\right]$, which will have $n=2$; other exponentially suppressed contributions will appear at $n=4,6,\ldots$. This is in contrast with other topological sectors, having explicit $\Theta$ dependence, where now the lowest level will already be exponentially suppressed. For example, for $n-2k=1$, the lowest value of $n$ which contributes is $n=1$, which corresponds to having only one instanton. This is the least suppressed contribution, and will play the analogous part of our perturbative series for this sector. The next contributions will occur at $n=3,5,\cdots=2\ell+1$, and correspond to instanton sectors of the type $\left[\mathcal{I}^{1+\ell}\bar{\mathcal{I}}^{\ell}\right]$. This discussion can be generalized for every topological sector, and is nicely summarized in the ``graded resurgence triangle'' found in \citep{Dunne:2012ae}.

The natural question one now needs to address is how to write the full transseries \textit{ansatz} for a problem with graded theta--angle. Assuming full independence between the different phases, and considering for simplicity of the argument that each sector will be described by an one--parameter transseries, this can be achieved by considering:
\begin{equation}
F\left( z,\left\{ \sigma_{i}\right\} \right)=\sum_{m \in \BZ} \rme^{\rmi m\Theta}\, F_{m} \left( z,\sigma_{m} \right),
\end{equation}
\noindent
where $F_{m}\left(z,\sigma_{m}\right)$ is the transseries for the topological $m$--sector. The lowest level of each transseries (equivalent to the ``usual'' perturbative series) will be proportional to $\rme^{-\left|m\right|Az}$, and the poles in the Borel plane for each sector are separated by the instanton/anti--instanton action $S_{\mathcal{I}\bar{\mathcal{I}}} = 2A$ (which corresponds to inserting an instanton/anti--instanton pair). Then
\begin{equation}
F_{m}\left(z,\sigma_{m}\right) = \rme^{-\left|m\right|Az}\, \sum_{n=0}^{+\infty} \sigma_{m}^{n}\, \rme^{- n S_{\CI\bar{\CI}} z}\, \Phi_{2n}^{\{m\}} \left(z\right),
\end{equation}
\noindent
where $\Phi_{2n}^{\{m\}}\left(z\right)$ has the familiar perturbative series expansion in the coupling $z$. The alien derivatives $\Delta_{\ell S_{\CI\bar{\CI}}}$, with $S_{\CI\bar{\CI}}=2A$, will be non--zero when acting on $F_{m}\left(z,\sigma_{m}\right)$, and the bridge equations over topological sectors may be written as
\begin{equation}
\dot{\Delta}_{\ell S_{\CI\bar{\CI}}} F = \sum_{m \in \BZ} S_{\ell}^{\{m\}} \left(\left\{ \sigma_{i}\right\} \right) \frac{\partial F}{\partial\sigma_{m}}.
\end{equation}
\noindent
Is is easy to see that the proportionality coefficients $S_{\ell}^{\{m\}}\left(\left\{ \sigma_{i}\right\} \right)$ must now be of the form
\be
S_{\ell}^{\{m\}} \left(\sigma_{m}\right) = \sum_{k\ge0} S_{\ell}^{\{m\}(k)} \sigma_{m}^{k},
\ee
\noindent
\textit{i.e.}, for each topological sector $\{m\}$ the coefficients can only depend on $\sigma_{m}$, with Stokes constants given by $S_{\ell}^{\{m\}(k)}$. Furthermore, the only surviving Stokes constant will be $S_{\ell}^{\{m\}(1-\ell)}$, for each sector $\{m\}$. We thus obtain, for each topological sector, the usual form of the bridge equations
\begin{equation}
\Delta_{\ell S_{\CI\bar{\CI}}} \Phi_{2n}^{\{m\}}(z) = S_{\ell}^{\{m\}(1-\ell)} \Phi_{2(n+\ell)}^{\{m\}}(z).
\end{equation}
\noindent
With the bridge equations in hand one may proceed to develop resurgent asymptotics and  median resummations, even for multi--parameter transseries, following the lines in \citep{Aniceto:2011nu} and in the present paper. It would be very interesting to put all these arguments in firmer ground, and develop the required technology as it applies to quantum theoretical problems with topological sectors.

Finally, let us mention that recently there has been some interest in applying strong--weak coupling duality to improve the resummation of perturbation theory within string theoretic contexts \citep{Sen:2013oza, Beem:2013hha, Banks:2013nga}. In this set--up, it would be very interesting to further investigate the existence of explicitly S--dual invariant median resummations.

%%%%%%%%%%%%%%%%%%%%%%%%%%%%%%%%%%%%%%%%%%%%%%%%%%%%%%%%%%%%%%%%%
\acknowledgments
We would like to thank Ricardo Couso--Santamar\'\i a, Gerald Dunne, Marcos Mari\~no, Ricardo Vaz, Marcel Vonk and, specially, to Mithat \"Unsal for useful discussions and/or comments. The authors would further like to thank CERN TH--Division for hospitality, where a part of this work was conducted. This research was partially supported by the FCT--Portugal grants PTDC/MAT/119689/\\ 2010 and EXCL/MAT-GEO/0222/2012.
%%%%%%%%%%%%%%%%%%%%%%%%%%%%%%%%%%%%%%%%%%%%%%%%%%%%%%%%%%%%%%%%%

\newpage

%%%%%%%%%%%%%%%%%%%%%%%%%%%%%%%%%%%%%%%%%%%%%%%%%%%%%%%%%%%%%%%%%
%%%%%%%%%%%%%%%%%%%%%%%%%%%%%%%%%%%%%%%%%%%%%%%%%%%%%%%%%%%%%%%%%
\appendix
%%%%%%%%%%%%%%%%%%%%%%%%%%%%%%%%%%%%%%%%%%%%%%%%%%%%%%%%%%%%%%%%%
%%%%%%%%%%%%%%%%%%%%%%%%%%%%%%%%%%%%%%%%%%%%%%%%%%%%%%%%%%%%%%%%%

%%%%%%%%%%%%%%%%%%%%%%%%%%%%%%%%%%%%%%%%%%%%%%%%%%%%%%%%%%%%%%%%%
%%%%%%%%%%%%%%%%%%%%%%%%%%%%%%%%%%%%%%%%%%%%%%%%%%%%%%%%%%%%%%%%%
\section{Structural Aspects of Resurgent Transseries\label{sec:App-Properties-of-med-resumm}}
%%%%%%%%%%%%%%%%%%%%%%%%%%%%%%%%%%%%%%%%%%%%%%%%%%%%%%%%%%%%%%%%%
%%%%%%%%%%%%%%%%%%%%%%%%%%%%%%%%%%%%%%%%%%%%%%%%%%%%%%%%%%%%%%%%%

This appendix addresses several structural aspects which may be deduced concerning arbitrary transseries, in particular many constraints on the associated set of Stokes constants. Let us begin by recalling the definition of lateral Borel resummation along some (singular) direction $\theta$, which may be found either in the main body of the text or in \citep{Aniceto:2011nu}. Given an asymptotic expansion of the form \eqref{eq:d} or \eqref{eq:pert-instanton-expansions-Phi_n},
\be
\Phi (z) \simeq z^{-\beta_{nm}} \sum_{g=0}^{+\infty} \frac{F_g}{z^g},
\label{asympseriesappA}
\ee
\noindent
one has
\begin{equation}
\CS_{\theta^{\pm}} \Phi(z) = \int_{0}^{\rme^{\rmi\theta}(\infty\pm\rmi\epsilon)} \rmd s\, \CB[\Phi](s)\, \rme^{-zs}.
\end{equation}
\noindent
With a simple change of variables one obtains
\begin{equation}
\CS_{\theta^{\pm}} \Phi(z) = \rme^{\rmi\theta} \int_{0}^{+\infty\pm\rmi\epsilon} \rmd s\, \CB[\Phi] (s\, \rme^{\rmi\theta})\, \rme^{-z s\,\rme^{\rmi\theta}} \equiv \CS_{0^{\pm}} \Phi^{(\theta)} (x),
\end{equation}
\noindent
where $x=z\rme^{\rmi\theta}$, and where we introduced the ``rotated'' asymptotic series $\Phi^{(\theta)} (x)$ which is obtained from the original expansion \eqref{asympseriesappA} by changing coefficients as
\begin{equation}
F_{g}^{(\theta)} = \rme^{2\rmi\theta\left(g+\beta_{nm}\right)} F_{g}.
\end{equation}
\noindent
The above relations interchange asymptotic series with singularities along $\theta$ to ``rotated'' asymptotic series with singularities along the positive real axis. In particular they show how generic cases may be reduced to our analysis, mostly along $\theta=0,\pi$. As such, we shall focus upon these directions in the following, setting $\theta \equiv 0,\pi$, and drop ``rotated'' superscripts.

Along the Borel real axis the usual complex conjugation operator, $\mathcal{H}F(z)\equiv\overline{F(z)}$, relates very naturally with the lateral Borel resummations. One finds, for \textit{real} $z$,
\begin{equation}
\mathcal{H} \circ \CS_{0^{\pm}} \Phi (z) = \int_{0}^{+\infty\mp\rmi\epsilon}\rmd s\, \overline{\CB[\Phi]}(s)\, \rme^{-zs},
\end{equation}
\noindent
and, as long as $z$ is real, it is also the case that $\overline{\CB[\Phi]} = \CB[\overline{\Phi}]$. In this case,
\begin{equation}
\mathcal{H}\circ\CS_{0^{\pm}} = \CS_{0^{\mp}}\circ\mathcal{H}.
\end{equation}
\noindent
Further assuming that $\beta_{nm}$ is an integer (we will soon lift this restriction), the same arguments allow one to find\footnote{Going back to our discussion in the first paragraph one may wonder if this relation can be generalized to \textit{any} singular direction $\theta$. In order to do so one needs to find a conjugation operator along the direction $\theta$, $\CH_\theta$, essentially reflecting the complex plane along that line which would naturally satisfy $\CH_\theta^2 = \boldsymbol{1}$ and which would further have to satisfy $\mathcal{H}_\theta \circ \CS_{\theta^{\pm}} = \CS_{\theta^{\mp}} \circ \mathcal{H}_\theta$. Then, a smiliar line of arguments goes through.}
\begin{equation}
\mathcal{H}\circ\CS_{\theta^{\pm}} = \CS_{\theta^{\mp}}\circ\mathcal{H}, \qquad \text{with } \theta=0,\pi.
\label{eq:App-Comm-rel-resumm-conjugation}
\end{equation}

Having understood how complex conjugation interplays with the lateral Borel resummations, one may now try to do the same on what concerns the Stokes automorphism \eqref{eq:stokesauto}. Using $\mathcal{H}^{2}=\boldsymbol{1}$, it is straightforward to check that for $\theta=0,\pi$ the commutation relation of complex conjugation and the Stokes automorphism is given by
\begin{equation}
\mathcal{H}\circ\underline{\mathfrak{S}}_{\theta} = \underline{\mathfrak{S}}_{\theta}^{-1}\circ\mathcal{H},
\label{eq:App-Comm-conjugation-stokes}
\end{equation}
\noindent
and obeys
\begin{equation}
\left(\mathcal{H}\circ\underline{\mathfrak{S}}_{\theta}\right)^{2} = \boldsymbol{1} = \left(\underline{\mathfrak{S}}_{\theta}^{-1}\circ\mathcal{H}\right)^{2}.
\end{equation}
\noindent
Next, as the Stokes automorphism relates to the alien derivative following
\eqref{eq:Stokes-aut-from-alien-derivatives}, we are still interested in exploring \eqref{eq:App-Comm-conjugation-stokes} within the alien calculus setting. Generically, and as follows from \eqref{eq:Stokes-aut-from-alien-derivatives}, one may write powers of the Stokes automorphism along a singular direction $\theta$ of the Borel plane as
\begin{equation}
\underline{\mathfrak{S}}_{\theta}^{\nu} = \exp \left\{ \nu \sum_{\omega\in\left\{ \omega_{\theta} \right\}} \rme^{-\omega z}\, \Delta_{\omega} \right\} \equiv \exp \left\{ \nu\, \dot{\Delta}_{\theta} \right\},
\end{equation}
\noindent
where we used the pointed alien derivative $\dot{\Delta}_{\theta}$. If we plug this expansion back in (\ref{eq:App-Comm-conjugation-stokes}), this condition may be written as
\begin{equation}
\sum_{n=1}^{+\infty}\frac{1}{n!}\,\mathcal{H}\circ\left(\dot{\Delta}_{\theta}\right)^{n}=\sum_{n=1}^{+\infty}\frac{\left(-1\right)^{n}}{n!}\left(\dot{\Delta}_{\theta}\right)^{n}\circ\mathcal{H},
\end{equation}
\noindent
which is obeyed if
\begin{equation}
\dot{\Delta}_{\theta}\circ\mathcal{H}=-\mathcal{H}\circ\dot{\Delta}_{\theta}.\label{eq:App-comm-conjugation-pointedalien}
\end{equation}
\noindent
In this case, condition (\ref{eq:App-Comm-conjugation-stokes}) may be generalized to
\begin{equation}
\mathcal{H}\circ\underline{\mathfrak{S}}_{\theta}^{\nu} = \underline{\mathfrak{S}}_{\theta}^{-\nu}\circ\mathcal{H}.
\label{eq:App-Comm-conjugation-stokestoanypower}
\end{equation}
\noindent
It is noteworthy to mention that one may lift the integer requirement on $\beta_{nm}$, or further consider asymptotic expansions with logarithmic sectors (as will be studied shortly). In these cases, one has to isolate the ``pure'' asymptotic series part, which induces the singularities in the complex Borel plane, from the extra factors which remain as they were. Then, the Stokes automorphism only acts on the asymptotic part. The conjugation operator, on the other hand, will act on all factors. Acting with these operators in the full transseries can become even more convoluted, as the transseries parameters $\sigma_{i}$ are in general complex. In any case, the procedure to analyze these properties at the level of the transseries is exemplified in the cases studied below.

We shall now make use of these relations to obtain structural properties of two--parameter transseries, starting off with their bridge equations \eqref{eq:Bridge-eqs-2-param}. Let us determine the implications of \eqref{eq:App-comm-conjugation-pointedalien} for the two--parameters transseries (\ref{eq:Ftransseries}--\ref{eq:pert-instanton-expansions-Phi_n}), further assuming that the coefficients in all asymptotic expansions in $F(z,\sigma_1,\sigma_2)$ are real---which we do throughout this appendix. For the direction $\theta=0$ ($\ell>0$) it follows
\begin{equation}
\mathcal{H}\circ\dot{\Delta}_{\ell A}F(z,\sigma_{1},\sigma_{2})=\overline{S}_{\ell}\left(\overline{\sigma}_{1},\overline{\sigma}_{2}\right)\frac{\partial\overline{F}}{\partial\overline{\sigma}_{1}}+\overline{\widetilde{S}}_{\ell}\left(\overline{\sigma}_{1},\overline{\sigma}_{2}\right)\frac{\partial\overline{F}}{\partial\overline{\sigma}_{2}},
\label{eq:App-H-after-pointed-F}
\end{equation}
\noindent
and
\begin{equation}
\dot{\Delta}_{\ell A}\circ\mathcal{H}F(z,\sigma_{1},\sigma_{2})=S_{\ell}\left(\overline{\sigma}_{1},\overline{\sigma}_{2}\right)\frac{\partial\overline{F}}{\partial\overline{\sigma}_{1}}+\widetilde{S}_{\ell}\left(\overline{\sigma}_{1},\overline{\sigma}_{2}\right)\frac{\partial\overline{F}}{\partial\overline{\sigma}_{2}}.\label{eq:App-pointed-after-H-F-only-theta-zero}
\end{equation}
\noindent
In this direction, using \eqref{eq:App-comm-conjugation-pointedalien} and the two above expressions, we find
\begin{equation}
\left(S_{\ell}\left(\overline{\sigma}_{1},\overline{\sigma}_{2}\right)+\overline{S}_{\ell}\left(\overline{\sigma}_{1},\overline{\sigma}_{2}\right)\right)\frac{\partial\overline{F}}{\partial\overline{\sigma}_{1}}=-\left(\widetilde{S}_{\ell}\left(\overline{\sigma}_{1},\overline{\sigma}_{2}\right)+\overline{\widetilde{S}}_{\ell}\left(\overline{\sigma}_{1},\overline{\sigma}_{2}\right)\right)\frac{\partial\overline{F}}{\partial\overline{\sigma}_{2}}.
\end{equation}
\noindent
Using the expansions of $S_{\ell}$ and $\widetilde{S}_{\ell}$ in terms of Stokes constants, as given by \eqref{eq:Expansion-of-Sell-2-param} and \eqref{eq:Expansion-of-Stildeell-2-param}, this implies
\begin{equation}
\re\, S_{\ell}^{(k)}=\re\,\widetilde{S}_{\ell}^{(k+\ell)}=\re\, S_{\ell}^{(0)}=0,\qquad\forall\, k,\ell\ge1.
\end{equation}
\noindent
In words, all Stokes constants associated with the $\theta=0$ direction are purely imaginary. 

As we turn to the $\theta=\pi$ direction, one sets\footnote{This allows us to consistently choose the usual branches of square--roots and logarithms, in direct comparison with results in \citep{Aniceto:2011nu, Schiappa:2013opa}. If we chose $z=\rme^{\rmi\pi}|z|$, we would have to use different branches to reach the same results.} $z=\rme^{-\rmi\pi}\left|z\right|$ and we thus find that $\mathcal{H}\Phi_{(n|m)}(z)\ne\Phi_{(n|m)}(z)$. As such, in this direction only \eqref{eq:App-H-after-pointed-F} remains valid. To be able to use \eqref{eq:App-comm-conjugation-pointedalien} notice that
\begin{equation}
\mathcal{H}F\left(z,\sigma_{1},\sigma_{2}\right)=\sum_{n,m=0}^{+\infty}\overline{\sigma}_{1}^{n}\overline{\sigma}_{2}^{m}\,\rme^{-(n-m)Az}\,\rme^{-2\pi\rmi\beta_{nm}}\,\Phi_{(n|m)}(z).
\end{equation}
\noindent
Assuming as usual that $\beta_{nm}=\left(n+m\right)\beta$, then
\begin{equation}
\mathcal{H}F\left(z,\sigma_{1},\sigma_{2}\right)=F\left(z,\overline{\sigma}_{1}\rme^{-2\pi\rmi\beta},\overline{\sigma}_{2}\rme^{-2\pi\rmi\beta}\right).
\end{equation}
\noindent
It follows
\begin{equation}
\mathcal{H}\left[\frac{\partial F}{\partial\sigma_{1}}\left(z,\sigma_{1},\sigma_{2}\right)\right]=\rme^{-2\pi\rmi\beta}\left.\frac{\partial F}{\partial\widetilde{\sigma}_{1}}\left(z,\widetilde{\sigma}_{1},\widetilde{\sigma}_{2}\right)\right|_{\widetilde{\sigma}_{i}=\overline{\sigma}_{i}\rme^{-2\pi\rmi\beta}},
\end{equation}
\noindent
and one may thus rewrite \eqref{eq:App-H-after-pointed-F} as
\begin{equation}
\mathcal{H}\circ\dot{\Delta}_{\ell A}F(z,\sigma_{1},\sigma_{2})=\rme^{-2\pi\rmi\beta}\left.\left[\overline{S}_{\ell}(\overline{\sigma}_{1},\overline{\sigma}_{2})\frac{\partial F(z,\widetilde{\sigma}_{1},\widetilde{\sigma}_{2})}{\partial\widetilde{\sigma}_{1}}+\overline{\widetilde{S}}_{\ell}(\overline{\sigma}_{1},\overline{\sigma}_{2})\frac{\partial F(z,\widetilde{\sigma}_{1},\widetilde{\sigma}_{2})}{\partial\widetilde{\sigma}_{1}}\right]\right|_{\widetilde{\sigma}_{i}=\overline{\sigma}_{i}\rme^{-2\pi\rmi\beta}},
\end{equation}
\noindent
and the equivalent of \eqref{eq:App-pointed-after-H-F-only-theta-zero} as
\begin{equation}
\dot{\Delta}_{\ell A}\circ\mathcal{H}F(z,\sigma_{1},\sigma_{2})=\left.\left[S_{\ell}(\widetilde{\sigma}_{1},\widetilde{\sigma}_{2})\frac{\partial F(z,\widetilde{\sigma}_{1},\widetilde{\sigma}_{2})}{\partial\widetilde{\sigma}_{1}}+\widetilde{S}_{\ell}(\widetilde{\sigma}_{1},\widetilde{\sigma}_{2})\frac{\partial F(z,\widetilde{\sigma}_{1},\widetilde{\sigma}_{2})}{\partial\widetilde{\sigma}_{1}}\right]\right|_{\widetilde{\sigma}_{i}=\overline{\sigma}_{i}\rme^{-2\pi\rmi\beta}}.
\end{equation}
\noindent
Using once again the expansions for the Stokes coefficients \eqref{eq:Expansion-of-Sell-2-param} and \eqref{eq:Expansion-of-Stildeell-2-param} into the commutation relation \eqref{eq:App-comm-conjugation-pointedalien}, one now finds that, for any $\ell>0$ and $k\ge0$,
\begin{eqnarray}
\widetilde{S}_{-\ell}^{(k)}+\overline{\widetilde{S}_{-\ell}^{(k)}}\,\rme^{2\pi\rmi\left(2k-2+\ell\right)\beta} & = & 0,\\
S_{-\ell}^{(k+\ell)}+\overline{S_{-\ell}^{(k+\ell)}}\,\rme^{2\pi\rmi\left(2k-2+\ell\right)\beta} & = & 0.
\end{eqnarray}

In summary, we found the constraints that \eqref{eq:App-comm-conjugation-pointedalien} imposes at the level of the bridge equations for a two--parameter transseries and, in particular, the constraints on the Stokes constants encoded in these bridge equations. In this process, we made a set of (very reasonable) assumptions:
\begin{itemize}
\item Assumed \eqref{eq:App-comm-conjugation-pointedalien} as the only solution of \eqref{eq:App-Comm-conjugation-stokes}; 
\item Started from a two--parameters transseries \textit{ansatz}, of the type (\ref{eq:Ftransseries}--\ref{eq:pert-instanton-expansions-Phi_n}); 
\item Both the asymptotic coefficients in the transseries and the instanton action are real; 
\item The exponent $\beta$ takes the (usual) form $\beta_{nm}=\left(n+m\right)\beta$, with $\beta$ a rational number. 
\end{itemize}
\noindent 
The constraints we thus found for the Stokes constants, for $\forall\, k\ge0,\ell\ge1$ and where we defined $\widetilde{S}_{\ell}^{(\ell)}=S_{-\ell}^{(\ell)}=0$, are:
\begin{eqnarray}
\re\, S_{\ell}^{(k)}=\re\,\widetilde{S}_{\ell}^{(k+\ell)} & = & 0,\\
\widetilde{S}_{-\ell}^{(k)}+\overline{\widetilde{S}_{-\ell}^{(k)}}\,\rme^{2\pi\rmi\left(2k-2+\ell\right)\beta} & = & 0,\\
S_{-\ell}^{(k+\ell)}+\overline{S_{-\ell}^{(k+\ell)}}\,\rme^{2\pi\rmi\left(2k-2+\ell\right)\beta} & = & 0.
\end{eqnarray}

%%%%%%%%%%%%%%%%%%%%%%%%%%%%%%%%%%%%%%%%%%%%%%%%%%%%%%%%%%%%%%%%%
\subsection*{The One--Parameter Transseries Example}
%%%%%%%%%%%%%%%%%%%%%%%%%%%%%%%%%%%%%%%%%%%%%%%%%%%%%%%%%%%%%%%%%

The previous analysis was done within the context of the two--parameters solutions, but the one--parameter transseries is just a particular case of these solutions, obtained by setting $\widetilde{S}=0$ and $\sigma_{2}=0$. In particular, the expansion \eqref{eq:Expansion-of-Sell-2-param} for the Stokes coefficients becomes
\begin{equation}
S_{-\ell}\left(\sigma_{1}\right)=S_{-\ell}^{(1+\ell)}\,\sigma_{1}^{1+\ell},\qquad\forall\,\ell\ge-1,
\end{equation}
\noindent
and the final constraints in the Stokes constants simply read, for $\ell\ge1$,
\begin{eqnarray}
\re\, S_{1}^{(0)} & = & 0,\\
S_{-\ell}^{(1+\ell)}+\overline{S_{-\ell}^{(1+\ell)}}\,\rme^{2\pi\rmi\ell\beta} & = & 0.
\end{eqnarray}
\noindent
In the case where $\beta=0$, this becomes 
\begin{equation}
\re\, S_{1}^{(0)}=\re\, S_{-\ell}^{(1+\ell)}=0.
\end{equation}

%%%%%%%%%%%%%%%%%%%%%%%%%%%%%%%%%%%%%%%%%%%%%%%%%%%%%%%%%%%%%%%%%
\subsection*{Generalization with Logarithmic Sectors}
%%%%%%%%%%%%%%%%%%%%%%%%%%%%%%%%%%%%%%%%%%%%%%%%%%%%%%%%%%%%%%%%%

The structure of the two--parameters transseries we are addressing, (\ref{eq:Ftransseries}--\ref{eq:pert-instanton-expansions-Phi_n}), has instanton actions $\pm A$. Of course more general structures are possible; in here we are illustrating our ideas with a setting inspired by the results in \citep{Aniceto:2011nu, Schiappa:2013opa}. One aspect of this setting, which is in fact generic for many other problems arising when considering non--linear differential equations, is resonance, \textit{i.e.}, the possibility that some instanton sectors will in fact \textit{not} have the usual exponential pre--factor because the combination of instanton actions canceled (see, \textit{e.g.}, \citep{Garoufalidis:2010ya, Aniceto:2011nu} for more detailed accounts of this phenomenon). To solve resonant problems, one needs to further introduce logarithmic sectors in the transseries structure, as was done for instance in the cases of the Painlev\'e I equation in \citep{Garoufalidis:2010ya, Aniceto:2011nu}
and the Painlev\'e II equation in \citep{Schiappa:2013opa}. For such cases we need a transseries \textit{ansatz} of the type \eqref{eq:Ftransseries} and \eqref{eq:IntroFn-def}, but where now
\begin{equation}
\Phi_{(n|m)}(z)=\sum_{k=0}^{k_{\text{max}}(n,m)}\log^{k} \left(z\right) \cdot z^{-\beta_{nm}^{[k]}}\, \Phi_{(n|m)}^{[k]}(z), \qquad \Phi_{(n|m)}^{[k]}(z) \simeq \sum_{g=0}^{+\infty} \frac{F_{g}^{(n|m)[k]}}{z^{g}},
\label{eq:AppA-Phinm-with-logs}
\end{equation}
\noindent
where the asymptotic coefficients $F_{g}^{(n|m)[k]}$ are taken to be real, and $k_{\text{max}}(n,m)=\min(n,m)-m\,\delta_{nm}$. To obtain more concrete results we shall still need a couple of assumptions which, in particular, are valid within the contexts of the Painlev\'e I and II equations. In general, the analysis will be model dependent but it will follow the very same strategy as below. We shall then further assume that
\begin{eqnarray}
\beta_{nm}^{[k]} & = & \beta_{n-k,m-k}^{[0]}, \qquad \beta_{nm}^{[0]}=\left(n+m\right)\beta,\\
\Phi_{(n|m)}^{[k]} & = & \left(\alpha\right)^{k}\, \frac{\left(m-n\right)^{k}}{k!}\, \Phi_{(n-k|m-k)}^{[0]}.
\end{eqnarray}
\noindent
The case of Painlev\'e I has $-2\alpha_{\text{PI}}=4/\sqrt{3}$ \citep{Aniceto:2011nu} and that of Painlev\'e II has\footnote{Comparing with \citep{Schiappa:2013opa} there would be a ${(-2)^{-k}}$ factor missing in \eqref{eq:AppA-Phinm-with-logs}, which is compensated with the removal of a similar ${(-2)^{k}}$ from $\Phi_{(n|m)}^{[k]}$ above. The results herein can then be directly compared to the ones of that paper.} $-2\alpha_{\text{PII}}=8$ \citep{Schiappa:2013opa}. Both cases have the same value of $\beta$, which we will be assuming from now on to be fixed as
\begin{equation}
\beta=\frac{1}{2}.
\end{equation}
\noindent
Using the fact that $\Phi_{(a|b)}^{[0]}=0$ if either $a$ or $b$ are negative, we can rewrite $\Phi_{(n|m)}(z)$ as
\begin{equation}
\Phi_{(n|m)}(z)=\sum_{k=0}^{+\infty} \alpha^{k}\, \frac{\left(m-n\right)^{k}}{k!}\, z^{-\beta_{n-k,m-k}^{[0]}}\, \log^{k} \left(z\right) \cdot \Phi_{(n-k|m-k)}^{[0]}(z).
\end{equation}
\noindent
Plugging these back into the transseries solution \eqref{eq:Ftransseries}, it may be rewritten as\footnote{The resonant transseries written in this form can also be used to write the Stokes transitions for resonant problems such as Painlev\'e I and II. Along the singular direction $\theta=0$ we can apply the Stokes automorphism to this transseries, and make use of \eqref{eq:2-param-Stokes0-in-Phi} to easily find a generalization of \eqref{eq:2-param-Stokes-trans-zero-first-form}, thus determining the Stokes transition for the resonant cases. The same can be done for the direction $\theta=\pi$.}
\begin{equation}
F(z,\sigma_{1},\sigma_{2})=\sum_{n=0}^{+\infty}\sum_{m=0}^{+\infty}\sigma_{1}^{n}\sigma_{2}^{m}\, z^{-\beta_{nm}^{[0]}}\,\rme^{-\left(n-m\right)Az}\,\rme^{\alpha\left(m-n\right)\sigma_{1}\sigma_{2}\,\log z}\,\Phi_{(n|m)}^{[0]}(z).
\end{equation}

With these results in hand, we may now proceed and address constraints on Stokes coefficients when in the presence of resonance. Along the positive real axis complex conjugation may be addressed very similarly to before, in which case (\ref{eq:App-comm-conjugation-pointedalien}) simply translates to the constraints
\begin{equation}
\re\, S_{\ell}^{(k)}=\re\,\widetilde{S}_{\ell}^{(k+\ell)}=0,\qquad\forall\, k\ge0,\ell\ge1.
\end{equation}

As we turn to the analysis of constraints arising from the negative real axis, things get a bit more intricate. Let us first determine the complex conjugate for $F(z,\sigma_{1},\sigma_{2})$, as depicted above, when $z=\rme^{-\rmi\pi}\left|z\right|$ (recall that $\beta_{nm}^{[0]}=\left(n+m\right)\beta$). One finds
\begin{eqnarray}
\mathcal{H}F(\widetilde{z}=\rme^{-\rmi\pi}\left|z\right|,\sigma_{1},\sigma_{2}) & = & \sum_{n=0}^{+\infty}\sum_{m=0}^{+\infty}\left(\overline{\sigma}_{1}\,\rme^{-2\pi\rmi\,\alpha\,\overline{\sigma}_{1}\overline{\sigma}_{2}}\,\rme^{-\alpha\,\overline{\sigma}_{1}\overline{\sigma}_{2}\,\log\widetilde{z}}\,\rme^{-2\pi\rmi\beta}\right)^{n}\times\nonumber \\
&& 
\times\left(\overline{\sigma}_{2}\,\rme^{2\pi\rmi\,\alpha\,\overline{\sigma}_{1}\overline{\sigma}_{2}}\,\rme^{\alpha\,\overline{\sigma}_{1}\overline{\sigma}_{2}\,\log\widetilde{z}}\,\rme^{-2\pi\rmi\beta}\right)^{m}\widetilde{z}^{-\beta_{nm}^{[0]}}\,\rme^{-\left(n-m\right)A\widetilde{z}}\,\Phi_{(n|m)}^{[0]}(\widetilde{z})=\nonumber \\
 & = & F(\widetilde{z},\widetilde{\sigma}_{1},\widetilde{\sigma}_{2}),
\end{eqnarray}
\noindent
where
\begin{eqnarray}
\widetilde{\sigma}_{1} & = & \overline{\sigma}_{1}\,\rme^{-2\pi\rmi\,\alpha\,\overline{\sigma}_{1}\overline{\sigma}_{2}}\,\rme^{-2\pi\rmi\beta},\\
\widetilde{\sigma}_{2} & = & \overline{\sigma}_{2}\,\rme^{2\pi\rmi\,\alpha\,\overline{\sigma}_{1}\overline{\sigma}_{2}}\,\rme^{-2\pi\rmi\beta}.
\end{eqnarray}
\noindent
The above identification between $\widetilde{\sigma}_{i}$ and $\overline{\sigma}_{i}$ was possible due to the fact that with $\beta=1/2$ the following property holds
\begin{equation}
\widetilde{\sigma}_{1}\widetilde{\sigma}_{2}=\overline{\sigma}_{1}\overline{\sigma}_{2}\,\rme^{-4\pi\rmi\beta}=\overline{\sigma}_{1}\overline{\sigma}_{2}.
\end{equation}%
\noindent
The inverse transformation is then given by
\begin{eqnarray}
\overline{\sigma}_{1} & = & \widetilde{\sigma}_{1}\,\rme^{2\pi\rmi\,\alpha\,\widetilde{\sigma}_{1}\widetilde{\sigma}_{2}}\,\rme^{2\pi\rmi\beta},\\
\overline{\sigma}_{2} & = & \widetilde{\sigma}_{2}\,\rme^{-2\pi\rmi\,\alpha\,\widetilde{\sigma}_{1}\widetilde{\sigma}_{2}}\,\rme^{2\pi\rmi\beta}.
\end{eqnarray}
\noindent
The constraints on the Stokes constants in the negative real axis are again obtained from enforcing the commutation relation \eqref{eq:App-comm-conjugation-pointedalien}. To do so, we need to determine $\dot{\Delta}_{-\ell A}\circ\mathcal{H}F(\widetilde{z},\sigma_{1},\sigma_{2})$ as well as $\mathcal{H}\circ\dot{\Delta}_{-\ell A}F(\widetilde{z},\sigma_{1},\sigma_{2})$. The former is now easily obtained. Using the complex conjugate for $F(z,\sigma_{1},\sigma_{2})$, together with the bridge equations \eqref{eq:Bridge-eqs-2-param}, we may write
\begin{equation}
\dot{\Delta}_{-\ell A}\circ\mathcal{H}F(\widetilde{z},\sigma_{1},\sigma_{2})=S_{-\ell}\left(\widetilde{\sigma}_{1},\widetilde{\sigma}_{2}\right)\frac{\partial F}{\partial\widetilde{\sigma}_{1}}\left(\widetilde{z},\widetilde{\sigma}_{1},\widetilde{\sigma}_{2}\right)+\widetilde{S}_{-\ell}\left(\widetilde{\sigma}_{1},\widetilde{\sigma}_{2}\right)\frac{\partial F}{\partial\widetilde{\sigma}_{2}}\left(\widetilde{z},\widetilde{\sigma}_{1},\widetilde{\sigma}_{2}\right).
\end{equation}
\noindent
The other term contributing to the commutation relation \eqref{eq:App-comm-conjugation-pointedalien}, $\mathcal{H}\circ\dot{\Delta}_{-\ell A}F(\widetilde{z},\sigma_{1},\sigma_{2})$, is also easily determined. In fact, we can write it as
\begin{equation}
\mathcal{H}\circ\dot{\Delta}_{-\ell A}F(\widetilde{z},\sigma_{1},\sigma_{2})=\overline{S}_{-\ell}\left(\overline{\sigma}_{1},\overline{\sigma}_{2}\right)\mathcal{H}\left[\frac{\partial F}{\partial\sigma_{1}}\right]+\overline{\widetilde{S}}_{-\ell}\left(\overline{\sigma}_{1},\overline{\sigma}_{2}\right)\mathcal{H}\left[\frac{\partial F}{\partial\sigma_{2}}\right].
\end{equation}
\noindent
In order to compare this to the previous term in the commutation relation, we need to use the following property
\begin{eqnarray}
\frac{\partial F(\widetilde{z},\widetilde{\sigma}_{1},\widetilde{\sigma}_{2})}{\partial\widetilde{\sigma}_{1}} & = & \rme^{2\pi\rmi\beta}\,\rme^{2\pi\rmi\,\alpha\,\widetilde{\sigma}_{1}\widetilde{\sigma}_{2}}\,\mathcal{H}\left[\frac{\partial F}{\partial\sigma_{1}}(\widetilde{z},\sigma_{1},\sigma_{2})\right]-\\
 &  & -2\pi\rmi\,\alpha\,\widetilde{\sigma}_{2}\,\sum_{n=0}^{+\infty}\sum_{m=0}^{+\infty}(m-n)\widetilde{\sigma}_{1}^{n}\widetilde{\sigma}_{2}^{m}{\rm e}^{\alpha(m-n)\widetilde{\sigma}_{1}\widetilde{\sigma}_{2}\log\widetilde{z}}\,{\rm e}^{-(n-m)Az}\,\widetilde{z}^{-\beta_{n,m}^{[0]}}\Phi_{(n|m)}^{[0]}(\widetilde{z}),\nonumber 
\end{eqnarray}
\noindent
alongside with the equivalent expression where one takes the derivative with respect to $\sigma_{2}$ instead. Finally, using the expansions for the Stokes coefficients \eqref{eq:Expansion-of-Sell-2-param} and \eqref{eq:Expansion-of-Stildeell-2-param}, the commutation relation \eqref{eq:App-comm-conjugation-pointedalien} applied to our transseries \textit{ansatz} yields
\begin{equation}
A+B=C,
\end{equation}
\noindent
where
\begin{eqnarray}
A & = & \sum_{n,m=0}^{+\infty}\sigma_{1}^{n+\ell-1}\sigma_{2}^{m-1}\left(n+\alpha\,\sigma_{1}\sigma_{2}\left(m-n\right)\log\widetilde{z}\right)\rme^{\alpha\left(m-n\right)\sigma_{1}\sigma_{2}\,\log\widetilde{z}}\, F^{(n|m)}\times\\
 &  & \qquad\qquad\qquad\qquad\qquad\qquad\qquad\qquad\qquad\qquad\times\,\sum_{k=1}^{+\infty}\left(\sigma_{1}\sigma_{2}\right)^{k}\, R_{-\ell}^{(k-1)}\left(\sigma_{1}\sigma_{2}\right),\nonumber \\
B & = & \sum_{n,m=0}^{+\infty}\sigma_{1}^{n+\ell-1}\sigma_{2}^{m-1}\left(m+\alpha\,\sigma_{1}\sigma_{2}\left(m-n\right)\log\widetilde{z}\right)\rme^{\alpha\left(m-n\right)\sigma_{1}\sigma_{2}\,\log\widetilde{z}}\, F^{(n|m)}\times\\
 &  & \qquad\qquad\qquad\qquad\qquad\qquad\qquad\qquad\qquad\qquad\times\,\sum_{k=0}^{+\infty}\left(\sigma_{1}\sigma_{2}\right)^{k}\,\widetilde{R}_{-\ell}^{(k+\ell-1)}\left(\sigma_{1}\sigma_{2}\right),\nonumber \\
C & = & -2\pi\rmi\,\alpha\sum_{n,m=0}^{+\infty}\sigma_{1}^{n+\ell-1}\sigma_{2}^{m-1}\left(m-n\right)\rme^{\alpha\left(m-n\right)\sigma_{1}\sigma_{2}\,\log\widetilde{z}}\,\rme^{-2\pi\rmi\ell\left(\beta-\alpha\,\sigma_{1}\sigma_{2}\right)}\, F^{(n|m)}\times\\
 &  & \qquad\qquad\qquad\qquad\qquad\qquad\qquad\qquad\qquad\qquad\times\,\sum_{k=1}^{+\infty}\left(\sigma_{1}\sigma_{2}\right)^{k}\,\left(\sigma_{1}\sigma_{2}\,\overline{S_{-\ell}^{(k+\ell)}}+\overline{\widetilde{S}_{-\ell}^{(k-1)}}\right).\nonumber 
\end{eqnarray}
\noindent
Above, we have used the notation
\begin{eqnarray}
R_{-\ell}^{(k)}(x) & = & S_{-\ell}^{(k+1+\ell)}+\overline{S_{-\ell}^{(k+1+\ell)}}\,\rme^{2\pi\rmi\ell\left(\beta+\alpha x\right)},\\
\widetilde{R}_{-\ell}^{(k)}(x) & = & \widetilde{S}_{-\ell}^{(k+1+\ell)}+\overline{\widetilde{S}_{-\ell}^{(k+1+\ell)}}\,\rme^{-2\pi\rmi\ell\left(\beta-\alpha x\right)}.
\end{eqnarray}

One may now apply the usual reasoning, \textit{i.e.}, compare equal powers of $\log\widetilde{z}$, of $x=\sigma_{1}\sigma_{2}$, and take into account the many different sectors $(n|m)$. The constraints one finally obtains on the Stokes constants are: %
\begin{eqnarray}
S_{-\ell}^{(\ell+1)}+\rme^{-2\pi\rmi\ell\beta}\,\overline{S_{-\ell}^{(\ell+1)}} & = & 2\pi\rmi\,\alpha\,\rme^{2\pi\rmi\ell\beta}\,\overline{\widetilde{S}_{-\ell}^{(0)}},\label{eq:App-constraints-with-logs-S-k=00003D00003D1}\\
S_{-\ell}^{(k+\ell)}+\rme^{-2\pi\rmi\ell\beta}\,\sum_{r=0}^{k-1}\frac{\left(2\pi\rmi\ell\alpha\right)^{r}}{r!}\,\overline{S_{-\ell}^{(k-r+\ell)}} & = & 2\pi\rmi\,\alpha\,\rme^{2\pi\rmi\ell\beta}\left\{ \frac{\left(2\pi\rmi\ell\alpha\right)^{k-1}}{(k-1)!}\,\overline{\widetilde{S}_{-\ell}^{(0)}}+\right.\label{eq:App-constraints-with-logs-S-kge2}\\
 &  & \qquad\quad\left.+\sum_{r=0}^{k-2}\frac{\left(2\pi\rmi\ell\alpha\right)^{r}}{r!}\left(\overline{S_{-\ell}^{(k-r+\ell-1)}}+\overline{\widetilde{S}_{-\ell}^{(k-r-1)}}\right)\right\} ,\nonumber \\
\widetilde{S}_{-\ell}^{(0)}+\rme^{2\pi\rmi\ell\beta}\,\overline{\widetilde{S}_{-\ell}^{(0)}} & = & 0,\label{eq:App-constraints-with-logs-Stilde-k=00003D00003D0}\\
\widetilde{S}_{-\ell}^{(1)}+\rme^{2\pi\rmi\ell\beta}\,\overline{\widetilde{S}_{-\ell}^{(1)}} & = & -2\pi\rmi\left(\ell+1\right)\alpha\,\rme^{2\pi\rmi\ell\beta}\,\overline{\widetilde{S}_{-\ell}^{(0)}},\label{eq:App-constraints-with-logs-Stilde-k=00003D00003D1}\\
\widetilde{S}_{-\ell}^{(k)}+\rme^{2\pi\rmi\ell\beta}\,\sum_{r=0}^{k}\frac{\left(2\pi\rmi\ell\alpha\right)^{r}}{r!}\,\overline{\widetilde{S}_{-\ell}^{(k-r)}} & = & -2\pi\rmi\,\alpha\,\rme^{2\pi\rmi\ell\beta}\left\{ \sum_{r=0}^{k-2}\frac{\left(2\pi\rmi\ell\alpha\right)^{r}}{r!}\,\overline{S_{-\ell}^{(k-r+\ell-1)}}+\right.\label{eq:App-constraints-with-logs-Stilde-kge2}\\
 &  & \qquad\quad\left.+\sum_{r=0}^{k-1}\frac{\left(2\pi\rmi\ell\alpha\right)^{r}}{r!}\,\overline{\widetilde{S}_{-\ell}^{(k-r-1)}}\right\} ,\nonumber 
\end{eqnarray}%
\noindent
where $\ell\ge1$, $k\ge2$ and $\beta=1/2$.

One interesting aspect of all these structural constraints is that, much like all the previous ones, they may be \textit{tested} in examples. Given how intricate the above relations are, these tests are actually rather non--trivial, in particular supporting the generality of \eqref{eq:App-comm-conjugation-pointedalien}. In the cases of the Painlev\'e I and II equations addressed in \citep{Aniceto:2011nu, Schiappa:2013opa} many relations between Stokes constants were written down; some arising from the study of the string genus expansion, others found ``experimentally'' via numerical work. In particular, the following relations were obtained in the aforementioned references: 
\begin{eqnarray}
S_{1}^{(0)}+(-1)^{-1/2}\,\widetilde{S}_{-1}^{(0)} & = & 0,\\
S_{2}^{(0)}-\widetilde{S}_{2}^{(0)} & = & 0,\\
S_{1}^{(1)}+(-1)^{1/2}\,\widetilde{S}_{-1}^{(1)}-2\pi\rmi\,\alpha\, S_{1}^{(0)} & = & 0,\\
\widetilde{S}_{1}^{(2)}+(-1)^{1/2}\, S_{-1}^{(2)}+\rmi\pi\,\alpha\, S_{1}^{(0)} & = & 0.
\end{eqnarray}
\noindent
One can derive one further relation, arising within the $\Phi_{(2|2)}$ sector, by the requirement that this sector has a genus expansion in the string coupling, similarly to what was done in \citep{Aniceto:2011nu}. This extra relation is
\begin{equation}
\rmi\, S_{1}^{(2)}+\widetilde{S}_{-1}^{(2)}-\rmi\pi\,\alpha\left\{ \rmi\,\widetilde{S}_{1}^{(2)}+2\rmi\, S_{1}^{(1)}+\frac{3\pi}{2}\,\alpha\, S_{1}^{(0)}\right\} =0.
\end{equation}
\noindent
Recalling that $S_{\ell}^{(k)},\widetilde{S}_{\ell}^{(k+\ell)}\in\rmi\mathbb{R}$, one can easily check that the above relations obey (\ref{eq:App-constraints-with-logs-S-k=00003D00003D1}) with $\ell=1$, (\ref{eq:App-constraints-with-logs-Stilde-k=00003D00003D0}) and (\ref{eq:App-constraints-with-logs-Stilde-k=00003D00003D1}) with $\ell=1$, and (\ref{eq:App-constraints-with-logs-Stilde-kge2}) with $k=2$, $\ell=1$. In this way, it is very interesting to finally realize that all the somewhat empirical relations in \citep{Aniceto:2011nu, Schiappa:2013opa} are in fact part of rather general structural constraints on resurgent transseries.

%%%%%%%%%%%%%%%%%%%%%%%%%%%%%%%%%%%%%%%%%%%%%%%%%%%%%%%%%%%%%%%%%
%%%%%%%%%%%%%%%%%%%%%%%%%%%%%%%%%%%%%%%%%%%%%%%%%%%%%%%%%%%%%%%%%
\section{Formulae for One--Parameter Transseries\label{sec:App-Formulae-One-Param-transs}}
%%%%%%%%%%%%%%%%%%%%%%%%%%%%%%%%%%%%%%%%%%%%%%%%%%%%%%%%%%%%%%%%%
%%%%%%%%%%%%%%%%%%%%%%%%%%%%%%%%%%%%%%%%%%%%%%%%%%%%%%%%%%%%%%%%%

In this appendix we focus on the case of a one--parameter transseries
\begin{equation}
F \left(z,\sigma\right) = \sum_{n=0}^{+\infty} \sigma^{n}\, \rme^{-nAz}\, \Phi_{n}\left(z\right),
\label{eq:appA-1-param-transs}
\end{equation}
\noindent
where there are two Stokes lines, at $\theta=0$ and $\theta=\pi$ \citep{Aniceto:2011nu}. We will study the Stokes automorphism and Stokes transitions in this setting, alongside with a discussion of how these results interplay with the cancelation of the nonperturbative ambiguity. In particular, we shall present rather general formulae and address the technicalities/results used in the main body of the text. To find the Stokes transitions one needs information on how each asymptotic series $\Phi_{n}$ crosses the Stokes line, \textit{i.e.}, how the Stokes automorphism acts on each $\Phi_{n}$. Some results along these lines were already obtained in \citep{Aniceto:2011nu}; in here we recall some of these relevant results as well as their respective generalizations, needed in the main text.

%%%%%%%%%%%%%%%%%%%%%%%%%%%%%%%%%%%%%%%%%%%%%%%%%%%%%%%%%%%%%%%%%
\subsection*{Stokes Automorphism at $\theta=0$}
%%%%%%%%%%%%%%%%%%%%%%%%%%%%%%%%%%%%%%%%%%%%%%%%%%%%%%%%%%%%%%%%%

The Stokes automorphism acting on each sector $\Phi_{n}$ may be completely determined up to the Stokes constants. A general one--parameter transseries has an infinite number of non--vanishing Stokes constants, $S_{1}$, $S_{-k}$ with $k\ge1$. At the Stokes line $\theta=0$ the Stokes automorphism acts as \citep{Aniceto:2011nu}
\begin{equation}
\underline{\mathfrak{S}}_{0} \Phi_{n} = \sum_{\ell=0}^{+\infty} \binom{n+\ell}{n}\, S_{1}^{\ell}\, \rme^{-\ell Az}\, \Phi_{n+\ell},
\end{equation}
\noindent
We shall be interested in determining a general power of the Stokes automorphism, $\underline{\mathfrak{S}}_{0}^{\nu}\Phi_{n}$. To do so, recall the definition of the Stokes automorphism in terms of alien derivatives \eqref{eq:Stokes-aut-from-alien-derivatives}. Now, in the $\theta=0$ direction of the complex Borel plane we have only one singularity, $\omega=A$, with $A$ being the instanton action. In this case, the Stokes automorphism becomes
\begin{equation}
\underline{\mathfrak{S}}_{0}^{\nu} = \exp \left\{ \nu\, \rme^{-Az} \Delta_{A} \right\},
\end{equation}
\noindent
and the bridge equations are
\begin{equation}
\Delta_{A}\Phi_{n} = S_{1} \left(n+1\right) \Phi_{n+1}.
\end{equation}
\noindent
Multiple derivatives are immediate to obtain,
\begin{equation}
\Delta_{A}^{k} \Phi_{n} = S_{1}^{k} \prod_{j=1}^{k} \left(n+j\right) \Phi_{n+k},
\end{equation}
\noindent
from where we can easily find $\underline{\mathfrak{S}}_{0}^{\nu}\Phi_{n}$ as
\begin{equation}
\underline{\mathfrak{S}}_{0}^{\nu} \Phi_{n} = \sum_{\ell=0}^{+\infty} \frac{\nu^{\ell}}{\ell!}\, \rme^{-\ell Az}\, \Delta_{A}^{\ell}\Phi_{n} = \sum_{\ell=0}^{+\infty} \binom{n+\ell}{n} \left(\nu S_{1}\right)^{\ell} \rme^{-\ell Az}\, \Phi_{n+\ell}.
\label{eq:App-1-param-stokes0-acting-phi-n}
\end{equation}
\noindent
Taking a general power $\nu$ of the Stokes automorphism exactly corresponds to multiplying the Stokes constant by this same number. This result is used in the main text.

%%%%%%%%%%%%%%%%%%%%%%%%%%%%%%%%%%%%%%%%%%%%%%%%%%%%%%%%%%%%%%%%%
\subsection*{Stokes Transition at $\theta=0$}
%%%%%%%%%%%%%%%%%%%%%%%%%%%%%%%%%%%%%%%%%%%%%%%%%%%%%%%%%%%%%%%%%

The Stokes transition of the one--parameter transseries (\ref{eq:appA-1-param-transs}), at $\theta=0$, is now very simple to compute
\begin{eqnarray}
\underline{\mathfrak{S}}_{0}^{\nu} F \left(z,\sigma\right) &=& \sum_{n=0}^{+\infty} \sigma^{n}\, \rme^{-nAz}\, \underline{\mathfrak{S}}_{0}^{\nu} \Phi_{n} = \sum_{n=0}^{+\infty} \sum_{\ell=0}^{+\infty} \binom{n+\ell}{n}\, \sigma^{n} \left(\nu S_{1}\right)^{\ell} \rme^{-(n+\ell)Az}\, \Phi_{n+\ell} = \nonumber \\
&=& \sum_{\ell=0}^{+\infty} \left(\sigma+\nu S_{1}\right)^{\ell} \rme^{-\ell Az}\, \Phi_{\ell} = F \left(z,\sigma+\nu S_{1}\right).
\end{eqnarray}
\noindent
This transition describes Stokes phenomenon across the singular line $\theta=0$, via a jump in the transseries parameter precisely given by the Stokes constant $S_{1}$. In the result above, we have already considered a general power of the Stokes automorphism; in the usual case one sets $\nu=1$.

%%%%%%%%%%%%%%%%%%%%%%%%%%%%%%%%%%%%%%%%%%%%%%%%%%%%%%%%%%%%%%%%%
\subsection*{Stokes Automorphism at $\theta=\pi$}
%%%%%%%%%%%%%%%%%%%%%%%%%%%%%%%%%%%%%%%%%%%%%%%%%%%%%%%%%%%%%%%%%

As we turn to $\theta=\pi$, the procedure will not be as straightforward. First, along $\theta=\pi$, the bridge equations take the form
\begin{equation}
\Delta_{-\ell A} \Phi_{n} = S_{-\ell} \left(n-\ell\right) \Phi_{n-\ell}, \qquad \ell\ge 1.
\label{eq:AppA-bridge-eqs-pi}
\end{equation}
\noindent
Recall that we always define $\Phi_{m}=0$ for $m<0$. These equations lead to the following expression for the Stokes automorphism
\begin{equation}
\underline{\mathfrak{S}}_{\pi} \Phi_{n} = \sum_{\ell=0}^{n} \underline{\Sigma} \left(n,\ell\right) \rme^{\ell Az}\, \Phi_{n-\ell},
\end{equation}
\noindent
where
\begin{equation}
\underline{\Sigma} \left(n,\ell\right) = \sum_{k=1}^{\ell} \frac{1}{k!} \sum_{\gamma\in\Gamma(k,\ell)} \left( \prod_{j=1}^{k} \left(n-\gamma_{j}\right) S_{-\mathbf{d}\gamma_{j}} \right) + \delta_{\ell,0}.
\end{equation}
\noindent
A proof of this result may be found in \citep{Aniceto:2011nu}. In the expression above, the sum over $\gamma\in\Gamma(k,\ell)$ is a sum over partitions $0=\gamma_{0}\le\gamma_{1}\le\cdots\le\gamma_{k}=\ell$, and we have further defined $\mathbf{d}\gamma_{j}=\gamma_{j}-\gamma_{j-1}$. In order to have a correct expression, one further needs to set $S_{0}=0$.

Once again, we are interested in generalizing the above result to a general power of the Stokes automorphism. Along this Stokes line one finds singularities at $\omega = - \ell A$, for $\ell\ge1$, and so the expression to consider is
\begin{equation}
\underline{\mathfrak{S}}_{\pi}^{\nu} = \exp \left\{ \nu\, \sum_{\ell=1}^{+\infty} \rme^{\ell Az} \Delta_{-\ell A} \right\}.
\label{eq:AppA-Stokes-Aut-pi-def}
\end{equation}
\noindent
Expanding the exponential and making use of the bridge equations, is is easy to see that also in this case the power $\nu$ translates into a multiplicative factor of $\nu$ for each and every Stokes constant\footnote{In order to see this, it is enough to realize that in the expansion of $\underline{\mathfrak{S}}_{\pi}^{\nu}$ we will find a factor of $\nu$ for each alien derivative. On the other hand, from the bridge equations \eqref{eq:AppA-bridge-eqs-pi}, each alien derivative essentially yields some Stokes constant. Thus, each Stokes constant appearing in the expansion of $\underline{\mathfrak{S}}_{\pi}^{\nu}$ has a factor of $\nu$.}. One finally obtains
\begin{equation}
\underline{\mathfrak{S}}_{\pi}^{\nu} \Phi_{n} = \sum_{\ell=0}^{n} \underline{\Sigma}_{\nu} \left(n,\ell\right) \rme^{\ell Az}\, \Phi_{n-\ell},
\end{equation}
\noindent
where
\begin{equation}
\underline{\Sigma}_{\nu} \left(n,\ell\right) = \sum_{k=1}^{\ell} \frac{\nu^{k}}{k!} \sum_{\gamma\in\Gamma(k,\ell)} \left( \prod_{j=1}^{k} \left(n-\gamma_{j}\right) S_{-\mathbf{d}\gamma_{j}}\right) + \delta_{\ell,0}.
\label{eq:AppA-coefficients-automorphism-at-pi-nu}
\end{equation}

%%%%%%%%%%%%%%%%%%%%%%%%%%%%%%%%%%%%%%%%%%%%%%%%%%%%%%%%%%%%%%%%%
\subsection*{Stokes Transition at $\theta=\pi$}
%%%%%%%%%%%%%%%%%%%%%%%%%%%%%%%%%%%%%%%%%%%%%%%%%%%%%%%%%%%%%%%%%

As compared to the case of $\theta=0$, it is already much harder to find an expression for the Stokes transition at $\theta=\pi$. Following the same reasoning as before, we find
\begin{eqnarray}
\underline{\mathfrak{S}}_{\pi}F(z,\sigma) &=& \sum_{n=0}^{+\infty} \sigma^{n}\, \rme^{-nAz}\, \underline{\mathfrak{S}}_{\pi} \Phi_{n} = \sum_{n=0}^{+\infty} \sigma^{n} \sum_{\ell=0}^{n} \underline{\Sigma} \left(n,n-\ell\right) \rme^{-\ell Az}\, \Phi_{\ell} = \nonumber \\
&=& \sum_{\ell=0}^{+\infty} \left( \sum_{n=0}^{+\infty} \sigma^{n+\ell}\, \underline{\Sigma} \left(n+\ell,n\right) \right) \rme^{-\ell Az}\, \Phi_{\ell}.
\label{eq:AppA-Stokes-Aut-for-transs}
\end{eqnarray}
\noindent
In parallel with what we did for the case of $\theta=0$, we would now like to find a function $\BS_\pi (\sigma)$ such that
\begin{equation}
\left( \BS_\pi (\sigma) \right)^{\ell} = \sum_{n=0}^{+\infty} \sigma^{n+\ell}\, \underline{\Sigma} \left(n+\ell,n\right).
\end{equation}
\noindent
A candidate for this function is 
\begin{equation}
\BS_\pi (\sigma) = \sum_{n=0}^{+\infty} \sigma^{n+1}\, \underline{\Sigma} \left(n+1,n\right).
\end{equation}
\noindent
We have checked this thoroughly from expanding (\ref{eq:AppA-Stokes-Aut-for-transs}) as follows:
\begin{eqnarray}
\underline{\mathfrak{S}}_{\pi} F &=& \Phi_{0} + \\
&&
\hspace{-40pt}
+ \left( 1 + \sigma S_{-1} + \sigma^{2} \left(S_{-2}+S_{-1}^{2}\right) + \sigma^{3} \left(S_{-3}+\frac{5}{2}S_{-1}S_{-2}+S_{-1}^{3} \right) + \cdots \right) \sigma\, \rme^{-Az}\, \Phi_{1} + \nonumber \\
&&
\hspace{-40pt}
+ \left( 1 + 2 \sigma S_{-1} + 2 \sigma^{2} \left(S_{-2}+\frac{3}{2}S_{-1}^{2}\right) + 2 \sigma^{3} \left(S_{-3}+\frac{7}{2}S_{-1}S_{-2}+2S_{-1}^{3} \right) + \cdots \right) \sigma^{2}\, \rme^{-2Az}\, \Phi_{2} + \nonumber \\
&&
\hspace{-40pt}
+ \left( 1 + 3 \sigma S_{-1} + 3 \sigma^{2} \left(S_{-2}+2S_{-1}^{2}\right) + 3 \sigma^{3} \left(S_{-3}+\frac{7}{2}S_{-1}S_{-2}+\frac{10}{3}S_{-1}^{3}\right) + \cdots \right) \sigma^{3}\, \rme^{-3Az}\, \Phi_{3} + \cdots. \nonumber 
\end{eqnarray}
\noindent
It is very simple to see that identifying the function $\BS_\pi (\sigma)$ with the appropriate terms on the second line above, then the third line follows by determining $\left( \BS_\pi (\sigma) \right)^2$, the fourth line by determining $\left( \BS_\pi (\sigma) \right)^3$, and so on. We have verified this structure computationally to higher orders; however a more exhaustive proof of this result is still under way. Nonetheless, based on strong symbolic computation evidence, the Stokes transition at $\theta=\pi$ can be written as
\begin{equation}
\underline{\mathfrak{S}}_{\pi} F \left(z,\sigma\right) = F \left(z,\BS_\pi (\sigma)\right).
\end{equation}

One can also generalize this result for an arbitrary power of the Stokes automorphism,
\begin{equation}
\underline{\mathfrak{S}}_{\pi}^{\nu} F(z,\sigma) = \sum_{\ell=0}^{+\infty} \left( \sum_{n=0}^{+\infty} \sigma^{n+\ell}\, \underline{\Sigma}_{\nu} \left(n+\ell,n\right) \right) \rme^{-\ell Az}\, \Phi_{\ell} = F \left(z,\BS_\pi^{(\nu)} (\sigma)\right),
\end{equation}
\noindent
where now
\begin{equation}
\BS_\pi^{(\nu)} (\sigma)=\sum_{n=0}^{+\infty} \sigma^{n+1}\, \underline{\Sigma}_{\nu} \left(n+1,n\right).
\end{equation}
\noindent
For $\nu=1$, this describes the Stokes transition at $\theta=\pi$ as a ``jump'' in the transseries parameter $\sigma \rightarrow \BS_\pi (\sigma)$. All of the Stokes constants $S_{-k}$, $k\ge1$ contribute to this transition.

%%%%%%%%%%%%%%%%%%%%%%%%%%%%%%%%%%%%%%%%%%%%%%%%%%%%%%%%%%%%%%%%%
\subsection*{Cancelation of the Nonperturbative Ambiguity}
%%%%%%%%%%%%%%%%%%%%%%%%%%%%%%%%%%%%%%%%%%%%%%%%%%%%%%%%%%%%%%%%%

Having understood the structure of transitions at the Stokes lines, $\theta=0$ and $\theta=\pi$, at the root of the ambiguity, we may now try to understand how is the ambiguity canceled. We discuss this point in the main text, but in here we still need to present the complete expressions associated to the nonperturbative ambiguity of the one--parameter transseries \eqref{eq:appA-1-param-transs}. In particular, along $\theta=0$, it is convenient to first write all formulae in terms of the real and imaginary parts of the multi--instanton sectors $F^{(n)} (z)$. In fact, along $\theta=0$  what we want to cancel is simply
\begin{eqnarray}
\im\, F &\equiv& \frac{1}{2\rmi} \left(\mathcal{S}_{0^+}-\mathcal{S}_{0^-}\right) F = \sum_{n=0}^{+\infty}\left(\im\, \sigma^{n}\, \re\, F^{(n)} + \re\, \sigma^{n}\, \im\, F^{(n)} \right),
\label{eq-app-imFgeneral}
\end{eqnarray}
\noindent
where we used
\begin{equation}
\im\, F^{(n)} := \frac{1}{2\rmi} \left(\mathcal{S}_{0^+}-\mathcal{S}_{0^-}\right) F^{(n)}, \qquad \re\, F^{(n)} := \frac{1}{2} \left(\mathcal{S}_{0^+}+\mathcal{S}_{0^-}\right) F^{(n)}.
\end{equation}
\noindent
Now one may very naturally relate $\im\, F^{(n)}$ with $\re\, F^{(n')}$ by making use of the Stokes automorphism in its most fundamental form \eqref{eq:stokesauto}. One simply has\footnote{Notice that this expression makes clear how ambiguities are associated to a non--trivial Stokes automorphism.}
\begin{equation}
\left(\mathcal{S}_{0^+}-\mathcal{S}_{0^-}\right) F^{(n)}(z) = - \mathcal{S}_{0^-} \circ \left(\mathbf{1}-\underline{\mathfrak{S}}_{0}\right) F^{(n)}(z),
\end{equation}
\noindent
where $\underline{\mathfrak{S}}_{0} F^{(n)}(z) = \rme^{-nAz}\, \underline{\mathfrak{S}}_{0} \Phi_{n}(z)$ was already computed in (\ref{eq:App-1-param-stokes0-acting-phi-n}). Explicitly, then,
\begin{equation}
\left( \mathcal{S}_{0^+}-\mathcal{S}_{0^-} \right) F^{(n)}(z) = \sum_{k=1}^{+\infty} \binom{n+k}{n}\, S_{1}^{k}\, \mathcal{S}_{0^-} F^{(n+k)}(z).
\end{equation}
\noindent
Using the relation
\begin{equation}
\mathcal{S}_{0^-} = \re - \frac{1}{2} \left(\mathcal{S}_{0^+}-\mathcal{S}_{0^-}\right)
\end{equation}
\noindent
in a recursive fashion, inside the previous expression, we can finally write the ambiguity of $F^{(n)}$ as an expansion of higher multi--instanton (real) contributions:
\begin{equation}
\im\, F^{(n)}(z) = \frac{1}{2\rmi}\, \sum_{k=1}^{+\infty} \binom{n+k}{n}\, \Omega(k)\, S_{1}^{k}\, \re\, F^{(n+k)},
\label{eq:App-1-param-im-part-F(n)}
\end{equation}
\noindent
with 
\begin{equation}
\Omega (k) = \sum_{r=1}^{k} \sum_{s=1}^{r} \binom{r}{s} (-1)^{s+1}\, \frac{s^{k}}{2^{r-1}}.
\end{equation}
\noindent
In particular note that $\Omega(2k)=0$, giving rise to an odd/even pattern in the structure above. Having this result in hand one may finally find a general expression for the imaginary part of any one--parameter transseries $F(z,\sigma)$ as
\begin{eqnarray}
\label{eq:app1imFfullformula}
\im\, F(z,\sigma) &=& \left( \frac{1}{2\rmi}\, S_{1} + \sigma_{\text{I}} \right) \re\, F^{(1)} + \\
&+&
\frac{1}{2\rmi} \sum_{n=2}^{+\infty} \left( \Omega(n)\, S_{1}^{n} + 2\rmi \sum_{r=0}^{\left[(n-1)/2\right]} \binom{n}{2r+1}\, (-1)^{r} \sigma_{\text{R}}^{n-(2r+1)} \sigma_{\text{I}}^{2r+1} \right. + \nonumber \\
&+&
\left.
\sum_{k=1}^{n-1} \binom{n}{k}\, \Omega(n-k)\, S_{1}^{n-k}\, \sum_{r=0}^{\left[k/2\right]} \binom{k}{2r} (-1)^{r} \sigma_{\text{R}}^{k-2r} \sigma_{\text{I}}^{2r} \right) \re\, F^{(n)}. \nonumber
\end{eqnarray}
\noindent
In this expression we use the following definitions: $\sigma = \sigma_{\text{R}} + \rmi\, \sigma_{\text{I}}$ with $\sigma_{\text{R}},\sigma_{\text{I}} \in \mathbb{R}$, and the usual notation of $\left[\bullet\right]$ for the integer part. The cancelation of the ambiguity, which in this case is the cancelation of this imaginary part, follows by solving $\im\, F(z,\sigma) = 0$. It can be checked that the condition already arising at first order, $\frac{1}{2\rmi}\, S_{1} + \sigma_{\text{I}} = 0$, actually solves this equation to \textit{all} orders, and we have done this using symbolic computation to very high order.

Similarly, one may obtain the real part of any one--parameter transseries $F(z,\sigma)$. Using the definition
\begin{equation}
\re\, F \equiv \frac{1}{2} \left(\mathcal{S}_{0^+}+\mathcal{S}_{0^-}\right) F = \sum_{n=0}^{+\infty} \left( \re\, \sigma^{n}\, \re\, F^{(n)} - \im\, \sigma^{n}\,\im\, F^{(n)} \right),
\end{equation}
\noindent
it follows the general result
\begin{eqnarray}
\re\, F(z,\sigma) &=& \re\, F^{(0)} + \sigma_{\text{R}}\, \re\, F^{(1)} + \sum_{n=2}^{+\infty} \left(\, \sum_{r=0}^{\left[n/2\right]} \binom{n}{2r}\, (-1)^{r} \sigma_{\text{R}}^{n-2r} \sigma_{\text{I}}^{2r} - \right. \\
&&
\left.
- \frac{1}{2\rmi}\, \sum_{k=1}^{n-1} \binom{n}{k}\, \Omega(n-k)\, S_{1}^{n-k} \sum_{r=0}^{\left[(k-1)/2\right]} \binom{k}{2r+1}\, (-1)^{r} \sigma_{\text{R}}^{k-2r-1} \sigma_{\text{I}}^{2r+1} \right) \re\, F^{(n)}. \nonumber 
\end{eqnarray}
\noindent
In particular, one may now impose the constraint arising from the cancelation of the ambiguity, namely $\sigma_{\text{I}} = \frac{\rmi}{2}\, S_{1}$. In this case, one may show that even further setting $\sigma_{\text{R}} = 0$ one will always find multi--instanton contributions to \textit{all} even orders in the final answer:
\begin{equation}
\re\, F(z,\sigma) = \re\, F^{(0)} + \sum_{n=1}^{+\infty} \left( \frac{1}{2^{2n}} - \sum_{k=0}^{n-1} \binom{2n}{2k+1}\, \frac{1}{2^{2(k+1)}}\, \Omega \left( 2(n-k)-1 \right) \right) S_{1}^{2n}\, \re\, F^{(2n)}. 
\end{equation}
\noindent
This is exactly the multi--instanton expansion for the median resummation. The discussion of these expressions is done in the main body of the paper.

%%%%%%%%%%%%%%%%%%%%%%%%%%%%%%%%%%%%%%%%%%%%%%%%%%%%%%%%%%%%%%%%%
%%%%%%%%%%%%%%%%%%%%%%%%%%%%%%%%%%%%%%%%%%%%%%%%%%%%%%%%%%%%%%%%%
\section{Formulae for Two--Parameter Transseries\label{sec:App-Formulae-for-Two--Parameters}}
%%%%%%%%%%%%%%%%%%%%%%%%%%%%%%%%%%%%%%%%%%%%%%%%%%%%%%%%%%%%%%%%%
%%%%%%%%%%%%%%%%%%%%%%%%%%%%%%%%%%%%%%%%%%%%%%%%%%%%%%%%%%%%%%%%%

This appendix generalizes results in the previous one to the case of a two--parameters transseries, of the form
\begin{equation}
F (z,\sigma_{1},\sigma_{2}) = \sum_{n,m=0}^{+\infty} \sigma_{1}^{n}\sigma_{2}^{m}\, \rme^{- \left(n-m\right) Az}\, \Phi_{(n|m)}(z),
\label{eq:app2parametersFtransseries}
\end{equation}
\noindent
including one ``physical'' and one ``generalized'' instanton sector. Note that from the point--of--view of the Stokes automorphism the actual asymptotic expansion of $\Phi_{(n|m)}(z)$ is not important; with or without logarithmic sectors the results that follow are unchanged. We have two Stokes lines, at $\theta=0$ and $\theta=\pi$, which will be rather similar due to the nature of the instanton actions being $\pm A$. On what concerns the Stokes automorphism, some preliminary results can already be found in \citep{Aniceto:2011nu} for the simplest cases. However, in order to determine the Stokes transitions for the full two--parameters transseries, one needs to know how the Stokes automorphism acts on a general sector $\Phi_{(n|m)}$. Happily, it turns out this can be done in much the same way as for the cases $m=0,1$ worked out in \citep{Aniceto:2011nu}. Before presenting the results, recall that the bridge equations for this case are given by
\begin{eqnarray}
\Delta_{\ell A} \Phi_{(n|m)} &=& \sum_{k=\max(0,\ell-1)}^{\min(m,n+\ell-1)} \left(n-k+\ell\right) S_{\ell}^{(k-\ell+1)}\, \Phi_{(n-k+\ell|m-k)} + \nonumber \\
&&
+ \sum_{k=\max(-\ell-1,0)}^{\min(m-\ell,n)} \left(m-k-\ell\right) \widetilde{S}_{\ell}^{(k+\ell+1)}\, \Phi_{(n-k|m-k-\ell)},
\end{eqnarray}
\noindent
valid for all $\ell\ne0$ (both singular directions $\theta=0$ and $\theta=\pi$ are contemplated). We now have two sets of Stokes constants, $S$ and $\widetilde{S}$.

%%%%%%%%%%%%%%%%%%%%%%%%%%%%%%%%%%%%%%%%%%%%%%%%%%%%%%%%%%%%%%%%%
\subsection*{Stokes Automorphism at $\theta=0$}
%%%%%%%%%%%%%%%%%%%%%%%%%%%%%%%%%%%%%%%%%%%%%%%%%%%%%%%%%%%%%%%%%

As compared to the previous appendix addressing one--parameter transseries, one now needs to consider all singularities $\omega=\ell A$ with $\ell\ge1$ already for the Stokes automorphism associated to the $\theta=0$ Stokes line. Directly computing an arbitrary power of this automorphism one has
\begin{eqnarray}
\underline{\mathfrak{S}}_{0}^{\nu} \Phi_{(n|m)} &=& \exp \left\{ \nu\, \sum_{\ell=1}^{+\infty} \rme^{-\ell Az} \Delta_{\ell A} \right\} \Phi_{(m|n)} = \sum_{k=0}^{+\infty} \frac{\nu^{k}}{k!} \left( \sum_{\ell=1}^{+\infty} \rme^{-\ell Az} \Delta_{\ell A} \right)^{k} \Phi_{(n|m)} = \nonumber \\
&=&
\left\{ 1 + \sum_{r=1}^{+\infty} \rme^{-rAz} \sum_{k=1}^{r} \frac{\nu^{k}}{k!} \sum_{\ell_{1}+\cdots+\ell_{k}=r} \left( \prod_{i=1}^{k}\Delta_{\ell_{i}A} \right) \right\} \Phi_{(n|m)},
\end{eqnarray}
\noindent
where the last sum is over $\ell_{i}$, $i=1,...,k$, positive integers. The difficulty, as always, rests in computing multiple alien derivatives; in particular we need to compute $\prod_{i=1}^{N} \Delta_{\ell_{i}A} \Phi_{(n|m)}$. For $N=1$ this is just given by the bridge equations which may now be conveniently rewritten as
\begin{equation}
\Delta_{\ell_{1}A} \Phi_{(n|m)} = \sum_{k=0}^{n+1} \left( k\, S_{\ell_{1}}^{(n-k+1)} + \left( m-n-\ell_{1}+k \right) \widetilde{S}_{\ell_{1}}^{(n-k+\ell_{1}+1)} \right) \Phi_{(k|m-n+k-\ell_{1})}.
\label{eq:AppB-one-alien-der-theta-zero}
\end{equation}
\noindent
Comparing both expressions above, one quickly realizes that once again the power $\nu$ of the Stokes automorphism gets simply translated into a multiplicative factor associated to each and every Stokes constant. The next step is to write a general expression for $\prod_{i=1}^{N}\Delta_{\ell_{i}A}\Phi_{(n|m)}$ and prove it by induction. The general expression will be
\begin{equation}
\prod_{i=1}^{N} \Delta_{\ell_{(N+1-i)}A} \Phi_{(n|m)} = \sum_{\ell=0}^{N+n} \sum_{\delta_{s}\in\Gamma\left(N,N+n+1-\ell\right)} \prod_{s=1}^{N} \Sigma_{0}^{(n|m)}(s)\, \Phi_{(\ell|m-n+\ell-\sum_{i=1}^{N}\ell_{i})},
\label{eq:AppB-prod-of-alien-ders-theta-zero}
\end{equation}
\noindent
with
\begin{equation}
\Sigma_{0}^{(n|m)} (s) = \left[ \left(m-\sum_{i=1}^{s}\ell_{i}+s+1-\delta_{s}\right) \widetilde{S}_{\ell_{s}}^{(\mathbf{d}\delta_{s}+\ell_{s})} + \left(s+n+1-\delta_{s}\right) S_{\ell_{s}}^{(\mathbf{d}\delta_{s})} \right] \Theta\left(s+n+1-\delta_{s}\right).
\label{eq:AppB-Coeff-Stokes-theta-zero}
\end{equation}
\noindent
The notation used above is the same as in \citep{Aniceto:2011nu} and as follows: the sum $\delta_{s}\in\Gamma(a,b)$ is over the partitions $1=\delta_{0}\le\delta_{1}\le\cdots\le\delta_{a}=b$; as in the one--parameter case $\mathbf{d}\delta_{s}=\delta_{s}-\delta_{s-1}$; and one sets $S_{0}^{(\ell)}=\widetilde{S}_{0}^{(\ell)}=S_{-\ell}^{(\ell)}=\widetilde{S}_{\ell}^{(\ell)}=0$ for any $\ell>0$. Finally, the function $\Theta(x)$ is the familiar Heaviside function, $\Theta(x)=1$ for $x\ge0$ and zero otherwise. It is straightforward to check that taking $N=1$ in our general expression \eqref{eq:AppB-prod-of-alien-ders-theta-zero} gives us the expected result \eqref{eq:AppB-one-alien-der-theta-zero}. Let us next consider \eqref{eq:AppB-prod-of-alien-ders-theta-zero} for general $N$, and check what follows once we act with one more
alien derivative. One has:
\begin{eqnarray}
\Delta_{\ell_{N+1}A} \prod_{i=1}^{N} \Delta_{\ell_{(N+1-i)}A} \Phi_{(n|m)} &=&  \sum_{\ell=0}^{N+n} \sum_{\delta_{s}\in\Gamma\left(N,N+n+1-\ell\right)} \prod_{s=1}^{N}\Sigma_{0}^{(n|m)}(s)\, \Delta_{\ell_{N+1}A} \Phi_{(\ell|m-n+\ell-\sum_{i=1}^{N}\ell_{i})} = \nonumber \\
&=&
\sum_{\ell=0}^{N+n} \sum_{\delta_{s}\in\Gamma\left(N,N+n+1-\ell\right)} \sum_{k=0}^{\ell+1}\, \prod_{s=1}^{N} \Sigma_{0}^{(n|m)} (s) \times \\
&&
\hspace{-70pt}
\times \left\{ k\, S_{\ell_{N+1}}^{(\ell-k+1)} + \left( m-n-\sum_{i=1}^{N+1}\ell_{i}+k \right) \widetilde{S}_{\ell_{N+1}}^{(\ell-k+\ell_{N+1}+1)} \right\} \Phi_{(k|m-n-\sum_{i=1}^{N+1}\ell_{i}+k)}.
\nonumber 
\end{eqnarray}
\noindent
Now make a change of variables $\tilde{\ell}=\ell+1$, and notice that taking $\tilde{\ell}=0$ gives a vanishing contribution (then we would also have $k=0$, and $\widetilde{S}_{\ell}^{(\ell)}=0$). Making use of the identity
\begin{equation}
\sum_{\tilde{\ell}=0}^{N+n+1}\, \sum_{k=0}^{\tilde{\ell}} = \sum_{k=0}^{N+1+n}\, \sum_{\tilde{\ell}=k}^{N+1+n},
\end{equation}
\noindent
one obtains
\begin{eqnarray}
\prod_{i=1}^{N+1} \Delta_{\ell_{(N+2-i)}A} \Phi_{(n|m)} &=& \sum_{k=0}^{N+n+1} \sum_{\tilde{\ell}=k}^{N+n+1} \sum_{\delta_{s}\in\Gamma\left(N,N+n+2-\tilde{\ell}\right)} \prod_{s=1}^{N} \Sigma_{0}^{(n|m)}(s) \times\\
&&
\hspace{-15pt}
\times \left\{ k\, S_{\ell_{N+1}}^{(\tilde{\ell}-k)} + \left( m-n-\sum_{i=1}^{N+1}\ell_{i}+k \right) \widetilde{S}_{\ell_{N+1}}^{(\tilde{\ell}-k+\ell_{N+1})} \right\} \Phi_{(k|m-n-\sum_{i=1}^{N+1}\ell_{i}+k)}. \nonumber 
\end{eqnarray}
\noindent
Finally we define $\delta_{N+1}\equiv N+n+2-k$, and perform the change of variables $\ell=N+n+2-\tilde{\ell}$ which takes values as $\ell=1,...,N+n+2-k=\delta_{N+1}$. It is then simple to get
\begin{eqnarray}
\prod_{i=1}^{N+1} \Delta_{\ell_{(N+2-i)}A} \Phi_{(n|m)} &=& \sum_{k=0}^{N+n+1} \sum_{\ell=1}^{\delta_{N+1}} \delta_{\delta_{N+1},N+n+2-k} \sum_{\delta_{s}\in\Gamma\left(N,\ell\right)} \prod_{s=1}^{N} \Sigma_{0}^{(n|m)}(s) \times \\
&&
\times \Bigg\{ \left(N+n+2-\delta_{N+1}\right) S_{\ell_{N+1}}^{(\delta_{N+1}-\ell)} + \nonumber \\
&& 
\hspace{-20pt}
+ \left( m+N+2-\delta_{N+1}-\sum_{i=1}^{N+1}\ell_{i} \right) \widetilde{S}_{\ell_{N+1}}^{(\delta_{N+1}-\ell+\ell_{N+1})} \Bigg\}\, \Phi_{(k|m-n-\sum_{i=1}^{N+1}\ell_{i}+k)}. \nonumber 
\end{eqnarray}
\noindent
Now if we evaluate \eqref{eq:AppB-Coeff-Stokes-theta-zero} for $s=N+1$, and noticing that $\Theta\left(N+n+2-\delta_{N+1}\right)=\Theta(k)=1$, and that $ $$\delta_{N}=\ell$, we finally obtain
\begin{eqnarray}
\prod_{i=1}^{N+1} \Delta_{\ell_{(N+2-i)}A} \Phi_{(n|m)} &=& \sum_{k=0}^{N+n+1} \sum_{\ell=1}^{\delta_{N+1}} \delta_{\delta_{N+1},N+n+2-k} \sum_{\delta_{s}\in\Gamma\left(N,\ell\right)} \prod_{s=1}^{N+1} \Sigma_{0}^{(n|m)}(s)\, \Phi_{(k|m-n-\sum_{i=1}^{N+1}\ell_{i}+k)} = \nonumber \\
&=& \sum_{k=0}^{N+n+1} \sum_{\delta_{s}\in\Gamma\left(N+1,N+n+2-k\right)} \prod_{s=1}^{N+1} \Sigma_{0}^{(n|m)}(s)\, \Phi_{(k|m-n-\sum_{i=1}^{N+1}\ell_{i}+k)}.
\end{eqnarray}
\noindent
This ends our proof of equation \eqref{eq:AppB-prod-of-alien-ders-theta-zero}.

Having this result in hand, one can finally write the Stokes automorphism as
\begin{equation}
\underline{\mathfrak{S}}_{0}^{\nu} \Phi_{(n|m)} = \Phi_{(n|m)} + \sum_{r=1}^{+\infty} \sum_{k=1}^{r} \frac{\rme^{-rAz}}{k!}\, \sum_{\ell=0}^{k+n} \sum_{\gamma_{i}\in\Gamma(k,r)} \sum_{\delta_{s}\in\Gamma\left(k,k+n+1-\ell\right)} \prod_{s=1}^{k} \Sigma_{0,\nu}^{(n|m)}(s)\, \Phi_{(\ell|m-n+\ell-r)}, \label{eq:2-param-Stokes0-in-Phi}
\end{equation}
\noindent
where we redefined the sum over the $\ell_{i}$ into a sum over partitions $\gamma_{i}=\ell_{1}+\cdots+\ell_{i}\in\Gamma(k,r)$, such that $\gamma_{0}=0$ and $\gamma_{i}>0$. The coefficients \eqref{eq:AppB-Coeff-Stokes-theta-zero} can also be now rewritten as (with $\nu=1$ being the usual case)
\begin{equation}
\Sigma_{0,\nu}^{(n|m)} (s) = \nu \left[ \left(m-\gamma_{s}+s+1-\delta_{s}\right) \widetilde{S}_{\mathbf{d}\gamma_{s}}^{(\mathbf{d}\delta_{s}+\mathbf{d}\gamma_{s})} + \left(s+n+1-\delta_{s}\right) S_{\mathbf{d}\gamma_{s}}^{(\mathbf{d}\delta_{s})} \right]\Theta \left(s+n+1-\delta_{s}\right).
\end{equation}
\noindent
Clearly, the result is now more complicated than in the one--parameter case.

%%%%%%%%%%%%%%%%%%%%%%%%%%%%%%%%%%%%%%%%%%%%%%%%%%%%%%%%%%%%%%%%%
\subsection*{Stokes Transition at $\theta=0$}
%%%%%%%%%%%%%%%%%%%%%%%%%%%%%%%%%%%%%%%%%%%%%%%%%%%%%%%%%%%%%%%%%

We now have the complete required information in order to construct the Stokes transition at $\theta=0$. Acting with the Stokes automorphism on the transseries itself, and using the results we just computed, one obtains
\begin{eqnarray}
\underline{\mathfrak{S}}_{0}^{\nu} F(z,\sigma_{1},\sigma_{2}) &=& \sum_{n,m=0}^{+\infty} \sigma_{1}^{n}\sigma_{2}^{m}\, \rme^{-\left(n-m\right)Az}\, \Phi_{(n|m)} + \label{eq:2-param-Stokes-trans-zero-first-form}\\
&&
\hspace{-70pt}
+\sum_{n,m=0}^{+\infty} \sigma_{1}^{n}\sigma_{2}^{m} \sum_{r=1}^{+\infty} \rme^{-\left(n+r-m\right)Az} \sum_{k=1}^{r} \sum_{\ell=0}^{k+n} \frac{1}{k!} 
\sum_{\gamma_{i}\in\Gamma(k,r)}\sum_{\delta_{s}\in\Gamma(k,k+n+1-\ell)} \prod_{s=1}^{k} \Sigma_{0,\nu}^{(n|m)}(s)\, \Phi_{(\ell|m-n-r+\ell)}. \nonumber 
\end{eqnarray}
\noindent
The first line of this expression may be included in the second one by simply introducing a factor $\delta_{k0}\delta_{r0}\delta_{\ell n}$. Further making two changes of variables, $\widetilde{m}=m+\ell-n-r$ and $\widetilde{n}=n+k$, recalling that $\Phi_{(a|b)}=0$ if either $a$ or $b$ is negative, and reshuffling the sums, it follows
\begin{eqnarray}
\underline{\mathfrak{S}}_{0}^{\nu} F &=& \sum_{\ell,\widetilde{m}=0}^{+\infty} \sigma_{1}^{\ell}\sigma_{2}^{\widetilde{m}}\, \rme^{-\left(\ell-\widetilde{m}\right)Az}\, \Phi_{(\ell|{\widetilde{m}})} \sum_{\widetilde{n}=\ell}^{+\infty} \sum_{k=0}^{\widetilde{n}} \sum_{r=k}^{+\infty} \sigma_{1}^{\widetilde{n}-\ell-k} \sigma_{2}^{\widetilde{n}-\ell-k+r} \times \\
&&
\times \left\{ \delta_{k0}\, \delta_{rk}\, \delta_{\widetilde{n},\ell+k} + \frac{1}{k!} \sum_{\gamma_{i}\in\Gamma(k,r)} \sum_{\delta_{s}\in\Gamma(k,\widetilde{n}+1-\ell)} \prod_{s=1}^{k} \Sigma_{0,\nu}^{(\widetilde{n}-k|\widetilde{m}+\widetilde{n}-k-\ell+r)}(s) \right\}. \nonumber 
\end{eqnarray}
\noindent
To get to our final expression, we still need to perform another change of variables, as $\widehat{n}=\widetilde{n}-\ell$ and $\widetilde{r}=r-k$. Dropping all tildes and hats (for ease of notation) we finally find
\begin{equation}
\underline{\mathfrak{S}}_{0}^{\nu} F = \sum_{m,\ell=0}^{+\infty} \sigma_{1}^{\ell}\sigma_{2}^{m}\, \BP_{0,\nu}^{(\ell|m)} (\sigma_{1},\sigma_{2})\, \rme^{-\left(\ell-m\right)Az}\, \Phi_{(\ell|m)},
\label{eq:App-Stokes-at-zero-2-param}
\end{equation}
\noindent
where the function implementing the transition on the parameters of the transseries is given by
\begin{eqnarray}
\BP_{0,\nu}^{(\ell|m)} (\sigma_{1},\sigma_{2}) &=& \sum_{n,r=0}^{+\infty} \sum_{k=0}^{n+\ell} \left(\sigma_{1}\sigma_{2}\right)^{n} \sigma_{1}^{-k}\sigma_{2}^{r}\, \times
\label{eq:App-Stokes-at-zero-2-param-extra} \\
&&
\times \left\{ \delta_{k0}\, \delta_{r0}\, \delta_{nk} + \frac{1}{k!} \sum_{\gamma_{i}\in\Gamma(k,r+k)} \sum_{\delta_{s}\in\Gamma(k,n+1)} \prod_{s=1}^{k} \Sigma_{0,\nu}^{(n+\ell-k|m+n+r)}(s) \right\}. \nonumber 
\end{eqnarray}

Now, the Stokes transition \eqref{eq:App-Stokes-at-zero-2-param} yields back a two--parameters transseries (as in \eqref{eq:app2parametersFtransseries}), in such a way that one may write
\begin{equation}
F \left(z,\widetilde{\sigma}_{1,\nu},\widetilde{\sigma}_{2,\nu}\right) = \underline{\mathfrak{S}}_{0}^{\nu} F \left(z,\sigma_{1},\sigma_{2}\right) = \sum_{n,m=0}^{+\infty} \sigma_{1}^{n}\sigma_{2}^{m}\, \BP_{0,\nu}^{(n|m)} (\sigma_{1},\sigma_{2})\, \rme^{-\left(n-m\right)Az}\, \Phi_{(n|m)}.
\end{equation}
\noindent
This means that the $\{ \widetilde{\sigma}_{i} \}$ have to satisfy
\begin{equation}
\widetilde{\sigma}_{1,\nu}^{n}\, \widetilde{\sigma}_{2,\nu}^{m} = \sigma_{1}^{n}\, \sigma_{2}^{m}\, \BP_{0,\nu}^{(n|m)} (\sigma_{1},\sigma_{2})
\end{equation}
\noindent
for every sector nonperturbative $(n|m)$. In particular, for the sectors $(1|0)$ and $(0|1)$ one has
\begin{eqnarray}
\widetilde{\sigma}_{1,\nu} &\equiv& \BS_{0,1}^{(\nu)} (\sigma_{1},\sigma_{2}) = \sigma_{1}\, \BP_{0,\nu}^{(1|0)} (\sigma_{1},\sigma_{2}), \\
\widetilde{\sigma}_{2,\nu} &\equiv& \BS_{0,2}^{(\nu)} (\sigma_{1},\sigma_{2}) = \sigma_{2}\, \BP_{0,\nu}^{(0|1)} (\sigma_{1},\sigma_{2}).
\end{eqnarray}
\noindent
Although we do not present an analytical proof, we have confirmed the validity of these relations with detailed symbolic computation evidence, further supporting that the following relation holds:
\begin{equation}
\BP_{0,\nu}^{(n|m)} (\sigma_{1},\sigma_{2}) = \left( \BP_{0,\nu}^{(1|0)} (\sigma_{1},\sigma_{2}) \right)^{n} \left( \BP_{0,\nu}^{(0|1)} (\sigma_{1},\sigma_{2}) \right)^{m}.
\end{equation}
\noindent
Then the Stokes transition in the direction $\theta=0$ can be finally written as
\begin{equation}
\underline{\mathfrak{S}}_{0}^{\nu} F \left(z,\sigma_{1},\sigma_{2}\right) = F \left( z, \BS_{0,1}^{(\nu)} (\sigma_{1},\sigma_{2}), \BS_{0,2}^{(\nu)} (\sigma_{1},\sigma_{2}) \right).
\label{eq:Stokes-transition-at-zero-2-param-strange-variables}
\end{equation}
\noindent
Notice that the structure is essentially the same as in the one--parameter case, it is only the functions implementing the transition which are now much harder to evaluate explicitly.

%%%%%%%%%%%%%%%%%%%%%%%%%%%%%%%%%%%%%%%%%%%%%%%%%%%%%%%%%%%%%%%%%
\subsection*{Stokes Automorphism at $\theta=\pi$}
%%%%%%%%%%%%%%%%%%%%%%%%%%%%%%%%%%%%%%%%%%%%%%%%%%%%%%%%%%%%%%%%%

Unlike the one--parameter case, where the directions $\theta=0$ and $\theta=\pi$ had distinct features, in this case, and due to the nature of the instanton actions as $\pm A$, the two directions are actually very similar. In particular, determining the Stokes automorphism in the $\theta=\pi$ singular direction follows an identical path to the $\theta=0$ case described above. One can notice a symmetry in every such expression, by changing $m\leftrightarrow n$, $S\leftrightarrow\widetilde{S}$ (and where $S_{\ell},\,\widetilde{S}_{\ell}$ now become $S_{-\ell},\widetilde{S}_{-\ell}$), the exponential of
negative powers to positive ones, and finally changing $\Phi_{(a|b)}\leftrightarrow\Phi_{(b|a)}$. As such, we will refrain from showing every step of the proof, and just state the end result.

The Stokes automorphism in the $\theta=\pi$ direction will be given by
\begin{equation}
\underline{\mathfrak{S}}_{\pi}^{\nu} \Phi_{(n|m)} = \Phi_{(n|m)} + \sum_{r=1}^{+\infty} \sum_{k=1}^{r} \frac{\rme^{rAz}}{k!} \sum_{\ell=0}^{k+m} \sum_{\gamma_{i}\in\Gamma(k,r)} \sum_{\delta_{s}\in\Gamma\left(k,k+m+1-\ell\right)} \prod_{s=1}^{k} \Sigma_{\pi,\nu}^{(n|m)}(s)\, \Phi_{(n-m+\ell-r|\ell)},
\end{equation}
\noindent
with
\begin{equation}
\Sigma_{\pi,\nu}^{(n|m)} (s) = \nu \left[ \left(n-\gamma_{s}+s+1-\delta_{s}\right) S_{-\mathbf{d}\gamma_{s}}^{(\mathbf{d}\delta_{s}+\mathbf{d}\gamma_{s})} + \left(s+m+1-\delta_{s}\right) \widetilde{S}_{-\mathbf{d}\gamma_{s}}^{(\mathbf{d}\delta_{s})} \right] \Theta \left(s+m+1-\delta_{s}\right).
\end{equation}

%%%%%%%%%%%%%%%%%%%%%%%%%%%%%%%%%%%%%%%%%%%%%%%%%%%%%%%%%%%%%%%%%
\subsection*{Stokes Transition at $\theta=\pi$}
%%%%%%%%%%%%%%%%%%%%%%%%%%%%%%%%%%%%%%%%%%%%%%%%%%%%%%%%%%%%%%%%%

As for the automorphism, also the Stokes transition along $\theta=\pi$ now follows in complete parallel with what was done for the direction $\theta=0$---one just has to use the results in the paragraphs above. Start with the action of the Stokes automorphism upon the transseries,
\begin{eqnarray}
\underline{\mathfrak{S}}_{\pi}^{\nu} F (z,\sigma_1,\sigma_2) &=& \sum_{n,m=0}^{+\infty} \sigma_{1}^{n}\sigma_{2}^{m}\, \rme^{-\left(n-m\right)Az}\, \Phi_{(n|m)}(z) + \\
&&
\hspace{-70pt}
+ \sum_{n,m=0}^{+\infty} \sigma_{1}^{n}\sigma_{2}^{m} \sum_{r=1}^{+\infty} \rme^{-\left(n-r-m\right)Az} \sum_{k=1}^{r} \sum_{\ell=0}^{k+m} \frac{1}{k!} \sum_{\gamma_{i}\in\Gamma(k,r)} \sum_{\delta_{s}\in\Gamma(k,k+m+1-\ell)} \prod_{s=1}^{k} \Sigma_{\pi,\nu}^{(n|m)} (s)\, \Phi_{(n-m+\ell-r|\ell)}. \nonumber 
\end{eqnarray}
\noindent
Now, akin to before, we perform consecutive changes of variables ($\widetilde{n}=n-m-r+\ell$, $\widetilde{m}=m+k$, $\widehat{m}=\widetilde{m}-\ell$ and $\widetilde{r}=r-k$) and reshuffle the sums. Then dropping tildes and hats for simplicity, we find the final result as
\begin{equation}
\underline{\mathfrak{S}}_{\pi}^{\nu} F = \sum_{n,\ell=0}^{+\infty} \sigma_{1}^{n}\sigma_{2}^{\ell}\, \BP_{\pi,\nu}^{(n|\ell)} (\sigma_{1},\sigma_{2})\, \rme^{-\left(n-\ell\right)Az}\, \Phi_{(n|\ell)},
\label{eq:App-Stokes-at-pi-2-param}
\end{equation}
\noindent
where the function implementing transition is now
\begin{eqnarray}
\BP_{\pi,\nu}^{(n|\ell)} (\sigma_{1},\sigma_{2}) &=& \sum_{m,r=0}^{+\infty} \sum_{k=0}^{m+\ell} \left(\sigma_{1}\sigma_{2}\right)^{m} \sigma_{1}^{r}\sigma_{2}^{-k} \times \label{eq:App-Stokes-at-pi-2-param-extra} \\
&&
\times \left\{ \delta_{k0}\, \delta_{r0}\, \delta_{mk} + \frac{1}{k!} \sum_{\gamma_{i}\in\Gamma(k,r+k)} \sum_{\delta_{s}\in\Gamma(k,m+1)} \prod_{s=1}^{k} \Sigma_{\pi,\nu}^{(n+m+r|m+\ell-k)}(s) \right\}. \nonumber
\end{eqnarray}

As for $\theta=0$, one may introduce more compact notation for this result. Defining
\be
F \left(z,\widetilde{\sigma}_{1,\nu},\widetilde{\sigma}_{2\nu}\right) = \underline{\mathfrak{S}}_{\pi}^{\nu} F  \left(z,\sigma_{1},\sigma_{2}\right),
\ee
\noindent
we may write
\be
\widetilde{\sigma}_{1,\nu}^{n} \widetilde{\sigma}_{2,\nu}^{m} = \sigma_{1}^{n} \sigma_{2}^{m}\, \BP_{\pi,\nu}^{(n|m)} (\sigma_{1},\sigma_{2})
\ee
\noindent
for every nonperturbative sector $(n|m)$. In particular, for the sectors $(1|0)$ and $(0|1)$ one has
\begin{eqnarray}
\widetilde{\sigma}_{1,\nu} &\equiv& \BS_{\pi,1}^{(\nu)} (\sigma_{1},\sigma_{2}) = \sigma_{1}\, \BP_{\pi,\nu}^{(1|0)} (\sigma_{1},\sigma_{2}), \\
\widetilde{\sigma}_{2,\nu} &\equiv& \BS_{\pi,2}^{(\nu)} (\sigma_{1},\sigma_{2}) = \sigma_{2}\, \BP_{\pi,\nu}^{(0|1)} (\sigma_{1},\sigma_{2}).
\end{eqnarray}
\noindent
Again, we have found strong computational evidence supporting this structure. The Stokes transition in the direction $\theta=\pi$ is then given by
\begin{equation}
\underline{\mathfrak{S}}_{\pi}^{\nu} F \left(z,\sigma_{1},\sigma_{2}\right) = F \left( z, \BS_{\pi,1}^{(\nu)} (\sigma_{1},\sigma_{2}), \BS_{\pi,2}^{(\nu)} (\sigma_{1},\sigma_{2}) \right).
\label{eq:Stokes-transition-at-pi-2-param-strange-variables}
\end{equation}

%%%%%%%%%%%%%%%%%%%%%%%%%%%%%%%%%%%%%%%%%%%%%%%%%%%%%%%%%%%%%%%%%
\subsection*{Cancelation of the Nonperturbative Ambiguity}
%%%%%%%%%%%%%%%%%%%%%%%%%%%%%%%%%%%%%%%%%%%%%%%%%%%%%%%%%%%%%%%%%

Having understood the structure of Stokes transitions at both Stokes lines, $\theta=0$ and $\theta=\pi$, also in the present two--parameters setting, we may now understand how is the ambiguity canceled. Again, this point is discussed at length in the main text; in here we wish to present the many exact/explicit formulae. Furthermore, the line of analysis will be similar to the one in the previous appendix, now having in mind the two--parameters transseries \eqref{eq:Ftransseries} or \eqref{eq:app2parametersFtransseries}. The ambiguity is essentially encoded in
\begin{eqnarray}
\im\, F &\equiv& \frac{1}{2\rmi} \left( \mathcal{S}_{0^+}-\mathcal{S}_{0^-} \right) F = \sum_{n,m=0}^{+\infty} \left( \im \left(\sigma_{1}^{n}\sigma_{2}^{m} \right) \re\, F^{(n|m)} + \re \left( \sigma_{1}^{n}\sigma_{2}^{m} \right) \im\, F^{(n|m)} \right),
\label{eq:App-imaginary-amb-for-2-param-transseries}
\end{eqnarray}
\noindent
with the usual
\begin{equation}
\im\, F^{(n|m)} := \frac{1}{2\rmi} \left(\mathcal{S}_{0^+}-\mathcal{S}_{0^-}\right) F^{(n|m)}, \qquad \re\, F^{(n|m)} := \frac{1}{2} \left(\mathcal{S}_{0^+}+\mathcal{S}_{0^-}\right) F^{(n|m)}.
\end{equation}
\noindent
One may similarly write the real contribution to the transseries as
\begin{equation}
\re\, F \equiv \frac{1}{2} \left( \mathcal{S}_{0^+}+\mathcal{S}_{0^-} \right) F = \sum_{n,m=0}^{+\infty} \left( \re \left(\sigma_{1}^{n}\sigma_{2}^{m} \right) \re\, F^{(n|m)} - \im \left( \sigma_{1}^{n}\sigma_{2}^{m} \right) \im\, F^{(n|m)} \right).
\end{equation}

The first thing to do is to use the Stokes automorphism to explicitly write the ambiguity of each sector, $\im\, F^{(n|m)}$, in terms of the real contributions which will later appear in the median resummation, $\re\, F^{(n'|m')}$. Very similarly to what we did in the previous appendix,
\begin{equation}
\left( \mathcal{S}_{0^+} - \mathcal{S}_{0^-} \right) F^{(n|m)}(z) = - \mathcal{S}_{0^-} \circ \left( \mathbf{1} - \underline{\mathfrak{S}}_{0} \right) F^{(n|m)}(z),
\end{equation}
\noindent
where $\underline{\mathfrak{S}}_{0} F^{(n|m)}(z) = \rme^{-\left(n-m\right)Az}\, \underline{\mathfrak{S}}_{0} \Phi_{(n|m)}(z)$ was already computed. Explicitly,
\begin{equation}
\left( \mathcal{S}_{0^+}-\mathcal{S}_{0^-} \right) F^{(n|m)}(z) = \sum_{r=1}^{+\infty}\sum_{s=0}^{n+r} \underline{\Sigma}^{(n|m)}(r,s)\, \mathcal{S}_{0^-}F_{(s|m-n+s-r)}(z),
\label{eq:App-Sp-Sm-of-F_(l|m)}
\end{equation}
\noindent
where we introduced
\begin{equation}
\underline{\Sigma}^{(n|m)}(r,s) = \sum_{k=\max(s-n,1)}^{r} \frac{1}{k!} \sum_{\gamma_{i}\in\Gamma(k,r)} \sum_{\delta_{p}\in\Gamma(k,k+n+1-s)} \prod_{p=1}^{k} \Sigma_{0}^{(n|m)}(p).
\end{equation}
\noindent
Using the relation
\begin{equation}
\mathcal{S}_{0^-} = \re - \frac{1}{2} \left(\mathcal{S}_{0^+}-\mathcal{S}_{0^-}\right),
\end{equation}
\noindent
we can in principle solve for the expression above recursively, much like it was done in the one--parameter case. However, we shall be more interested in specific examples as in here they are more illuminating than a closed form expression. Let us determine the contributions to the nonperturbative ambiguity for the perturbative series $F^{(0|0)}$. Starting from \eqref{eq:App-Sp-Sm-of-F_(l|m)}, we find
\begin{equation}
2\rmi\, \im\, F^{(0|0)} = \sum_{r=1}^{+\infty} (S_{1}^{(0)})^{r} \left( \re\, F^{(r|0)} - \frac{1}{2} \left(\mathcal{S}_{0^+}-\mathcal{S}_{0^-}\right) F^{(r|0)} \right),
\end{equation}
\noindent
where we made use of the fact that $\underline{\Sigma}^{(0|0)}(r,r)=(S_{1}^{(0)})^{r}$. It is clear that $\im\, F^{(0|0)}$ will have contributions arising from the imaginary part of the instanton series, $\im\, F^{(n|0)}$, which may then be determined in much the same way
\begin{equation}
2\rmi\, \im\, F^{(n|0)} = \sum_{r=1}^{+\infty} (S_{1}^{(0)})^{r}\, \binom{n+r}{r} \left( \re\, F^{(n+r|0)} - \frac{1}{2} \left(\mathcal{S}_{0^+}-\mathcal{S}_{0^-}\right) F^{(n+r|0)} \right).
\end{equation}
\noindent
It is very easy to see that this is completely analogous to what happened in the one--parameter case studied before: the contributions to $\im\, F^{(0|0)}$ are only of the type $\re\, F^{(n|0)}$. Proceeding with the analogy, the tendency at this point would be to assume that the same cancelation would then occur within these terms. However, the mechanism canceling the nonperturbative ambiguity of the two--parameters transseries is not as straightforward as within the one--parameter case. Write the expansion of the transseries ambiguity $\im\, F$ in real terms, $\re\, F^{(n|m)}$, and focus on all the contributions to one such term, as these will need to cancel between each other. For example, let us determine all factors contributing to $\re\, F^{(\alpha|0)}$. From \eqref{eq:App-Sp-Sm-of-F_(l|m)} we see that such terms will arise from $\im\, F^{(n|m)}$ whenever
\begin{equation}
2\rmi\, \im\, F^{(n|m)} \approx \underline{\Sigma}^{(\alpha|0)} (m-n+\alpha,\alpha)\, \mathcal{S}_{0^-}F^{(\alpha|0)}, \qquad m \ge 1+n-\alpha.
\end{equation}
\noindent
But it is not only the original contributions in \eqref{eq:App-imaginary-amb-for-2-param-transseries} which will contribute at this level. Analyzing \eqref{eq:App-Sp-Sm-of-F_(l|m)} and noting that
\begin{equation}
\mathcal{S}_{0^-} F^{(a|b)} = \re\, F^{(a|b)} - \rmi\,\im\, F^{(a|b)},
\end{equation}
\noindent
there will be terms within each $\mathcal{S}_{0^-} F^{(a|b)} $, coming from the $\im\, F^{(a|b)}$, which will again contribute to the original term $\re\, F^{(\alpha|0)}$. A  detailed analysis of these results is presented in the main text.

%%%%%%%%%%%%%%%%%%%%%%%%%%%%%%%%%%%%%%%%%%%%%%%%%%%%%%%%%%%%%%%%%
\subsection*{Contributions to the Ambiguity in Table \ref{tab:Cancelations-2-parameters}}
%%%%%%%%%%%%%%%%%%%%%%%%%%%%%%%%%%%%%%%%%%%%%%%%%%%%%%%%%%%%%%%%%

Finally, we list the non--zero coefficients in the expansions $\im\,  F^{(n|m)} \sim \sum_{a,b} C_{(n|m)}^{(a|b)}\, \re\, F^{(a|b)}$, which appear in table \ref{tab:Cancelations-2-parameters}, up to $a=4$, $b=2$. The non--zero coefficients from the term $\im\, F^{(0|0)}$ are:
\begin{eqnarray}
\label{eq:Coefficients-of-expansion-imF00-2-parameter}
C_{(0|0)}^{(1|0)} & = & \frac{1}{2\rmi}\, S_{1}^{(0)}, \\
C_{(0|0)}^{(3|0)} & = & -\frac{1}{4\rmi} \left(S_{1}^{(0)}\right)^{3}.
\end{eqnarray}
\noindent
From the term $\re\, \sigma_{1}\, \im\, F^{(1|0)}$ we find:
\begin{eqnarray}
C_{(1|0)}^{(2|0)} & = & \frac{1}{\rmi}\, S_{1}^{(0)}\, \re\, \sigma_{1}, \\
C_{(1|0)}^{(4|0)} & = & -\frac{1}{\rmi} \left(S_{1}^{(0)}\right)^{3} \re\, \sigma_{1}.\label{eq:Coefficients-of-expansion-imF10-2-parameter}
\end{eqnarray}
\noindent
The next term appearing in the table is $\re\, \sigma_{2}\, \im\, F^{(0|1)}$, whose non--zero coefficients are:
\begin{eqnarray}
C_{(0|1)}^{(1|0)} & = & \frac{1}{2\rmi}\, S_{2}^{(0)}\, \re\, \sigma_{2}, \\
C_{(0|1)}^{(2|0)} & = & -\frac{1}{4\rmi} \left(S_{1}^{(0)}\right)^{2} S_{1}^{(1)}\, \re\, \sigma_{2}, \\
C_{(0|1)}^{(3|0)} & = & -\frac{3}{4\rmi} \left(S_{1}^{(0)}\right)^{2} S_{2}^{(0)}\, \re\, \sigma_{2}, \label{eq:Coefficients-of-expansion-imF01-2-parameter} \\
C_{(0|1)}^{(4|0)} & = & \frac{1}{\rmi} \left(S_{1}^{(0)}\right)^{4} S_{1}^{(1)}\, \re\, \sigma_{2}, \\
C_{(0|1)}^{(1|1)} & = & \frac{1}{2\rmi}\, S_{1}^{(0)}\, \re\, \sigma_{2}, \\
C_{(0|1)}^{(3|1)} & = & -\frac{1}{4\rmi} \left(S_{1}^{(0)}\right)^{3} \re\, \sigma_{2}. 
\end{eqnarray}
\noindent
From $\re\, \sigma_{1}^{2}\, \im\, F^{(2|0)}$ we have the following non--zero coefficient:
\begin{eqnarray}
C_{(2|0)}^{(3|0)} & = & \frac{3}{2\rmi}\, S_{1}^{(0)}\, \re\, \sigma_{1}^{2}.\label{eq:Coefficients-of-expansion-imF20-2-parameter}
\end{eqnarray}
\noindent
The term $\re \left(\sigma_{1}\sigma_{2}\right) \im\, F^{(1|1)}$ contributes with:
\begin{eqnarray}
C_{(1|1)}^{(1|0)} & = & \frac{1}{2\rmi}\, S_{1}^{(1)}\, \re \left( \sigma_{1}\sigma_{2} \right), \\
C_{(1|1)}^{(2|0)} & = & \frac{1}{\rmi}\, S_{2}^{(0)}\, \re \left( \sigma_{1}\sigma_{2} \right), \\
C_{(1|1)}^{(3|0)} & = & -\frac{3}{2\rmi} \left(S_{1}^{(0)}\right)^{2} S_{1}^{(1)}\, \re \left( \sigma_{1}\sigma_{2} \right),
\label{eq:Coefficients-of-expansion-imF11-2-parameter} \\
C_{(1|1)}^{(4|0)} & = & -\frac{3}{\rmi} \left(S_{1}^{(0)}\right)^{2} S_{2}^{(0)}\, \re \left( \sigma_{1}\sigma_{2} \right), \\
C_{(1|1)}^{(2|1)} & = & \frac{1}{\rmi}\, S_{1}^{(0)}\, \re \left( \sigma_{1}\sigma_{2} \right), \\
C_{(1|1)}^{(4|1)} & = & -\frac{1}{\rmi} \left(S_{1}^{(0)}\right)^{3} \re \left( \sigma_{1}\sigma_{2} \right). 
\end{eqnarray}
\noindent
For $\re\, \sigma_{2}^{2}\, \im\, F^{(0|2)}$ we find:
\begin{eqnarray}
C_{(0|2)}^{(1|0)} & = & -\frac{1}{24\rmi} \left( 2 \left(S_{1}^{(0)}\right)^{2} S_{1}^{(2)} + S_{1}^{(0)}S_{1}^{(1)} \left( 2\widetilde{S}_{1}^{(2)}+S_{1}^{(1)} \right) -  12 S_{3}^{(0)} \right) \re\, \sigma_{2}^{2}, \\
C_{(0|2)}^{(2|0)} & = & -\frac{1}{4\rmi}\, S_{1}^{(0)} \left( 2 S_{1}^{(1)}S_{2}^{(0)} + S_{1}^{(0)}S_{2}^{(1)} + S_{2}^{(0)}\widetilde{S}_{1}^{(2)} \right) \re\, \sigma_{2}^{2}, \\
C_{(0|2)}^{(3|0)} & = & \frac{1}{8\rmi}\, S_{1}^{(0)} \left( 4 \left(S_{1}^{(0)}\right)^{3} S_{1}^{(2)}  -6 \left(S_{2}^{(0)}\right)^{2} - 6 S_{1}^{(0)}S_{3}^{(0)} + \right. \nonumber \\
&&
\hspace{120pt}
\left. + \left(S_{1}^{(0)}\right)^{2} S_{1}^{(1)} \left( 4\widetilde{S}_{1}^{(2)} + 5S_{1}^{(1)} \right) \right) \re\, \sigma_{2}^{2}, \\
C_{(0|2)}^{(4|0)} & = & \frac{1}{\rmi} \left(S_{1}^{(0)}\right)^{3} \left( 4S_{1}^{(1)}S_{2}^{(0)} + S_{1}^{(0)}S_{2}^{(1)} + S_{2}^{(0)}\widetilde{S}_{1}^{(2)} \right) \re\, \sigma_{2}^{2}, \\
C_{(0|2)}^{(0|1)} & = & \frac{1}{2\rmi}\, \widetilde{S}_{1}^{(2)}\, \re\, \sigma_{2}^{2},
\label{eq:Coefficients-of-expansion-imF02-2-parameter} \\
C_{(0|2)}^{(1|1)} & = & \frac{1}{2\rmi}\, S_{2}^{(0)}\, \re\, \sigma_{2}^{2}, \\
C_{(0|2)}^{(2|1)} & = & -\frac{1}{4\rmi} \left(S_{1}^{(0)}\right)^{2} \left(S_{1}^{(1)} + \widetilde{S}_{1}^{(2)}\right) \re\, \sigma_{2}^{2}, \\
C_{(0|2)}^{(3|1)} & = & -\frac{3}{4\rmi} \left(S_{1}^{(0)}\right)^{2} S_{2}^{(0)}\, \re\, \sigma_{2}^{2}, \\
C_{(0|2)}^{(4|1)} & = & \frac{1}{2\rmi} \left(S_{1}^{(0)}\right)^{4} \left( 2S_{1}^{(1)}+\widetilde{S}_{1}^{(2)} \right) \re\, \sigma_{2}^{2}, \\
C_{(0|2)}^{(1|2)} & = & \frac{1}{2\rmi}\, S_{1}^{(0)}\, \re\, \sigma_{2}^{2}, \\
C_{(0|2)}^{(3|2)} & = & -\frac{1}{4\rmi} \left(S_{1}^{(0)}\right)^{3} \re\, \sigma_{2}^{2}. 
\end{eqnarray}
\noindent
The next term of interest is $\re \left(\sigma_{1}^{2}\sigma_{2}\right) \im\, F^{(2|1)}$, which contributes with the non--zero coefficients:
\begin{eqnarray}
C_{(2|1)}^{(2|0)} & = & \frac{1}{\rmi}\, S_{1}^{(1)}\, \re\left(\sigma_{1}^{2}\sigma_{2}\right), \\
C_{(2|1)}^{(3|0)} & = & \frac{3}{2\rmi}\, S_{2}^{(0)}\, \re\left(\sigma_{1}^{2}\sigma_{2}\right), \\
C_{(2|1)}^{(4|0)} & = & -\frac{9}{2\rmi} \left(S_{1}^{(0)}\right)^{2} S_{1}^{(1)}\, \re\left(\sigma_{1}^{2}\sigma_{2}\right),
\label{eq:Coefficients-of-expansion-imF21-2-parameter} \\
C_{(2|1)}^{(3|1)} & = & \frac{3}{2\rmi}\, S_{1}^{(0)}\, \re\left(\sigma_{1}^{2}\sigma_{2}\right). 
\end{eqnarray}
\noindent
The last term appearing in table \ref{tab:Cancelations-2-parameters} is $\re\left(\sigma_{1}\sigma_{2}^{2}\right) \im\, F^{(1|2)}$, for which we have:
\begin{eqnarray}
C_{(1|2)}^{(1|0)} & = & \frac{1}{2\rmi}\, S_{2}^{(1)}\, \re\left(\sigma_{1}\sigma_{2}^{2}\right), \\
C_{(1|2)}^{(2|0)} & = & \frac{1}{12\rmi} \left(-8\left(S_{1}^{(0)}\right)^{2}S_{1}^{(2)}+12S_{3}^{(0)}-S_{1}^{(0)}S_{1}^{(1)}\left(5\widetilde{S}_{1}^{(2)}+7S_{1}^{(1)}\right)\right) \re\left(\sigma_{1}\sigma_{2}^{2}\right), \\
C_{(1|2)}^{(3|0)} & = & -\frac{3}{4\rmi}\, S_{1}^{(0)} \left(4S_{1}^{(1)}S_{2}^{(0)}+2S_{1}^{(0)}S_{2}^{(1)}+S_{2}^{(0)}\widetilde{S}_{1}^{(2)}\right) \re\left(\sigma_{1}\sigma_{2}^{2}\right), \\
C_{(1|2)}^{(4|0)} & = & \frac{1}{2\rmi}\, S_{1}^{(0)} \left(8\left(S_{1}^{(0)}\right)^{3}S_{1}^{(2)}-6\left(S_{2}^{(0)}\right)^{2}-6S_{1}^{(0)}S_{3}^{(0)}- \right. \nonumber \\
&&
\hspace{120pt}
\left.
-\left(S_{1}^{(0)}\right)^{2}S_{1}^{(1)}\left(6\widetilde{S}_{1}^{(2)}+13S_{1}^{(1)}\right)\right) \re\left(\sigma_{1}\sigma_{2}^{2}\right), \\
C_{(1|2)}^{(1|1)} & = & \frac{1}{2\rmi} \left(S_{1}^{(1)}+\widetilde{S}_{1}^{(2)}\right) \re \left(\sigma_{1}\sigma_{2}^{2}\right), \\
C_{(1|2)}^{(2|1)} & = & \frac{1}{2\rmi}\, S_{2}^{(0)}\, \re \left(\sigma_{1}\sigma_{2}^{2}\right), \\
C_{(1|2)}^{(3|1)} & = & -\frac{3}{4\rmi} \left(S_{1}^{(0)}\right)^{2} \left(2S_{1}^{(1)}+\widetilde{S}_{1}^{(2)}\right) \re \left(\sigma_{1}\sigma_{2}^{2}\right), \\
C_{(1|2)}^{(4|1)} & = & -\frac{3}{\rmi} \left(S_{1}^{(0)}\right)^{2} S_{2}^{(0)}\, \re\left(\sigma_{1}\sigma_{2}^{2}\right), \\
C_{(1|2)}^{(2|2)} & = & \frac{1}{\rmi}\, S_{1}^{(0)}\, \re\left(\sigma_{1}\sigma_{2}^{2}\right), \\
C_{(1|2)}^{(4|2)} & = & -\frac{1}{\rmi} \left(S_{1}^{(0)}\right)^{3} \re \left(\sigma_{1}\sigma_{2}^{2}\right).
\label{eq:Coefficients-of-expansion-imF12-2-parameter}
\end{eqnarray}
\noindent
Finally, the last term we analyzed was $\re \left(\sigma_{1}^{2}\sigma_{2}^{2}\right) \im\, F^{(2|2)}$, for which we have:
\begin{eqnarray}
\label{eq:Coefficients-of-expansion-imF22-2-parameter-alpha}
C_{(2|2)}^{(2|0)} & = & \frac{1}{2\rmi}\, S_{1}^{(2)}\, \re\left(\sigma_{1}^{2}\sigma_{2}^{2}\right), \\
C_{(2|2)}^{(3|0)} & = & \frac{1}{2\rmi} \left(-5\left(S_{1}^{(0)}\right)^{2}S_{1}^{(2)}+3S_{3}^{(0)}-\frac{1}{4}S_{1}^{(0)}S_{1}^{(1)}\left(19S_{1}^{(1)}+8\widetilde{S}_{1}^{(2)}\right)\right) \re\left(\sigma_{1}^{2}\sigma_{2}^{2}\right), \\
C_{(2|2)}^{(4|0)} & = & -\frac{3}{2\rmi}\, S_{1}^{(0)} \left(3S_{1}^{(0)}S_{2}^{(1)}+S_{2}^{(0)}\left(\widetilde{S}_{1}^{(2)}+6S_{1}^{(1)}\right)\right) \re \left(\sigma_{1}^{2}\sigma_{2}^{2}\right), \\
C_{(2|2)}^{(2|1)} & = & \frac{1}{2\rmi} \left(\widetilde{S}_{1}^{(2)}+2S_{1}^{(1)}\right) \re \left(\sigma_{1}^{2}\sigma_{2}^{2}\right), \\
C_{(2|2)}^{(3|1)} & = & \frac{3}{2\rmi}\, S_{2}^{(0)}\, \re \left(\sigma_{1}^{2}\sigma_{2}^{2}\right), \\
C_{(2|2)}^{(4|1)} & = & -\frac{3}{2\rmi} \left(S_{1}^{(0)}\right)^{2} \left(\widetilde{S}_{1}^{(2)}+3S_{1}^{(1)}\right) \re \left(\sigma_{1}^{2}\sigma_{2}^{2}\right), \\
C_{(2|2)}^{(3|2)} & = & \frac{3}{2\rmi}\, S_{1}^{(0)}\, \re \left(\sigma_{1}^{2}\sigma_{2}^{2}\right).
\label{eq:Coefficients-of-expansion-imF22-2-parameter}
\end{eqnarray}

\newpage

%%%%%%%%%%%%%%%%%%%%%%%%%%%%%%%%%%%%%%%%%%%%%%%%%%%%%%%%%%%%%%%%%
%%%%%%%%%%%%%%%%%%%%%%%%%%%%%%%%%%%%%%%%%%%%%%%%%%%%%%%%%%%%%%%%%

\bibliographystyle{my-JHEP-3}
\bibliography{Ambiguity_bib}

\end{document}